\documentclass[a4paper,11pt]{article}
\pdfoutput=1 % if your are submitting a pdflatex (i.e. if you have
\usepackage{jcappub} % for details on the use of the package, please
\usepackage{array}    
\usepackage[T1]{fontenc} % if needed
\usepackage{url}
\usepackage{hyperref}
\usepackage{latexsym}
\usepackage{amssymb}
\usepackage{graphicx}
\usepackage{color}

\usepackage{mathptmx}       % selects Times Roman as basic font
\usepackage{helvet}         % selects Helvetica as sans-serif font
\usepackage{courier}        % selects Courier as typewriter font
\usepackage{type1cm}        % activate if the above 3 fonts are
\usepackage{makeidx}         % allows index generation
\usepackage{multicol}        % used for the two-column index
\usepackage[bottom]{footmisc}% place

\title{Perturbations in Tachyon Dark Energy and their Effect on Matter Clustering}

\author[a]{Avinash Singh}
\author[a]{H. K. Jassal}
\author[a,b]{Manabendra Sharma}

\affiliation[a]{Indian Institute of Science Education and Research,
				Mohali,\\ SAS Nagar, Mohali 140306, Punjab, India.}
\affiliation[b]{Institute for Theoretical Physics \& Cosmology,
				Zhejiang University of Technology,\\
				 Hangzhou, 310023, China.}

\emailAdd{avinashsingh@iisermohali.ac.in}
\emailAdd{hkjassal@iisermohali.ac.in}
\emailAdd{manabendra@zjut.edu.cn}

\abstract{A non-canonical scalar tachyon field is a viable
candidate for dark energy and has been found to be in good agreement
with observational data. 
Background data alone cannot
completely rule out degeneracy between this model and others.
To further constrain  the parameters, apart from the  distance
measurements, we study perturbations in tachyon scalar field and
how they affect matter clustering. 
We consider two tachyon potentials for this study, an inverse square
potential and an exponential potential.  
We study the evolution of the gravitational potential, matter
density contrast and dark energy density contrast, and compare 
them with the evolution in the $\Lambda CDM $ model. 
Although perturbations in dark energy at sub-Hubble scales are
negligible in comparison  with matter perturbations, they cannot be
ignored at Hubble and super-Hubble scales ($\lambda_p > 1000$ Mpc).
We also study the evolution of growth function and growth rate of
matter, and find that the growth rate is significantly suppressed
in dark energy dominated era with respect to the growth rate for $\Lambda
CDM$ model. 
A comparison of these models with Redshift Space Distortion
growth rate data is presented by way of calculating $f\sigma_8(z)$.
There is a tension of $2.9\sigma$ ($2.26\sigma$ ) between growth rate data and
Planck-2015 (Planck-2018) Cosmic Microwave Background Radiation data for $\Lambda
CDM$ model.   
We present constraints on free parameters of these models and show that 
perturbations in tachyon scalar field reduce this tension
between different data sets.}

\makeindex

\begin{document}
\maketitle
\flushbottom

\section{Introduction}
    \label{sec::intro}
Cosmological observations, which include observation of Supernova Type
Ia~\cite{Suzuki2012,Perlmutter1997,Perlmutter1999,Riess1998},
Baryon Acoustic Oscillations~\cite{Seo2003,Percival2007,sca2013,Blake2012},
Cosmic Microwave Background~\cite{Ade:2013sjv,planck2016}, etc.,
indicate a late-time acceleration of the Universe.
This acceleration can  be explained by considering the energy
density of the Universe to be dominated by a  negative pressure
medium~\cite{planck2016}. 
One of the main goals of modern cosmology is to explain, whether
the equation of state parameter $w$ is constant or a dynamical quantity. 
There is a large number of models which are able to describe the
acceleration.
The most intuitive is the cosmological constant model ($\Lambda CDM$ model)
~\cite{Carroll1992,Carroll:2000fy},
with the equation of state parameter $w=-1$, in which a constant $\Lambda$ representing vacuum energy
density, is understood to be the reason of the late-time acceleration. 
Although this model shows good agreement with the observations
~\cite{planck2016,Suzuki2012,Blake2012}, it suffers from theoretical
problems like the fine-tuning problem and
the coincidence problem~\cite{Carroll:2000fy,Weinberg1989,Paddy2003,Peebles2003}  . 
On the other hand observations do not rule out $w\neq-1$ and in
general the equation of state parameter can be a function of the scale
factor.

Dynamical dark energy models are an alternative to $\Lambda CDM$ model
and can have  an evolving equation of state parameter.
These models include the barotropic fluid models, canonical and non-canonical
scalar field models, etc.
A varying, fluid dark energy equation of state parameter is considered to be
a function of redshift or scale factor.
There are two parameters, the present day value of the equation of state
parameter, $w_0$, and the value of its derivative, $w^\prime_0$.
Detailed studies of the background evolution and constraints on the parameters for 
these models have been done 
in~\cite{JBP2005,Efstathiou1999,Lee2005,Tripathi:2016slv,
Sangwan:2017kxi,ZHENG2017,Rezaei2017,Yang2018}.
Quintessence scalar field is also a potential candidate for dark energy.
Using a slow rolling potential, the late-time accelerated
expansion can be achieved.  
The background cosmology in the presence of the canonical scalar field 
has been studied  in ~\cite{Ratra:1987rm,Linder:2007wa,Huterer:2006mv,Zlatev:1998tr,
Copeland:1997et,Sangwan:2018zpz,Watson:2003kk,Scherrer:2007pu}.
In~\cite{unniJassalSesh2008}, it was  shown that a homogeneous
quintessence field with inhomogeneous matter is inconsistent with
observation.
Therefore the scalar field must be perturbed in the course of
evolution of the Universe. 
The perturbations in the quintessence field, its dynamics, and its
effect on the evolution of matter clustering have been studied
in~\cite{unniJassalSesh2008,Jassal2009,Jassal2010,Jassal2012}. 

A potential alternative to the canonical scalar field and the fluid
model is a non-canonical scalar field model known as the tachyon model. 
Tachyon scalar field arises as a decay mode of D-branes in string
theory~\cite{asen2002a,asen2002b,asen2002c}.  
The background cosmology for this model has been studied
in~\cite{BJP2003,Paddy2002,Calcagni2006} and it is potentially 
a good candidate for dark energy. 
Tachyon scalar field has also been used to explain 
inflation~\cite{FAIRBAIRN2002,KofmanLinde2002,Feinstein2002,RezazadehKaramiHashemi2017,
Nandinii_et_all2015,Fei_et_al2017,GaoGongFei2018,Nndinii_et_al2018}.
Since its equation of state becomes dust like in the course of time, it is also
considered  a viable candidate for dark
matter~\cite{PadyChoudhury2002,asen2002b,asen2002c,SugimotoTerashima2002,Das2004,Davies2004,Makukov2016}. 
The tachyon model is in good agreement
with current observations \cite{AAJ2019}; data puts tight constraints on 
cosmological parameters and reduces the fine-tuning problem.
It can not however completely distinguish this model from the $\Lambda
CDM$ and other models.  
Perturbation in dark energy can potentially break the degeneracy
between models, for instance via the Integrated Sachs-Wolf Effect (ISW
effect) as it affects the low $l$ CMB angular power
spectrum~\cite{WellerLewis2003,BeanDor2004}. 
    
In this paper, we analyze the dynamics and nature of tachyon
perturbations and their effect on the evolution of matter perturbations.
We begin with a homogeneous tachyon scalar field and allow it to get
perturbed, as the matter clustering grows with time. 
In this analysis, we consider two tachyon potentials, an inverse
square potential and an exponential potential, and solve linearized
Einstein's equations. 
The clustering of dark energy is a scale dependent phenomena, it is higher
at larger scales, just opposite to the matter clustering which is higher at 
shorter scales.
Dark energy perturbations are insignificant with respect to matter clustering
at sub-Hubble scales, and dark energy can be considered homogeneous. 
At Hubble and super-Hubble scales, dark energy perturbations are significant
when compared with the matter perturbation.
However, as the present value of the equation of state $w_{\phi 0} \to -1$,
it can be considered homogeneous and this model coincides with the $\Lambda CDM$ model.

We also study the linear growth rate $f(z)$ of matter clustering for these
models and compare our theoretical computation with the redshift space
distortion (RSD) data.
We find that initially, in matter dominated era, growth rate is
higher for tachyon model than it is for $\Lambda CDM$ model, but in
dark energy dominated era the situation is opposite. 
This makes tachyon model a better alternate to fit growth rate data.
We use the `Gold-2017' RSD data compiled and tabulated
in~\cite{Nesseris2017} with some additional data from~\cite{Zhao_etal2019}. 
The growth rate measurements from RSD provide the value of $f\sigma_8(z)$,
where $\sigma_8(z)$ is the root mean square fluctuation in the matter power
spectrum in a sphere of radius $8~h^{-1} Mpc$.  
In~\cite{Nesseris2017}, it has been shown that there is a tension of $>3\sigma$
between `Gold-2017' and Planck-2015 data for $\Lambda CDM$ model.
We find that this tension still exists between the RSD data we use and
Planck-2018 data for $\Lambda CDM$ model.
We show that, for tachyon models, this tension is reduced when
equation of state parameter $w_{\phi 0}$ is larger than $-1$ and dark energy
is allowed to get perturbed.

In section~\ref{sec::BGcosmology} we present the equations for 
background tachyon model and introduce two potentials.
Perturbations in the tachyon scalar field and the matter part are
introduced in  section~\ref{sec::prtbd_eq}.
We have discussed our numerical approach in section~\ref{sec::num_methods}, 
and the results of our analysis have been shown in section~\ref{sec::resultDiscussion}.
Finally, we summarize our results in  section~\ref{sec::summaryConclusion}.
%%%%%%%%%%%%%%%%%%%%%%%%%%%%%%%%%%%%%%%%%%%%%%%%%%%%%%%%%%%%%%%%%%%%%%%%%%%%%%%%%%%%%%%%%
\section{Homogeneous Tachyon Background}
  \label{sec::BGcosmology}
The background evolution of a spatially flat homogeneous and isotropic
Universe is described by the metric
      \begin{equation}
	ds^2 = -dt^2 + a^2(t)[dx^2 + dy^2 + dz^2].
	\label{eq:flrwmetric}
      \end{equation}
Here, $a(t)$ is the scale factor of expansion. 
For a system of pressureless matter and tachyon scalar field, the dynamics
of background is completely governed by Friedmann equations
      \begin{equation}
		\label{eq:friedmanneq}
	    	\left( \frac{\dot{a}}{a} \right )^2 = \frac{8\pi G}{3}\rho , \, \,
			\frac{\ddot{a}}{a} = -\frac{4\pi G}{3}(\rho + 3P),
      \end{equation}
where $\rho = \rho_m + \rho_\phi$ is the total energy density of the
Universe. 
The relativistic component of energy density $\rho_r \propto a^{-4}$
is negligible and hence we do not include it.
The energy density of the matter component is given by $\rho_m \propto
a^{-3}$. 
The tachyon scalar field is described by a Lagrangian
      \begin{equation}
		\label{eq:Lagrangian}
		L_\phi = -V(\phi)\sqrt{1+g^{\mu \nu}\partial_\mu \phi \partial_\nu \phi},
      \end{equation}
where $V(\phi)$ is an arbitrary potential.
For the tachyon field, the energy density and pressure are given by
      \begin{equation}
	\rho_\phi = \frac{V({\phi})}{\sqrt{1-\dot{\phi}^2}} \,  , \, \,
	   p_\phi = -V(\phi)\sqrt{1-\dot{\phi}^2}.
      \end{equation}
The equation of state parameter for the tachyon scalar field can then
be written as 
      \begin{equation}
		  \label{eq:eqofstate}
		  w_\phi = \frac{p_\phi}{\rho_\phi}=\dot{\phi}^2 - 1,
      \end{equation}
and the  dynamics of the tachyon scalar field are governed by equation
\begin{equation}
\label{eq:scalar_dynamics}
\ddot{\phi} = -(1-\dot{\phi}^2) \left [ 3H\dot{\phi} + \frac{1}{V(\phi)}\frac{dV}{d\phi}\right].
\end{equation}
We work with two different scalar field potentials, one is the inverse
square potential  
	\begin{equation}
	  \label{eq:inpotential}
	  V(\phi)=\frac{n}{4\pi G} \left(1-\frac{2}{3n}\right)^{1/2}\phi^{-2},
	\end{equation}
here $n$ is a real number defines the amplitude of this potential.
The exponential potential given by
	\begin{equation}
	  \label{eq:expopotential}
	  V(\phi) = V_a exp \left(-\phi/\phi_a \right),
	\end{equation}
where amplitude $V_a$ and $\phi_a$ are parameters.
The background cosmology have been studied with these tachyon potentials in
~\cite{AAJ2019,BJP2003,Paddy2002} and these are found to be  suitable candidates
to generate late time  acceleration.
The study of cosmological dynamics and the stability analysis have been done
in~~\cite{Copeland2005,Calcagni2006,Aguirregabiria2004} for these potentials.
	
%%%%%%%%%%%%%%%%%%%%%%%%%%%%%%%%%%%%%%%%%%%%%%%%%%%%%%%%%%%%%%%%%%%%%%%%%%%%%%%%%%%%%%%%%%%%%%%%%%%%%%%%%%%%%%%%%%%%%%%%%%%%%%%%%%%%%%%%%%%%%%%%%%%%%%%%%%%%%%%%%%%%
\section{Perturbation in Tachyon Scalar Field} 
    \label{sec::prtbd_eq} 
We consider the perturbed FLRW metric to study the perturbations in the matter
and the scalar field. 
If there are no anisotropic components, in the spatial part of 
energy-momentum tensor, i.e. $T^i_j=0$ if $i\neq j$, then the
perturbations can be described by a line element in longitudinal gauge
of the form 
	  \begin{equation}
	    \label{eq:metric_prtbd}
	    ds^2 = -(1 + 2\Phi)dt^2 + a^2(t)(1 - 2\Phi)[dx^2 + dy^2 + dz^2],
	  \end{equation}
where $\Phi$ is the scalar perturbation.
In the Newtonian limit, the metric perturbation $\Phi$ represents the effective
gravitational potential.  
The dynamical equation for this scalar perturbation $\Phi$ can be
derived by solving perturbed Einstein's equation $\delta G^\mu_\nu =
8\pi G \delta T^\mu_\nu$. 
Here, the perturbed energy-momentum tensor $\delta T^\mu_\nu$ consists
of two parts,  one for the matter component $\delta
T^\mu_{\nu_{(matter)}}$ and other for the  scalar field $\delta
T^\mu_{\nu_{(\phi)}}$. 
We consider matter  as a perfect fluid with energy-momentum tensor
	  \begin{equation}
	    \label{eq:enmom_mtr}
	    T^\mu_{\nu_{(matter)}} = (\rho + p)u^\mu u_\nu + p g ^\mu _\nu.
	  \end{equation}
Here $\rho$, $p$ and $u^\mu$ are energy density, pressure and four
velocity respectively. 
The perturbations in the matter field are defined by 
	  \begin{equation}
	    \begin{aligned}
	      \rho(t,\vec{x}) &= \bar{\rho}(t) + \delta \rho(t,\vec{x}),\\
	      p(t,\vec{x}) &= \bar{p}(t) + \delta p (t,\vec{x}), \\
	      u^\mu &= {\bar{u}^\mu} + \delta u^\mu,
	    \end{aligned}
	  \end{equation}
where $\bar{u}^\mu = \{1,0,0,0 \}$, $\bar{\rho}(t)$ and $\bar{p}(t)$
are the average values of their respective quantities and $\delta
u^\mu$ is the peculiar velocity. 
Substituting these values in equation~(\ref{eq:enmom_mtr}), 
the components of the perturbed energy-momentum tensor of matter are 
	  \begin{equation}
	    \label{eq:prtbd_tmu_mtr}
	    \begin{aligned}
	      \delta T^0_0 &= -\delta \rho, \\
	      \delta T^i_0 &= (\bar{\rho} + \bar{p})\delta u^i, \\
	      \delta T^i_j &= \delta p\delta^i_j.
	    \end{aligned}
	  \end{equation}  
The energy-momentum tensor for the tachyon field can be derived from
	  \begin{equation}
		 T^{\mu}_{\nu (\phi)}   = \frac{V(\phi)\partial^\mu \phi \partial_\nu \phi}{\sqrt{1 + g^{\alpha \beta}\partial_\alpha \phi \partial_\beta \phi}} + L_\phi g^\mu_\nu,
	    \label{eq:enmom_scfield}
	  \end{equation}
where for tachyon scalar field the Lagrangian $L_\phi$ is given by
equation~(\ref{eq:Lagrangian}).
We define the perturbation in the scalar field as
	  \begin{equation}
	    \phi (t,\vec{x}) = \bar{\phi}(t) + \delta \phi(t,\vec{x}).
	  \end{equation}  
Here $\bar{\phi}(t)$ is the average background field. 
Using equation~(\ref{eq:enmom_scfield}) with the metric element 
of longitudinal gauge from equation~(\ref{eq:metric_prtbd}),
components of perturbed energy-momentum tensor for tachyon scalar
field can be calculated:
	    \begin{equation}
	      \label{eq:prtbd_tmu_scfield}
	      \begin{aligned}
		\delta T^0_0 & =-\delta \rho_\phi = -\frac{\left( \frac{\partial V }{\partial \phi}\right)_{\bar{\phi}}\delta \phi}{\sqrt{1-\dot{\bar{\phi}}^2}} 
				+ \frac{1}{2}\frac{V(\bar{\phi})}{\sqrt{1-\dot{\bar{\phi}}^2}}\left ( \frac{2\Phi \dot{\bar{\phi}}^2 - 2\dot{\bar{\phi}}\dot{\delta \phi} }{1-\dot{\bar{\phi}}^2} \right),\\
		\delta T^i_j & = \delta p_\phi \delta^i_j = -V(\bar{\phi})\sqrt{1-\dot{\bar{\phi}}^2} \left ( \frac{\Phi \dot{\bar{\phi}}^2 
				- \dot{\delta \phi} \dot{\bar{\phi}}}{1-\dot{\bar{\phi}}^2} \right ) \delta^i_j 
				- \left ( \frac{\partial V}{\partial \phi }\right )_{\bar{\phi}} \delta \phi \sqrt{1-\dot{\bar{\phi}}^2} \, \delta^i_j, \\
		\delta T^0_i & = (\rho_\phi + p_\phi)\delta u_i = \frac{V(\bar{\phi})}{\sqrt{1-\dot{\bar{\phi}}^2}} \dot{\bar{\phi}} \delta \phi_{,i}. \\
	      \end{aligned}
	    \end{equation} 
We can now solve perturbed Einstein's equation $\delta G^\mu_\nu 
= 8\pi G \delta T^\mu_\nu$;
where the perturbed energy-momentum tensor are given by equations~(\ref{eq:enmom_mtr}) 
and (\ref{eq:enmom_scfield}).
Components of the perturbed Einstein tensor $\delta G^\mu_\nu$ 
can be calculated using line element~(\ref{eq:metric_prtbd}).
We retain the terms in the solution of perturbed Einstein's equations up to
first (or linear) order in all perturbed quantities.
We  then transform these linearized Einstein equations into the
Fourier space or the $k-$ space, 
where the perturbed quantities of both the spaces are related by the equation
	\begin{equation}
		A(\vec{x},t) = \int d^3 k A(\vec{k},t)e^{i\vec{k}.\vec{x}}.
	\end{equation}   
Here, $\vec{k}$ is the wave vector.

In longitudinal gauge, the Fourier transformed Einstein's equations
are given by 
    \begin{equation}
      \label{eq:Eineq_1}
	3\frac{\dot{a}^2}{a^2}\Phi + 3\frac{\dot{a}}{a}\dot{\Phi} + \frac{k^2\Phi}{a^2}  
		= -4\pi G \left [\delta \rho_m + \delta \rho_\phi \right ],
    \end{equation}
    \begin{equation}
	\label{eq:Eineq_2}
    	\ddot{\Phi} + 4\frac{\dot{a}}{a}\dot{\Phi} + \left ( 2\frac{\ddot{a}}{a} + 
		\frac{\dot{a}^2}{a^2} \right )\Phi  
		= 4\pi G \left [ -V(\bar{\phi})\sqrt{1-\dot{\bar{\phi}}^2}
		\left ( \frac{\Phi \dot{\bar{\phi}}^2 - \dot{\delta \phi} \dot{\bar{\phi}}}
		{1-\dot{\bar{\phi}}^2} \right ) - \left ( \frac{\partial V}{\partial \phi }\right )_{\bar{\phi}}
		\delta \phi \sqrt{1-\dot{\bar{\phi}}^2} \right ],     
    \end{equation}
    \begin{equation}
	\label{eq:Eineq_3}
	\dot{\Phi} + \frac{\dot{a}}{a}\Phi  = 4\pi G \left (\bar{\rho} a^{-3}v_m 
		    + \frac{V(\bar{\phi})}{\sqrt{1-\dot{\bar{\phi}}^2}} \dot{\bar{\phi}} \delta \phi_{,i} \right),
    \end{equation}
where $v_m$ represents the potential for the matter peculiar velocity,
i.e., $\delta u_i = \triangledown_i v_m $.
Here although we have used the same symbol for quantities 
$\Phi, \delta \phi, \delta \rho_m$ and $v_m$,
as they are in real physical space, they represent the Fourier components
of respective quantities in $k^{th}$ mode of perturbation.
The wave number is given by $k = 2\pi /\lambda_p$, where $\lambda_p $ 
is the comoving length of the perturbation.
Therefore, the Einstein's equations given above represent the evolution of 
the $k^{th}$ mode of perturbations.
Equation~(\ref{eq:Eineq_2}) is the dynamical equation for metric
perturbation $\Phi$.
Since matter is pressureless, the dynamics of metric
perturbation $\Phi$ is driven only by perturbation in the scalar field. 
Here, in these equations, there are two unknown perturbed quantities,
$\Phi$ and $\delta \phi$. 
Once these two are determined, then other perturbed quantities like
$\delta \rho_m$ and $v_m$ can be calculated from
equation~(\ref{eq:Eineq_1}) and (\ref{eq:Eineq_3}). 
The dynamical equation for the perturbed tachyon scalar field $\delta
\phi$ can be derived by solving the Euler-Lagrangian equation 
using the Lagrangian function~(\ref{eq:Lagrangian}) for the perturbed
scalar field, and in the Fourier space for $k^{th}$ mode, it is given by
  \begin{equation}
    \label{eq:deltaphi}
    \begin{split}
      \frac{\ddot{\delta \phi}}{(1-\dot{\bar{\phi}}^2)} + & \left[3H + \frac{2\dot{\bar{\phi}} \ddot{\bar{\phi}}}{(1-\dot{\bar{\phi}}^2)^2} \right ]
	      \dot{\delta \phi} + \left [3H\dot{\bar{\phi}} \frac{V^\prime}{V } + \frac{k^2}{a^2} + \frac{\ddot{\bar{\phi}}}{(1-\dot{\bar{\phi}}^2)}
	      \left( \frac{V^\prime }{V}\right) + \frac{V^{\prime \prime }}{V} \right ]\delta \phi \\
      - & \left [ 12H\dot{\bar{\phi}} + \frac{2(2+\dot{\bar{\phi}}^2 )\ddot{\bar{\phi}}}{(1-\dot{\bar{\phi}}^2)} + \frac{2V^\prime}{V}
	      + \frac{2\dot{\bar{\phi}}^4 \ddot{\bar{\phi}}}{(1-\dot{\bar{\phi}}^2)^2} \right ]\Phi + \frac{3\dot{\bar{\phi}}^3 - 4\dot{\bar{\phi}}}
	      {(1-\dot{\bar{\phi}}^2)} \dot{\Phi} = 0,
    \end{split}
   \end{equation}
where the prime represents the derivative with respect to the
background scalar field $\bar{\phi}$. 
The coupled equations~(\ref{eq:Eineq_2}) and (\ref{eq:deltaphi})
form a closed system of equations.
Solving these equations together with the background equations,
we can find the quantities $\Phi$ and $\delta \phi$ and then
the respective fractional density contrasts
$\delta = \delta \rho /\bar{\rho}$ of $k^{th}$ mode for matter 
and tachyon scalar field can be computed from the following equations 
      \begin{equation}
	\label{eq:densityContrasts}
	\begin{aligned}
	\delta_\phi &= \frac{V^\prime(\bar{\phi})}{V(\bar{\phi})} \delta \phi 
	             - \left( \frac{\Phi \dot{\bar{\phi}}^2 -
	             	\dot{\bar{\phi}}\dot{\delta \phi}}{1- {\dot{\phi}}^2} \right),\\ 
	\delta_m &= -\frac{1}{4\pi G \rho_m a^{-3}} \left[ 3\frac{\dot{a}^2}{a^2}\Phi 
		  + 3\frac{\dot{a}}{a}\dot{\Phi} + \frac{k^2\Phi}{a^2} \right]
		  -\frac{1}{\rho_m a^{-3}} \left[ \frac{V^\prime (\bar{\phi})\delta \phi}
		  {\sqrt{1-\dot{\bar{\phi}}^2}} - \frac{V(\bar{\phi})}{\sqrt{1-\dot{\bar{\phi}}^2}}
		  \left( \frac{ \Phi \dot{\bar{\phi}}^2 - \dot{\bar{\phi}}\dot{\delta \phi}}{1- {\dot{\phi}}^2} \right )\right]\\
		&=  -\frac{1}{4\pi G \rho_m a^{-3}} \left[ 3\frac{\dot{a}^2}{a^2}\Phi 
		  + 3\frac{\dot{a}}{a}\dot{\Phi} + \frac{k^2\Phi}{a^2} \right]
		  -\frac{\delta_\phi}{\rho_m a^{-3}}\frac{V(\bar{\phi})}{\sqrt{1-\dot{\bar{\phi}}^2}}. \\
	\end{aligned}
      \end{equation}
To calculate matter density contrast $\delta_m = \delta \rho_m/\rho_m$
we have used equation~(\ref{eq:Eineq_1}).
We can see from the above equations that the density contrasts of matter
and dark energy are coupled with each other. 

The growth of structure, quantified by the linear growth function
$D^+_m$, defined as  
\begin{equation}
	D^+_m = \frac{\delta_m}{\delta_{m0}},
	\label{eq:growthfuction}
\end{equation}
The quantity  $\delta_{m0}$ is the  present value of matter density
contrast, and the growth rate, defined as
\begin{equation}
	f= \frac{d~ ln~ \delta}{d~ ln~ a}.
	\label{eq:growthrate}
\end{equation}

%%%%%%%%%%%%%%%%%%%%%%%%%%%%%%%%%%%%%%%%%%%%%%%%%%%%%%%%%%%%%%%%%%%%%%%%%%%%%%%%%%%%%%%%%%%  
\section{Numerical Approach and Methodology}
    \label{sec::num_methods}
To solve for $a,~\phi,~\Phi$ and $\delta \phi$, we need four equations. 
We choose two background equations, first of the Friedmann
equations~(\ref{eq:friedmanneq})
and the dynamical equation of scalar field~(\ref{eq:scalar_dynamics}).
The third equation is the dynamical equation of the perturbed scalar
field, equation~(\ref{eq:deltaphi}) 
and the fourth one is the dynamical equation for the metric perturbation,
the second equation of Einstein's equations~(\ref{eq:Eineq_2}).
We rewrite these equations in the dimensionless form by introducing
the following variables
  \begin{equation}
    \label{eq:variables}
    x=tH_0, \,\,\, y=\frac{a}{a_{in}},\, \, \, \psi = \frac{\phi}{\phi_{in}}, \,\,\,
  \Phi_N = \frac{\Phi}{\Phi_{in}}, \,\,\, \delta \psi = \frac{\delta \phi}{\Phi_{in}\phi_{in}},  
  \end{equation}
to above equations to solve them.
Derivatives are defined with respect to x as
   \begin{equation}
      y^\prime = \frac{dy}{dx}, \,\,\,\psi^\prime = \frac{d\psi}{dx},
      \,\,\, \Phi_N^\prime = \frac{d\Phi_N}{dx}.
   \end{equation}
    
%%%%%%%%%%%%%%%%%%%%%%%%%%%%%%%%%%%%%%%%%%%%%%%%%%%%%%%%%%%%%%%%%
\subsection{Dimensionless Equations for Inverse Square Potential}
    \label{sub::dimless_invrse}
In terms of the above dimensionless variables~[\ref{eq:variables}],
the background equations~[\ref{eq:friedmanneq}] and [\ref{eq:scalar_dynamics}] 
with inverse square potential~[\ref{eq:inpotential}], take the form  
    \begin{equation}
      \label{eq:bg_invrse_dimless_scale}
      y^\prime = y \left[ \Omega_{m_{in}}y^{-3} 
		+ \frac{\frac{2n}{3}(1-\frac{2}{3n})^{1/2}\psi^{-2}}{\phi_{in}^2H_0^2 \sqrt{1-\phi_{in}^2H_0^2\psi^{\prime 2}}} \right ]^{1/2},
    \end{equation}  
    \begin{equation}
      \label{eq:bg_invrse_dimless_phi}
      \psi^{\prime \prime} = \left ( 1-\phi_{in}^2H_0^2\psi^{\prime 2} \right )
			     \left [ \frac{2}{\phi_{in}^2H_0^2 \psi} - 3 \frac{y^\prime}{y} \psi^\prime \right ],
    \end{equation}
where $\Omega_{m_{in}}$ can be linked to the present matter density parameter
$\Omega_{m0}$ using the relation
      \begin{equation}
		\label{eq:omm}
		\Omega_m = \frac{\Omega_{m0}}{(H/H_0)^2}\left( \frac{a}{a_0}\right)^{-3}.
      \end{equation}
Here, $a_0$ is the present day value of the scale factor.
To solve the above background equations, we need values of 
the parameters $\Omega_{m_{in}}$, $C_n$, and $\phi_{in}H_0$.
Here $C_n = \frac{2n}{3}(1-\frac{2}{3n})^{1/2}$ is the amplitude of
the potential. 
    
Using the variables defined in equation~(\ref{eq:variables}),
with inverse square potential~(\ref{eq:inpotential}), the dynamical 
equation for metric perturbation
$\Phi$, equation~(\ref{eq:Eineq_2}),
and the dynamical equation of perturbed scalar field $\delta \phi$,
equation~(\ref{eq:deltaphi}) takes the form
    \begin{equation}
      \label{eq:prtb_invrse_potentl}
      \begin{split}
	\Phi_N^{\prime \prime} + 4\frac{y^\prime}{y}\Phi_N^\prime  
	      &+ \left \{ 2 \frac{y^{\prime\prime}}{y} + \left(\frac{y^\prime}{y} \right)^2 \right \}\Phi_N \\
	      &= n(1-\frac{2}{3n})^{1/2}\left [ \frac{2\delta \psi}{\phi_{in}^2H_0^2\psi^3}\sqrt{1-\phi_{in}^2H_0^2\psi^{\prime 2}} 
	      - \frac{\Phi_N \psi^{\prime 2} - \psi^\prime \delta \psi^\prime}{\psi^2 \sqrt{1-\phi_{in}^2H_0^2\psi^{\prime 2}} }\right ],    
      \end{split}
    \end{equation}
    \begin{equation}
      \label{eq:prtb_invrse_delta}
      \begin{split}
	\frac{\delta \psi^{\prime\prime}}{\left(1-\phi_{in}^2H_0^2\psi^{\prime 2}\right)} &+ \left[ 3\frac{y^\prime}{y}
			    + \frac{2\phi_{in}^2H_0^2\psi^\prime\psi^{\prime\prime}} 
			    {\left(1-\phi_{in}^2H_0^2\psi^{\prime 2}\right)^2} \right ] \delta \psi^\prime \\
			 &+ \left [-6\frac{y^\prime}{y}\frac{\psi^\prime}{\psi} + \frac{k^2}{a_{in}^2H_0^2y^2}
			    - \frac{2\psi^{\prime\prime}}{\psi\left(1-\phi_{in}^2H_0^2\psi^{\prime 2}\right)}
			    + \frac{6}{\phi_{in}^2H_0^2\psi^2} \right ] \delta\psi \\
			 &- \left[ 12\frac{y^\prime}{y}\psi^\prime 
 			    + \frac{2\left(2+\phi_{in}^2H_0^2\psi^{\prime 2}\right)\psi{\prime\prime}}
 			    {\left(1-\phi_{in}^2H_0^2\psi^{\prime 2}\right)} -\frac{4}{\phi_{in}^2H_0^2\psi} 
 			    + \frac{2\phi_{in}^4H_0^4\psi^{\prime4}\psi^{\prime\prime}}
 			    {\left(1-\phi_{in}^2H_0^2\psi^{\prime 2}\right)^2}\right]\Phi_N \\
 			 &+ \left[ \frac{3\phi_{in}^2H_0^2\psi^{\prime 3} 
			    - 4\psi^\prime}{\left(1-\phi_{in}^2H_0^2\psi^{\prime 2}\right)} \right]\Phi_N^\prime \\
			 &= 0.
      \end{split}
    \end{equation}

On solving the perturbation equations along with the background using
the above initial conditions, 
we can find the values of $\Phi_N$ and $\delta \psi$ as a functions
of redshift or scale factor.
Subsequently, the values of density parameters can be calculated using equations
	\begin{equation}
	  \label{eq:in_dimless_contrst}
	  \begin{aligned}
	  \frac{\delta_\phi}{\Phi_{in}} &= -2\frac{\delta \psi}{\psi} - \phi_{in}^2H_0^2 \left(\frac{\psi^{\prime 2}\Phi_N 
	  	- \psi^\prime \delta \psi^\prime}{1-\phi_{in}^2H_0^2\psi^{\prime 2}}   \right),\\
	  \frac{\delta_m}{\Phi_{in}} &= \frac{-2}{\Omega_{m_{in}}y^{-3}} \left [ \frac{y^{\prime 2}}{y^2}\Phi_N  
				      + \frac{y^{\prime }}{y}\Phi_N^{\prime} 
				      + \frac{k^2/H_0^2}{2a_{in}^2y^2}\Phi_N\right]
				      -  \frac{\delta_\phi/\Phi_{in}}{\Omega_{m_{in}}y^{-3}} 
					\frac{\frac{2n}{3}(1-2/3n)^{1/2}\psi^{-2}}{\phi_{in}^2H_0^2\sqrt{1-\phi_{in}^2H_0^2\psi^{\prime 2}}}.	 
	  \end{aligned}
	\end{equation}
To derive the above equations we have substituted dimensionless variables
defined in equation~(\ref{eq:variables}) to equation~(\ref{eq:densityContrasts}). 
%%%%%%%%%%%%%%%%%%%%%%%%%%%%%%%%%%%%%%%%%%%%%%%%%%%%%%%%%%%%%%%
\subsection{Dimensionless Equations for Exponential Potential}
    \label{sub::dimless_expo}    
In terms of the variables defined in equation~(\ref{eq:variables}),
the background equations for exponential
potential~(\ref{eq:expopotential}) can be written as
\begin{equation}
  \label{eq:bg_expo_dimless_scale}
	  y^\prime = y \left[ \Omega_{m_{in}}y^{-3} + \frac{\frac{V_a}{\rho_{cr}}e^{-\frac{\phi_{in}}{\phi_a}\psi}}{\sqrt{1-\phi_{in}^2H_0^2\psi^{\prime 2}}} \right ]^{1/2},
\end{equation} 
\begin{equation}
	\label{eq:bg_expo_dimless_phi}
	  \psi^{\prime \prime} = \left ( 1-\phi_{in}^2H_0^2\psi^{\prime 2} \right )
			     \left [ \frac{\phi_{in}/\phi_a}{\phi_{in}^2H_0^2} - 3 \frac{y^\prime}{y} \psi^\prime \right ].
\end{equation}
To solve these background equations, we need value of parameters
$\Omega_{m_{in}}$, $V_a/\rho_{cr}$, $\phi_{in}H_0$ and $\phi_{in}/\phi_a$.
On introducing variables defined in equation~(\ref{eq:variables}),
with exponential potential,
equations~(\ref{eq:Eineq_2}) and (\ref{eq:deltaphi}) for perturbed
quantities $\Phi$ and $\delta \phi$ are
\begin{equation}
  \label{eq:prtb_expo_potentl}
  \begin{split}
	\Phi_N^{\prime \prime} + 4\frac{y^\prime}{y}\Phi_N^\prime  
	      &+ \left \{ 2 \frac{y^{\prime\prime}}{y} + \left(\frac{y^\prime}{y} \right)^2 \right \}\Phi_N \\
	      &= \frac{3}{2}\frac{V_ae^{-\frac{\phi_{in}}{a}\psi} }{\rho_{cr}}\sqrt{1-\phi_{in}^2H_0^2\psi^{\prime 2}}
		\left [ \frac{\phi_{in}}{\phi_a}\delta\psi
	      - \frac{\phi_{in}^2H_0^2(\Phi_N \psi^{\prime 2} - \psi^\prime \delta \psi^\prime)}{1-\phi_{in}^2H_0^2\psi^{\prime 2}} \right ],   
    \end{split}
\end{equation}
    
\begin{equation}
      \label{eq:prtb_expo_delta}
      \begin{split}
	\frac{\delta \psi^{\prime\prime}}{\left(1-\phi_{in}^2H_0^2\psi^{\prime 2}\right)} &+ \left[ 3\frac{y^\prime}{y}
			    + \frac{2\phi_{in}^2H_0^2\psi^\prime\psi^{\prime\prime}} 
			    {\left(1-\phi_{in}^2H_0^2\psi^{\prime 2}\right)^2} \right ] \delta \psi^\prime \\
			 &+ \left [-3\frac{\phi_{in}}{\phi_a}\frac{y^\prime}{y}\psi^{\prime} + \frac{K^2}{a_{in}^2H_0^2y^2}
			    - \frac{\frac{\phi_{in}}{\phi_a}\psi^{\prime\prime}}{\left(1-\phi_{in}^2H_0^2\psi^{\prime 2}\right)}
			    + \frac{\left(\frac{\phi_{in}}{\phi_a}\right)^2}{\phi_{in}^2H_0^2} \right ] \delta\psi \\
			 &- \left[ 12\frac{y^\prime}{y}\frac{\psi^\prime}{\psi} 
 			    + \frac{2\left(2+\phi_{in}^2H_0^2\psi^{\prime 2}\right)\psi{\prime\prime}}
 			    {\left(1-\phi_{in}^2H_0^2\psi^{\prime 2}\right)} -\frac{2\frac{\phi_{in}}{\phi_a}}{\phi_{in}^2H_0^2} 
 			    + \frac{2\phi_{in}^4H_0^4\psi^{\prime4}\psi^{\prime\prime}}
 			    {\left(1-\phi_{in}^2H_0^2\psi^{\prime 2}\right)^2}\right]\Phi_N \\
 			 &+ \left[ \frac{3\phi_{in}^2H_0^2\psi^{\prime 3} 
			    - 4\psi^\prime}{\left(1-\phi_{in}^2H_0^2\psi^{\prime 2}\right)} \right]\Phi_N^\prime \\
			 &= 0.
  \end{split}
\end{equation}
    
In terms of the dimensionless variables, defined in equation~(\ref{eq:variables}), 
the equation for density parameters (\ref{eq:densityContrasts}) for exponential
potential takes the form    
\begin{equation}
	\begin{aligned}
	  \frac{\delta_\phi}{\Phi_{in}} &= -\frac{\phi_{in}}{\phi_a}\delta \psi - \phi_{in}^2H_0^2 \left(\frac{\psi^{\prime 2}\Phi_N 
	  	- \psi^\prime \delta \psi^\prime }{1-\phi_{in}^2H_0^2\psi^{\prime 2}}  \right),\\
	  \frac{\delta_m}{\Phi_{in}} &= \frac{-2}{\Omega_{m_{in}}y^{-3}} \left [ \frac{y^{\prime 2}}{y^2}\Phi_N 
				      + \frac{y^{\prime }}{y}\Phi_N^{\prime} 
				     + \frac{k^2/H_0^2}{2a_{in}^2y^2}\Phi_N\right]
				    - \left[ \frac{\delta_\phi/\Phi_{in}}{\Omega_{m_{in}}y^{-3}} 
				     \frac{\frac{V_a}{\rho_{cr}}e^{-\frac{\phi_{in}}{\phi_a}\psi}}{\sqrt{1-\phi_{in}^2H_0^2\psi^{\prime 2}}} \right].	 
	\end{aligned}
\end{equation}

%%%%%%%%%%%%%%%%%%%%%%%%%%%%%%%%%%%%%%%%%%%%%%%%%%%%%%%%%%%%%%%%%%%%%%%%%%%%%%%%%%%%%%%%%%%%%%%%%%%%%%%%%%%%%%%%%%%%%%%%%%%%%%%%%%%%%%%%%%%%%%%%%%%%%%%%%%%%%%%%%%%%
  \section{Results and Discussion}
      \label{sec::resultDiscussion}
      
We evolve the perturbation equations from redshift $z=1000$ to the present day.
The main assumption we have  made  is that the dark energy field is
	initially homogeneous.
	Equation~(\ref{eq:densityContrasts}) suggests  that for this assumption to be valid we need
	not only to consider ${\delta \phi}_{in} =0$, but also $\dot{\bar{\phi}}_{in}=0$ or equivalently
	an initial equation of state parameter of dark energy $w_{\phi_{in}}=-1$.
	Therefore the  analysis, along with constraints on the free parameters we are 
	providing, are subject to this assumption.   
For background equations, our initial conditions are
      \begin{equation}
	  y_{in} = 1, \, \, \, \psi_{in} = 1,
	  \label{eq:initial_condi_1}
      \end{equation}
and $\psi_{in}^\prime$ can be calculated using relation
    \begin{equation}
	    \psi^\prime = \frac{\dot{\phi}}{\phi_{in} H_0} = \frac{\sqrt{1+w_\phi}}{\phi_{in} H_0}
	    \label{eq:initial_condi_2}.
    \end{equation}
    \begin{figure}[t]
      \centering
      \includegraphics[width=0.45\textwidth]{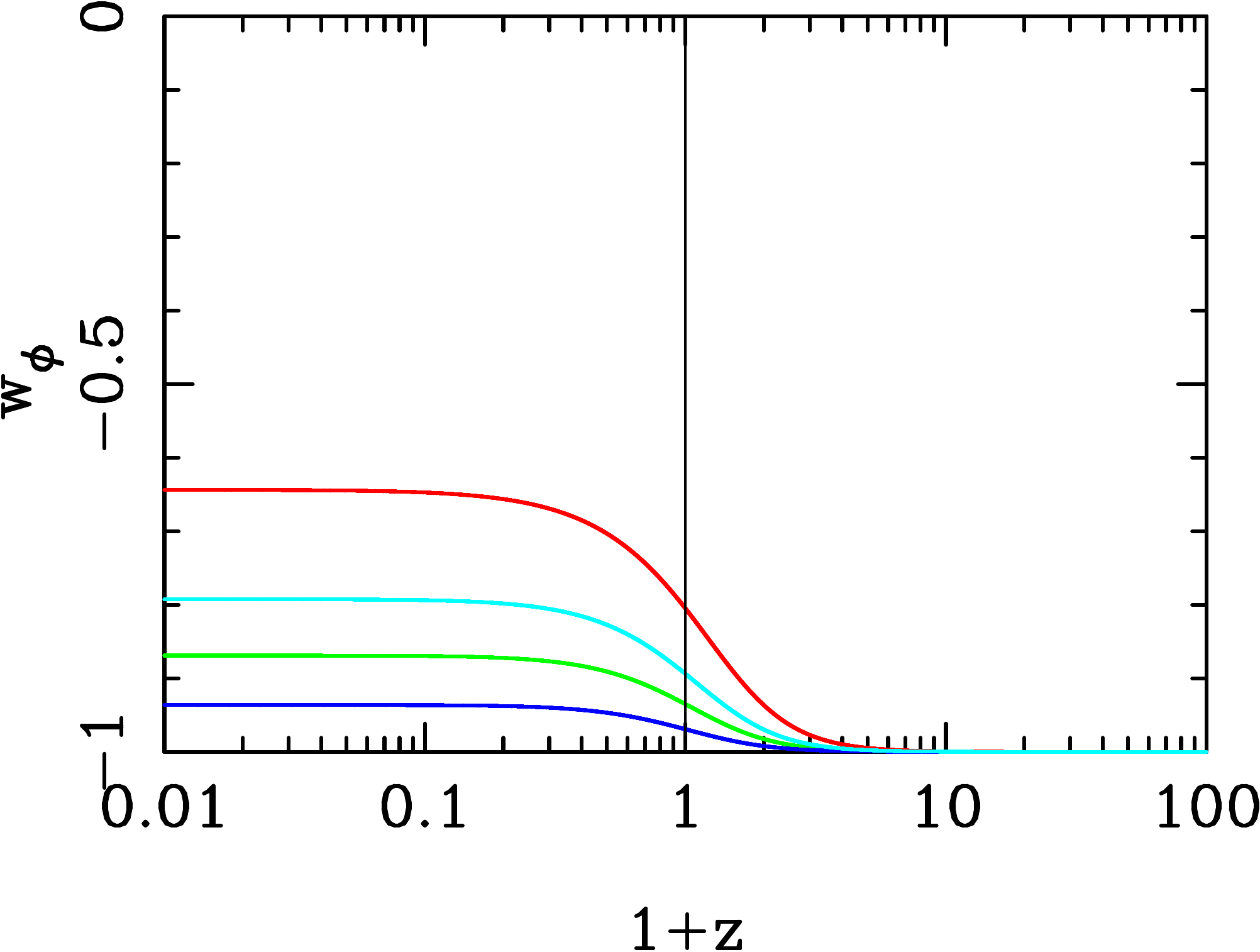}
      \includegraphics[width=0.45\textwidth]{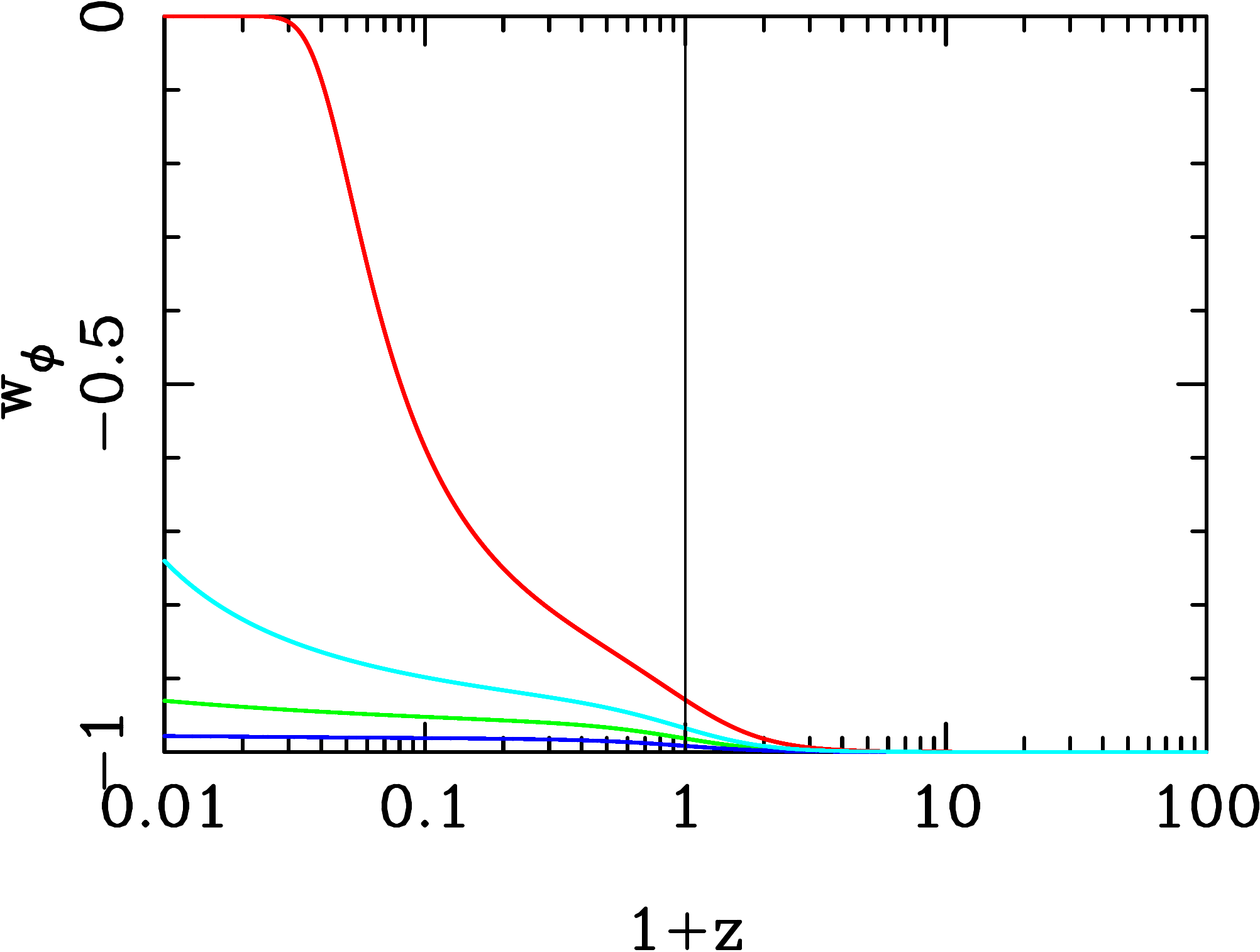}\\
      \includegraphics[width=0.45\textwidth]{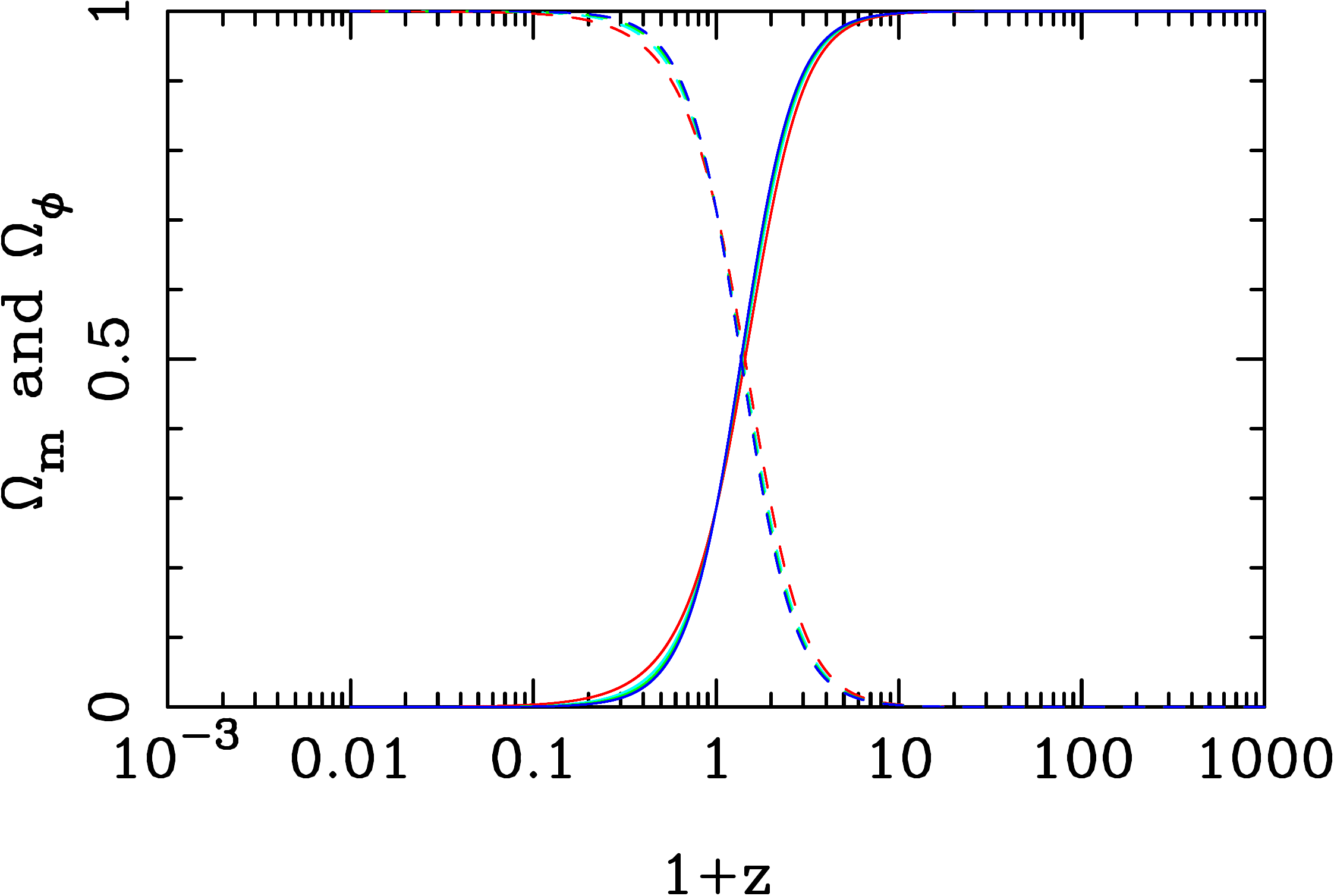}
      \includegraphics[width=0.45\textwidth]{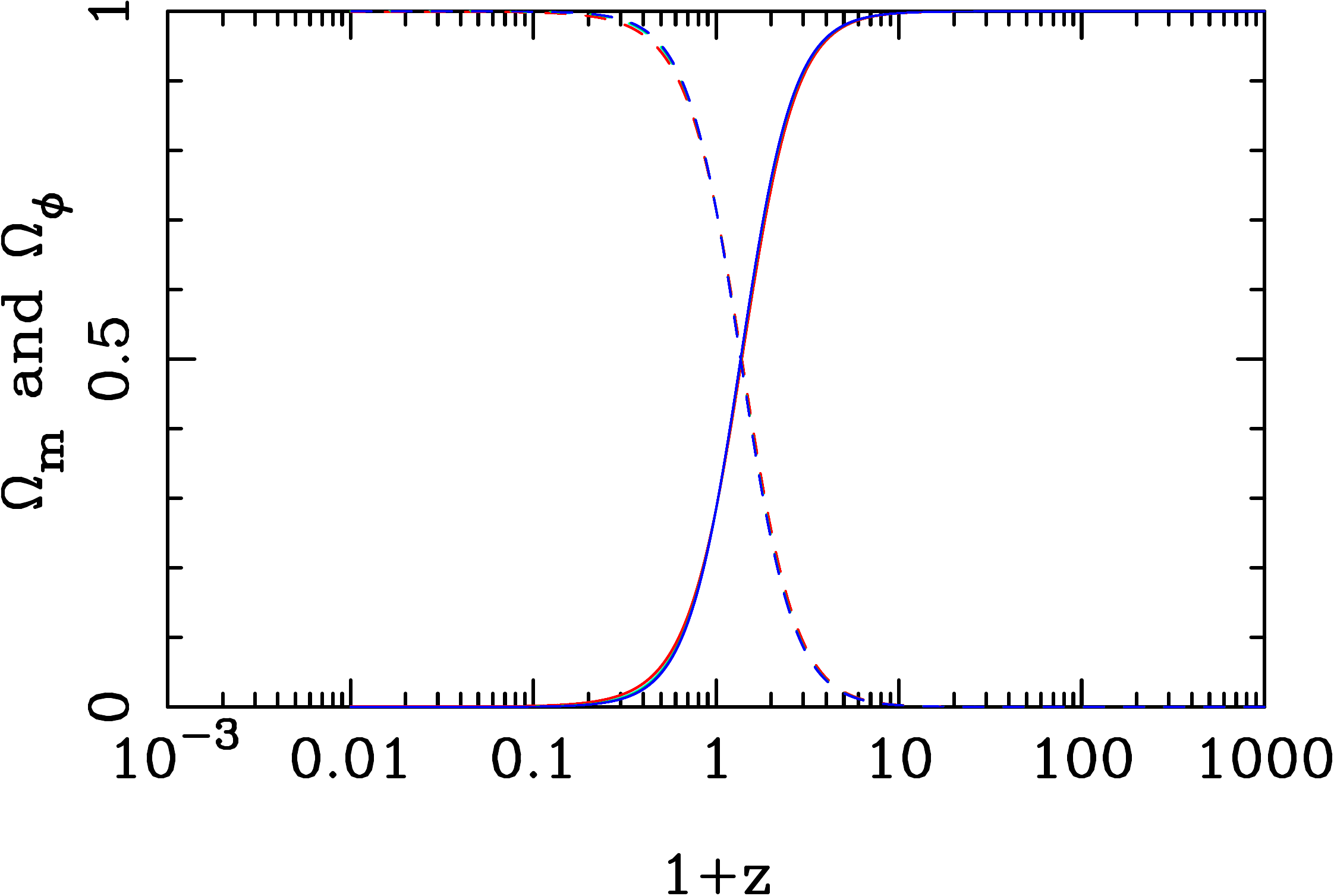}\\
      \caption{The top panels of this figure show evolution of
        the equation of state of dark energy and the bottom panels show the evolution of the density parameters.
        The column-1 and 2 are for the  inverse square potential and the exponential potential respectively.
        Red, sky-blue, green and blue colours represent
        $\phi_{in}H_0=$ ~1.0,~1.5,~2.0 and ~3.0. In the bottom row, the solid
        line is for $\Omega_m$ and the dashed line is for
        $\Omega_\phi$. Parameters $C_n$ and $V_a/\rho_{cr}$ are tuned
        for each value of $\phi_{in}H_0$ to get $\Omega_{m0}=0.285$.}
      \label{fig:inbg}
    \end{figure}
    \begin{figure}[t]
	    \centering
      \includegraphics[width=0.45\textwidth]{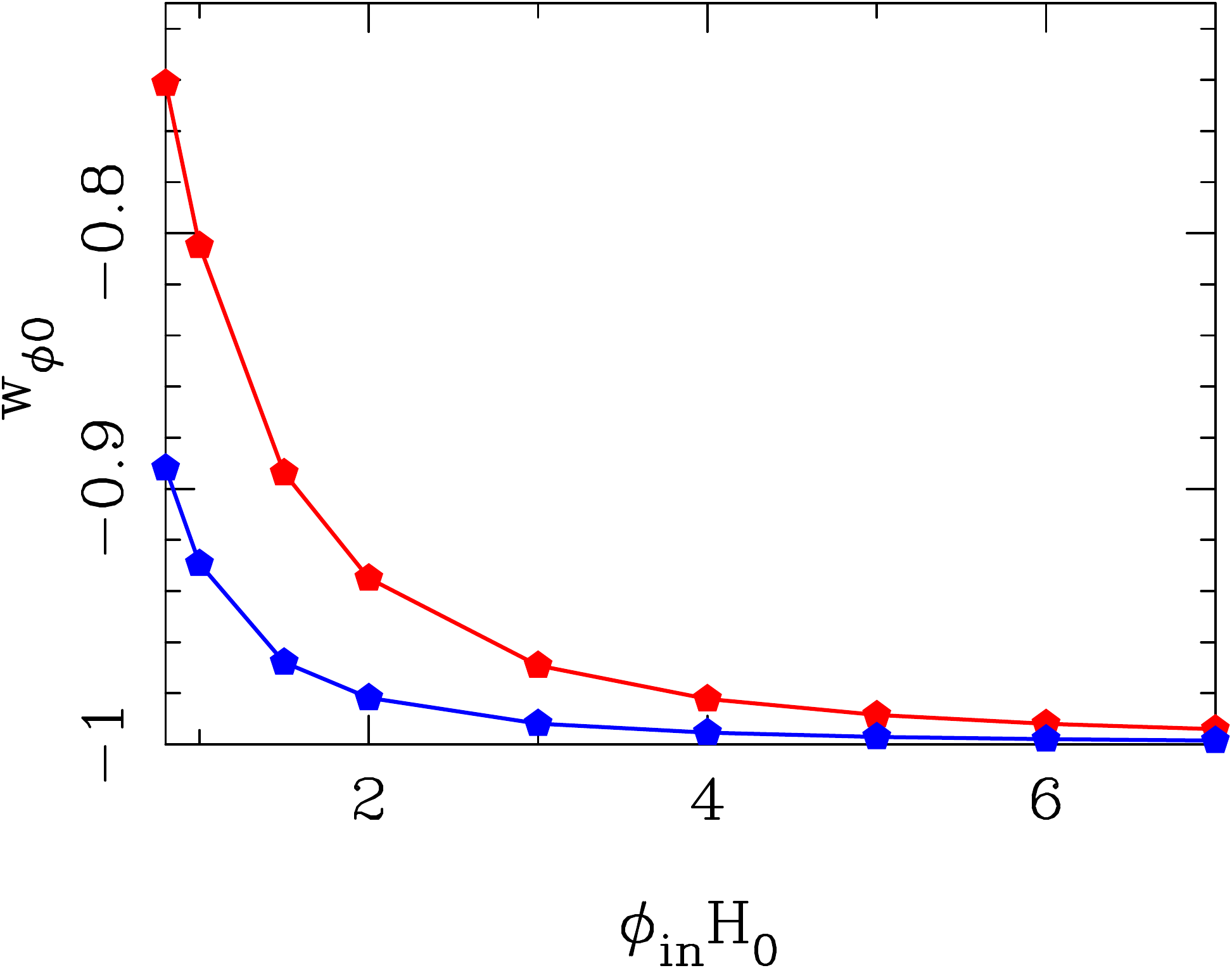}
      \includegraphics[width=0.45\textwidth]{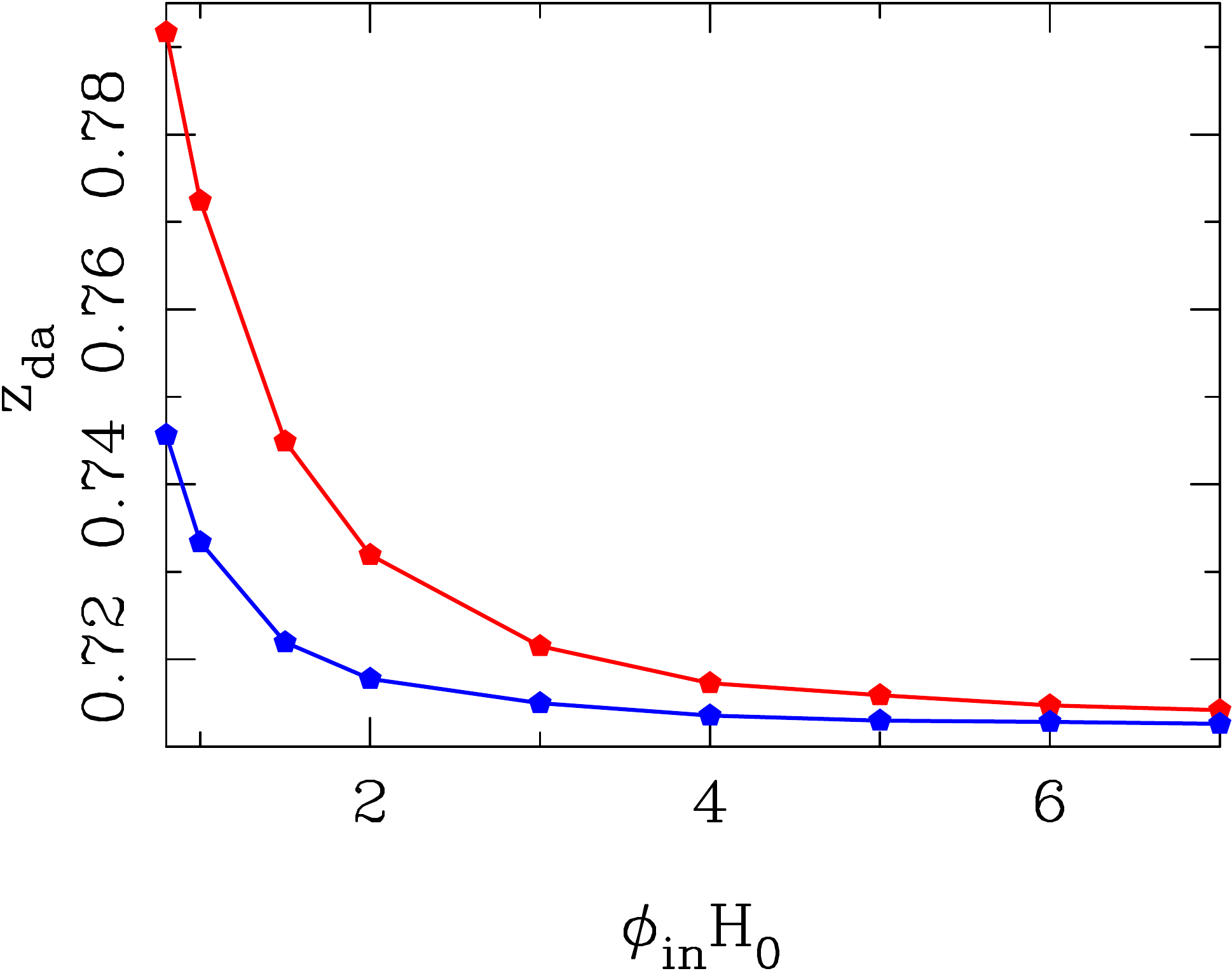}
      \caption{In this figure we show the dependence of the
        present day value of the equation of state parameter, $w_{\phi
          0}$ (plot on the left), and the deceleration  to
        acceleration transition redshift, $z_{dz}$ (in the right
        panel) on $\phi_{in}H_0$. The red curve is for inverse square
        potential and blue cure is for the exponential
        potential. The values of $C_n$ and
        $V_a/\rho_{cr}$  are the same as in~\ref{fig:inbg}.} 
      \label{fig:hph0VSwp0}
    \end{figure}

In~\cite{AAJ2019}, it has been shown that with the potentials mentioned in section~\ref{sec::BGcosmology},
the constraint on matter density contrast is $\Omega_{m0}=0.285^{+0.023}_{-0.022}$
at $3\sigma$ confidence. 
On the other hand, background data puts only a lower bound
$\phi_0H_0\gtrsim 0.775$ and all larger values are allowed.
Here, $\phi_0$ is the value of the scalar field at present, i.e.,
$(\bar{\phi})_0$.
Constraint on $w_{\phi 0}$ depends on the value of $\phi_0H_0$, as
they are correlated quantities.  
The tachyon scalar field starts evolution only in the near past, this
allow us to assume $\phi_{in}H_0 \approx \phi_0H_0$~\cite{BJP2003}. 
In this paper, we have done our analysis for the best fit value of
$\Omega_{m0}$ and other parameters have been varied.
In the case of the  exponential potential, differences  due to
the change in the parameter $\phi_{in}/\phi_a$ can be restored by
scaling $\phi_{in}H_0$ appropriately~\cite{AAJ2019}.
We have fixed the value of this parameter at $\phi_{in}/\phi_a = 1$.

The evolution of the equation of state of dark energy and the density
parameters are shown in figure~\ref{fig:inbg} for both the potentials.
Red, sky-blue, green and blue colours represent
$\phi_{in}H_0=~1.0,~1.5,~2.0$ and $3.0$. 
For each value of $\phi_{in}H_0$, we need to tune the amplitude of potential,
 $C_n = \frac{2n}{3}(1-2/3n)^{1/2}$ for the inverse square potential 
and $V_a/\rho_{cr}$ for the exponential potential, such that the
present value of the matter density parameter matches $\Omega_{m0}=0.285$.
We can see that the equation of state parameter for both the
potentials remains at $-1$ in the matter dominated era,
and starts evolving as the dark energy begins to dominate. 
In the right panel of figure~\ref{fig:hph0VSwp0}, we can see that
the deceleration to acceleration transition redshift, $z_{da}$,
is higher for smaller value of $\phi_{in}H_0$ and gradually decreases as we increase
this parameter.  
Hence for smaller values of $\phi_{in}H_0$, the value of equation of
state parameter begin to deviate, or start increasing, from -1 earlier.
That is the reason why $w_{\phi 0}$ is larger for these values than
it is for the larger value of $\phi_{in}H_0$. 
For larger $\phi_{in}H_0$, the value of $w_{\phi 0}$ is closer to $-1$. 
This correlation can be seen in the left panel of figure~\ref{fig:hph0VSwp0}. 
We can see that for a given value of $\phi_{in}H_0$, $w_{\phi 0}$
relatively  closer to $-1$  for the exponential
potential than it is for the inverse square potential.
The reason for this is that the transition from decelerated to accelerated
expansion, for a fixed value of $\phi_{in}H_0$, occurs earlier for the 
inverse square potential than for the exponential potential.
For example, for $\phi_{in}H_0=2.0$ the value of the transition redshift
$z_{da}= 0.732$ for the inverse square potential and $z_{da}= 0.717$
for the exponential potential.
Comparing the panels of figure~\ref{fig:hph0VSwp0}, we can see that
there is a linear relation between $w_{\phi 0}$ and $z_{da}$. 

The future evolution of  $w_{\phi}$ can be seen in figure~\ref{fig:inbg}, 
and it is clear that the $w_{\phi}$ for the inverse square potential
becomes constant in future, as for this potential, the equation of
state asymptotically approaches  $w_\phi = 2/3n
-1$~\cite{BJP2003,Paddy2002,Calcagni2006}. 
Whereas for the exponential potential, the equation of state increases
to $w_{\phi}=0$ (dust like).
For smaller values of $\phi_{in}H_0$, it evolves faster and approaches
$w_{\phi}=0$ relatively earlier than for  larger values of
$\phi_{in}H_0$. 
Since in future the  dominating component is dark energy,
the effective equation of state of the Universe depends only on $w_{\phi}$.
For the exponential potential, when $w_{\phi}$ becomes larger than $-1/3$,
the Universe once again goes to a decelerating phase. 
Hence, for the exponential potential, there is no future horizon
problem for tachyon model of dark
energy~\cite{BJP2003,Paddy2002,Calcagni2006}.   

The perturbation in the scalar field at initial (at $z=1000$) is
assumed to be negligibly small, compared to $\Phi$ and $\delta_m$.   
The scalar field can initially be assumed to be homogeneous, and our
initial conditions for perturbation are 
 \begin{equation}
	\label{eq:initial_prt}
	\Phi_{N_{in}} = 1, \, \, \delta \psi_{in} = 0, \, \, \delta \psi_{in}^\prime = 0.	
 \end{equation}
In~\cite{unniJassalSesh2008}, it was shown that the gravitational potential
does not evolve in the matter dominated era, and starts to decay when 
dark energy begins to dominate.
This fact allows us to assume $\Phi_{N_{in}}^\prime(k)=0$, for all scales.
    \begin{figure}[t]
      \centering
      \includegraphics[width=0.45\textwidth]
	      {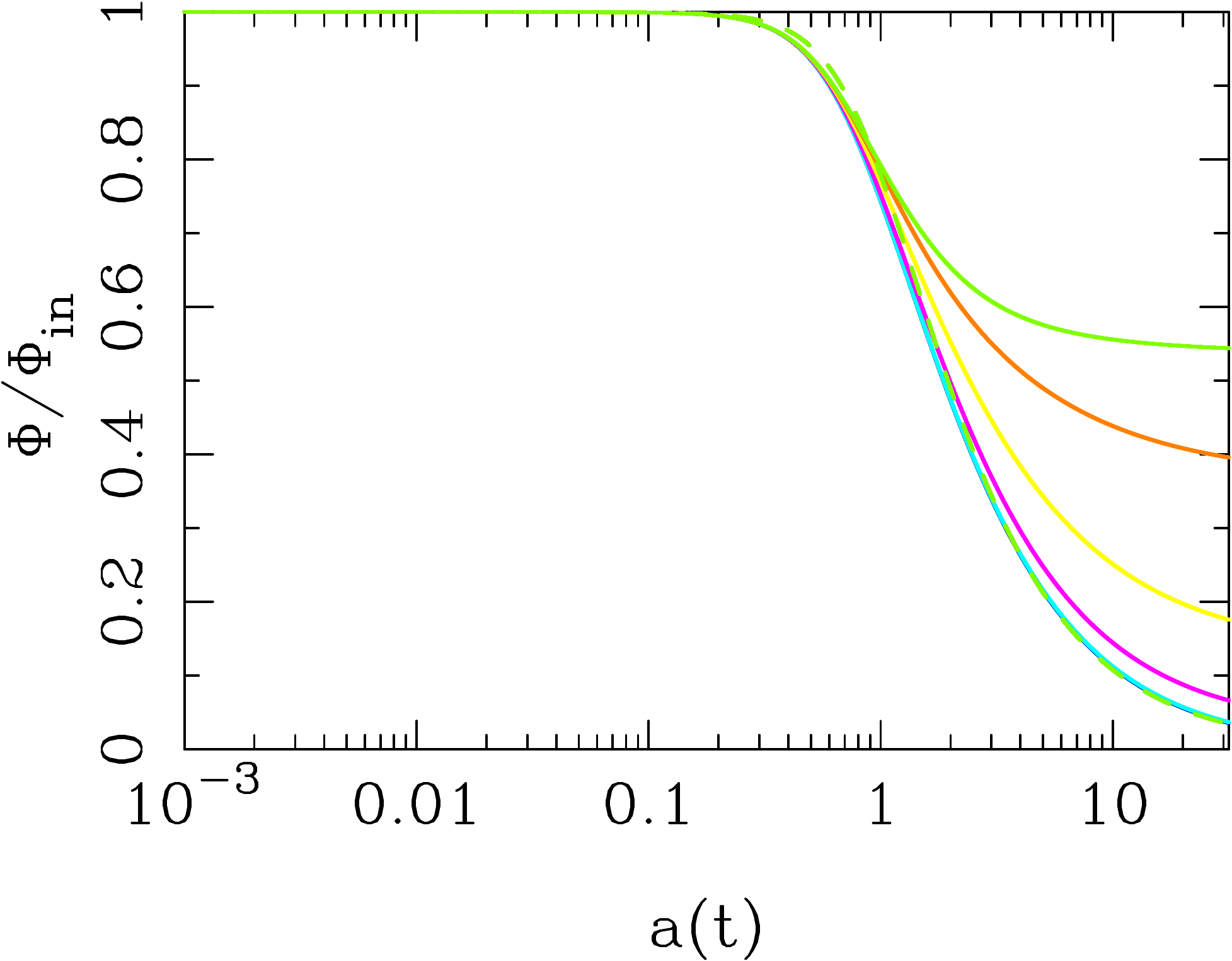}
      \includegraphics[width=0.45\textwidth]
	      {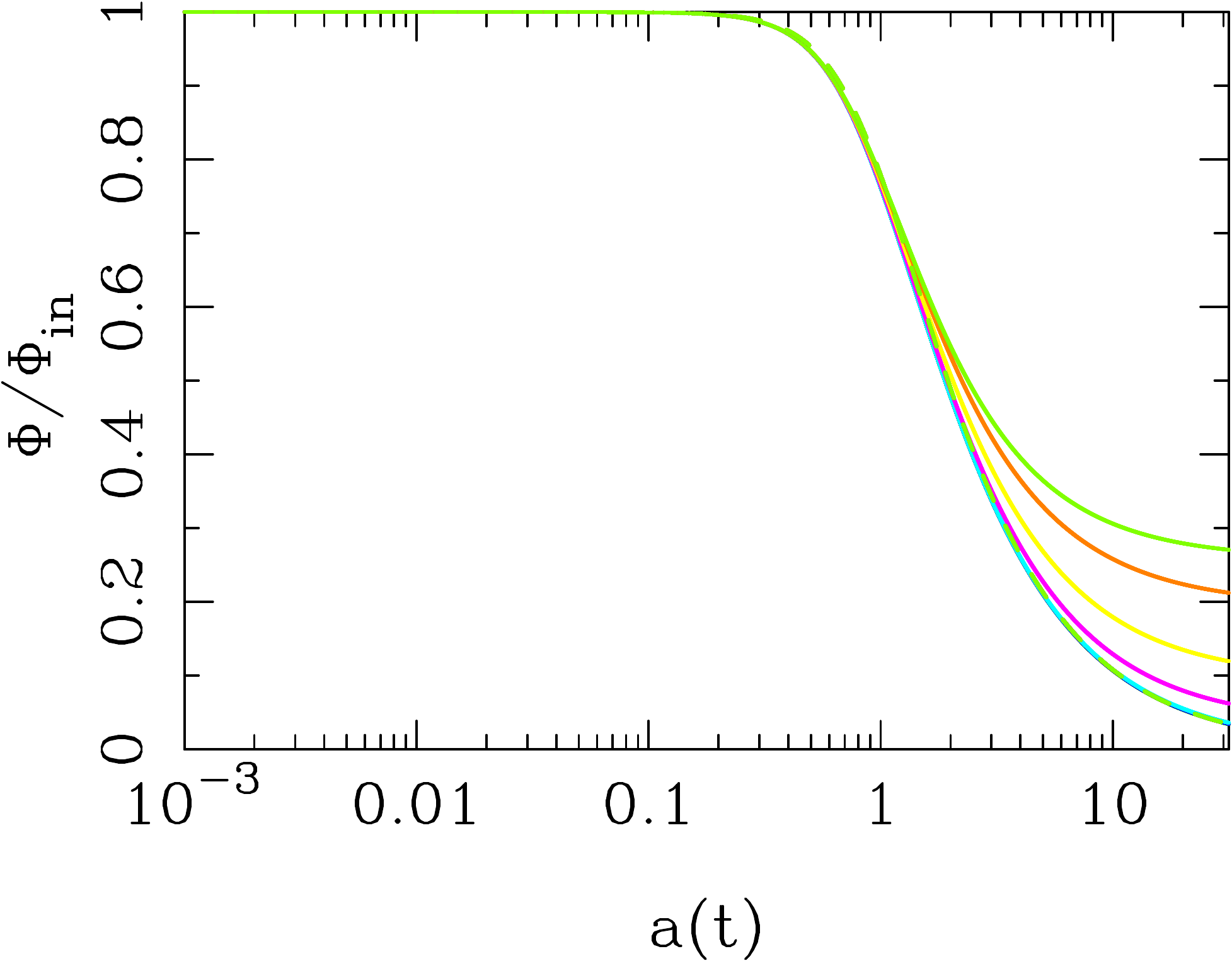}\\
      \includegraphics[width=0.45\textwidth]
	      {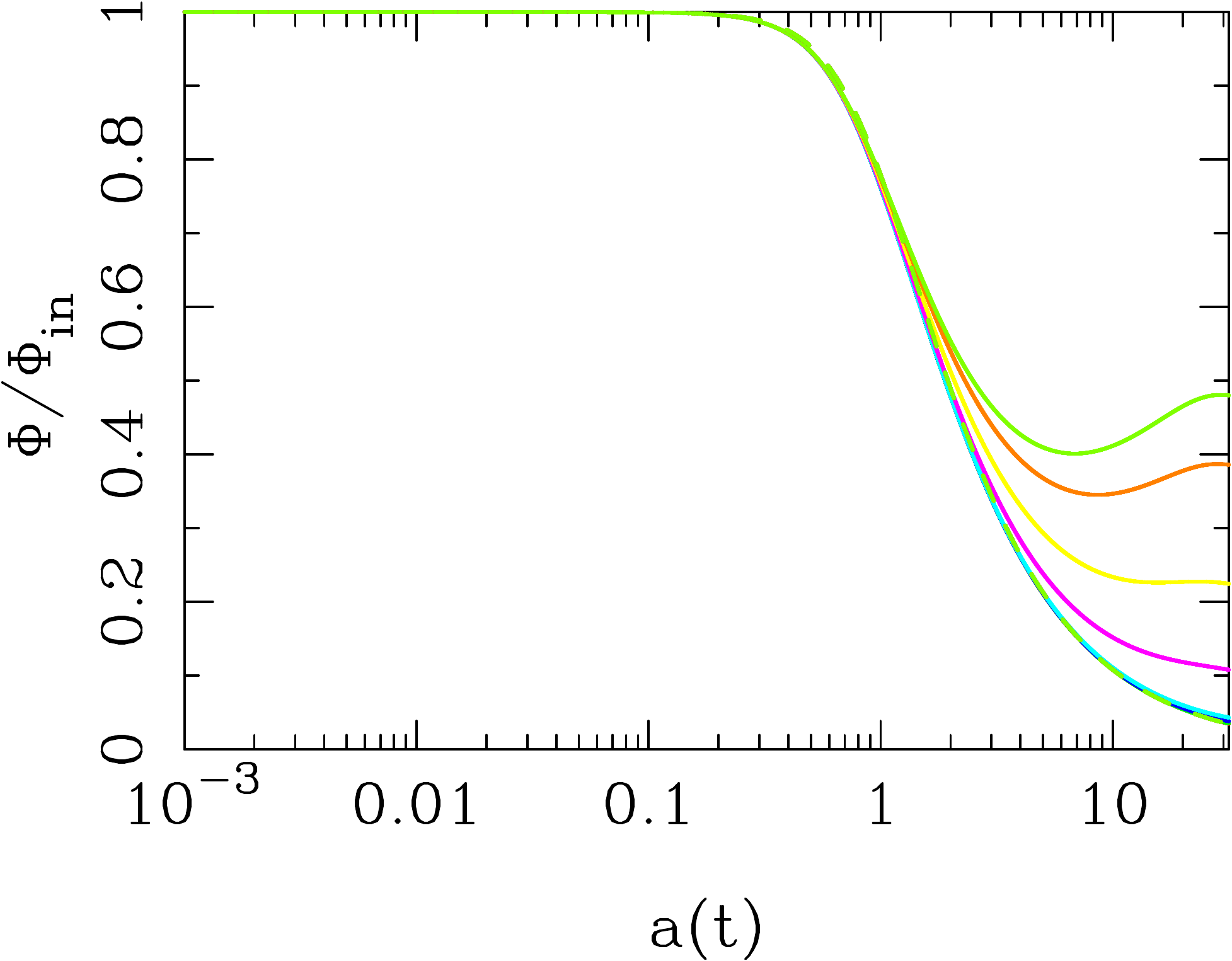}
      \includegraphics[width=0.45\textwidth]
	      {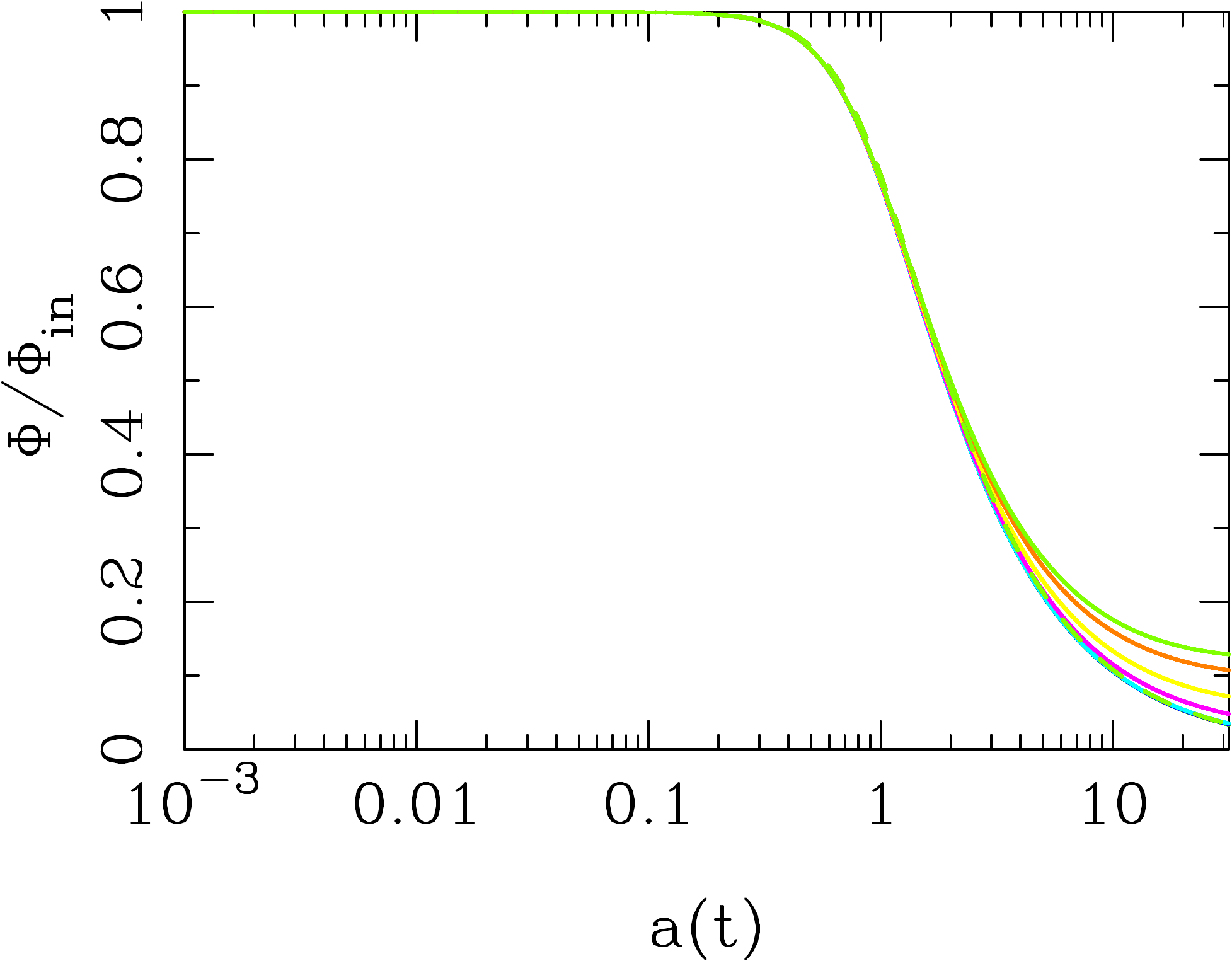}
      \caption{Here we show the evolution of gravitational potential
        with scale factor. The top row is for the inverse square
        potential and the second row is for the exponential
        potential. Solid lines correspond to the respective scalar
        field potential whereas the dashed lines are for $\Lambda CDM$
        model. The plots on the left show for $\phi_{in}H_0=~1.0$ and
        those on the right correspond to  $\phi_{in}H_0= ~2.0$
        respectively, for the same values of $C_n$ and $V_a/\rho_{cr}$
        as in the previous two figures. The red, green, blue,
        sky-blue, pink, yellow, orange and light-green  colours
        represents the scales of perturbation $\lambda_p= ~50,~100,
        ~500, ~1000, ~5000, ~10000,~20000 ~\text{and} ~50000~Mpc$
        respectively.} 
      \label{fig:gvptnl}
    \end{figure}
In figure~\ref{fig:gvptnl}, we show the evolution of the gravitational
potential with the scale factor.
The gravitational potential is normalized to its initial value; solid
lines are for tachyon models and dashed lines are for $\Lambda CDM$
model.  
Different colours represent different length scales of the
perturbation, $\lambda_p$, from $50~Mpc$ to $5\times 10^4~Mpc$.
We solve the set of required equations for each of these fixed scales,
introduced using the dimensionless ratio $\bar{k}=kc/H_0$, where
$k=2\pi/\lambda_p$; with $H_0=70~Km s^{-1}Mpc^{-1}$ and
$c=2.99\times10^5~Km s^{-1}$.  
The gravitational potential remains a constant during the matter
dominated era. 
As dark energy starts to dominate the energy budget, gravitational
potential falls at all length scales.  
We can see that for $\Lambda CDM$ model, the gravitational potential
falls more rapidly and  at the same rate at all scales.
For tachyon models, the gravitational potential falls more rapidly at a
smaller scales. 
At super-Hubble scales, its decay slows down in future.
In the bottom left panel of figure~\ref{fig:gvptnl}, is can be seen that for
the exponential potential, the gravitational potential at super-Hubble scales
in future first rises and then become constant.
However, as we increase the value of parameter $\phi_{in}H_0$ (because $w_{\phi 0}\to -1$),
this effect of scale dependence decreases, 
and the difference with respect to the $\Lambda CDM$ model also decreases.
The model with exponential potential is more sensitive
to the value of the parameter $\phi_{in}H_0$, as we can see that
increasing this parameter from $~1$ to $~2$ decreases the scale dependence
effect more significantly. 

\begin{figure}[t]
  \centering
  \includegraphics[width=0.45\textwidth]
	  {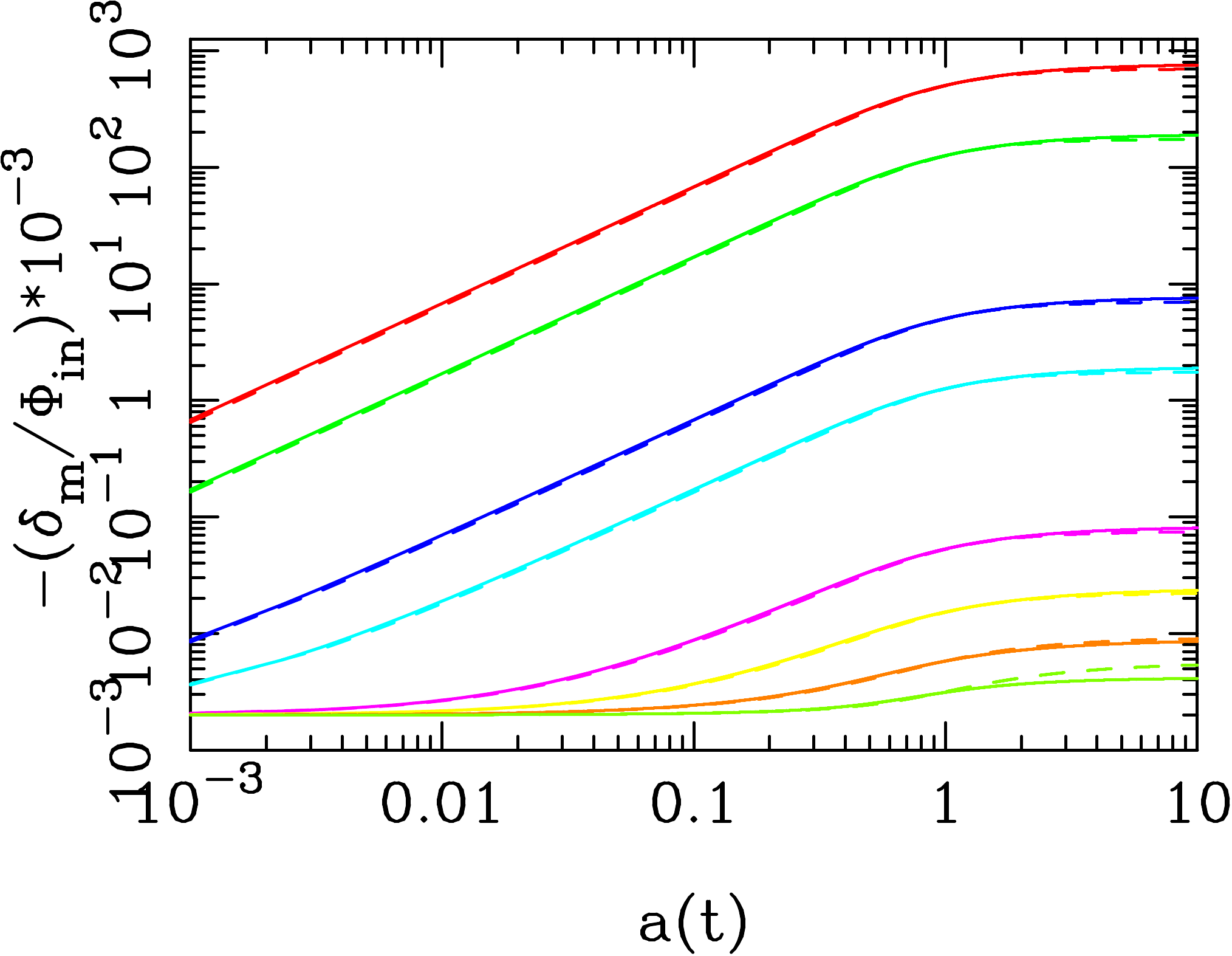}
  \includegraphics[width=0.45\textwidth]
	  {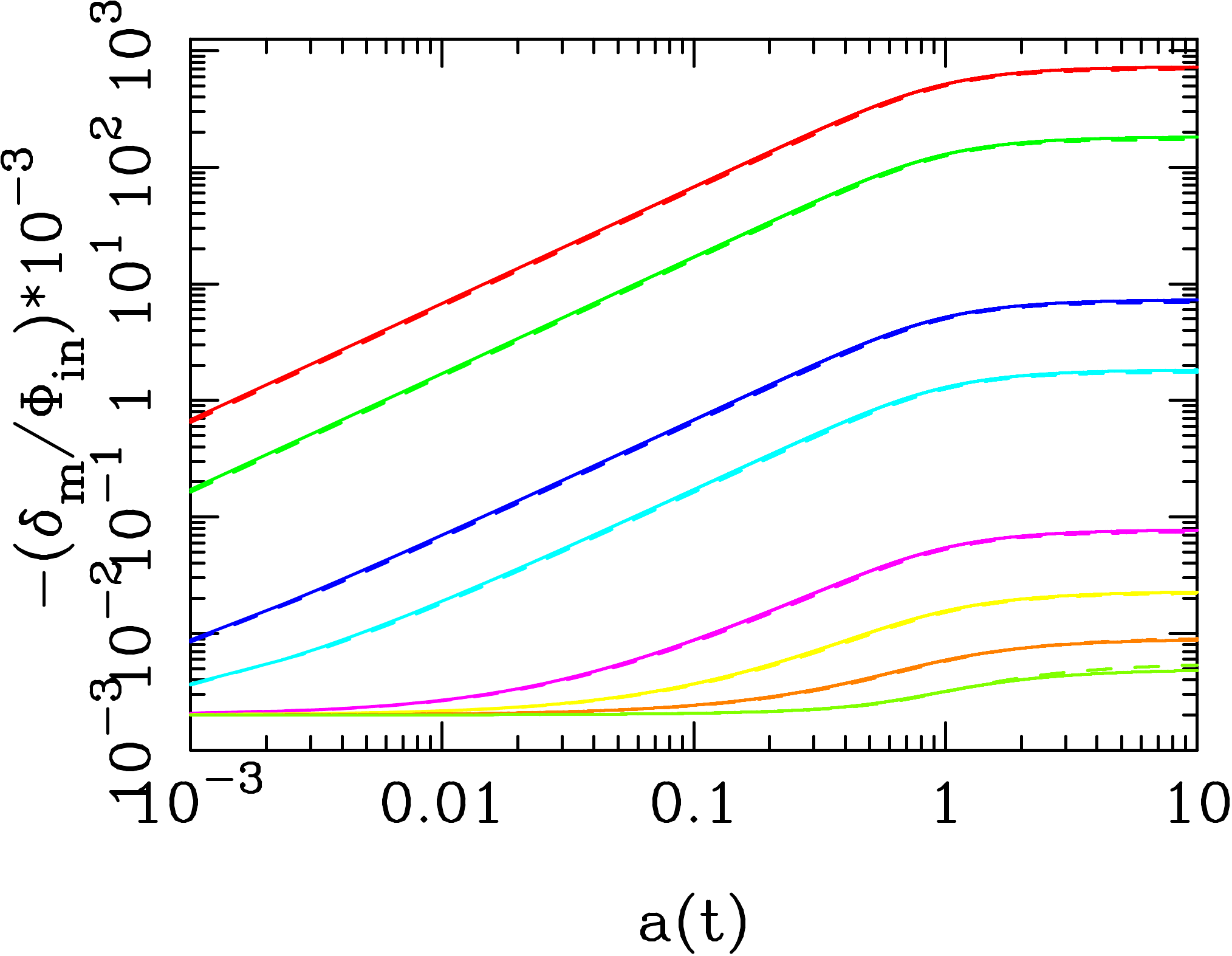}\\
  \includegraphics[width=0.45\textwidth]
	  {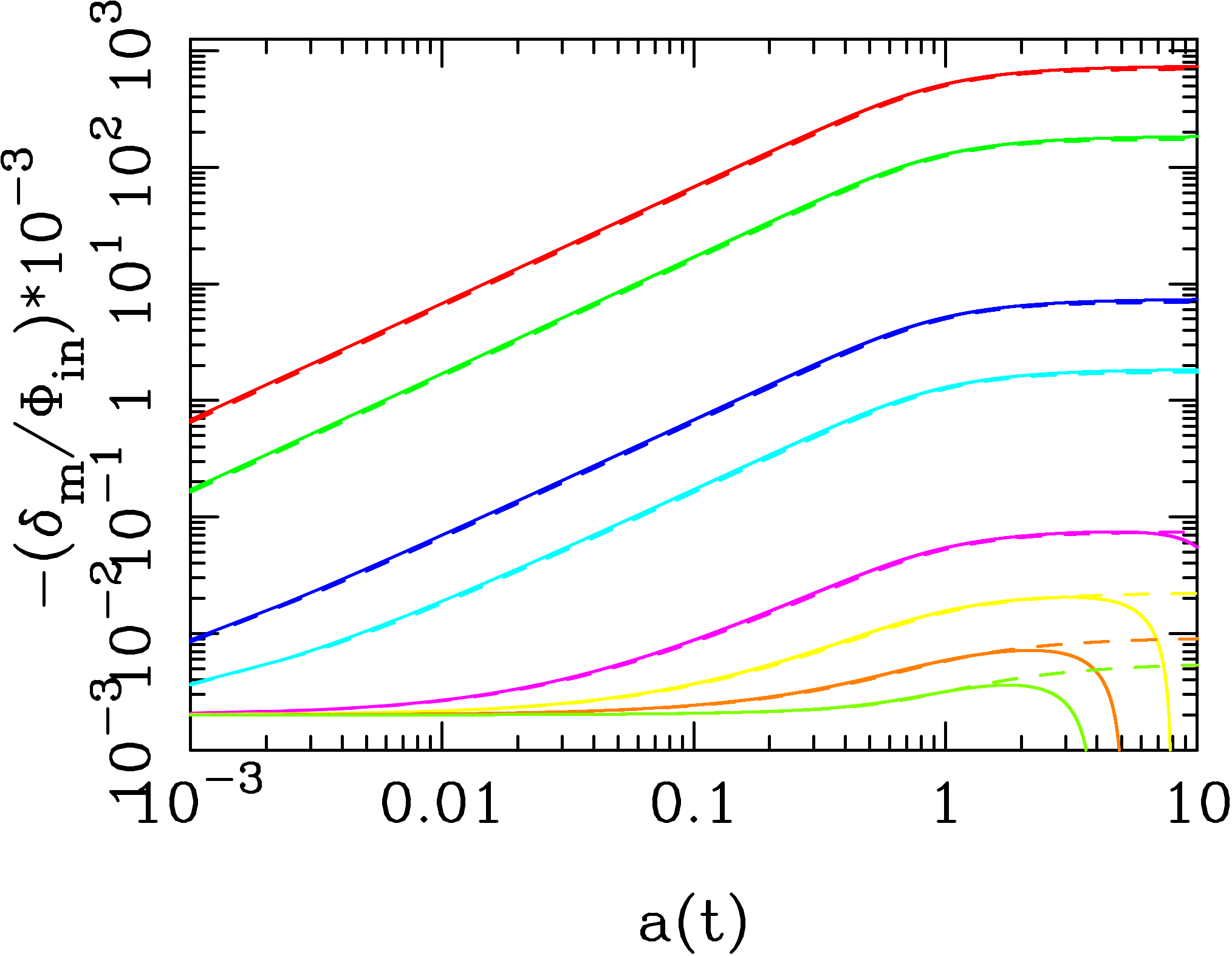}
  \includegraphics[width=0.45\textwidth]
	  {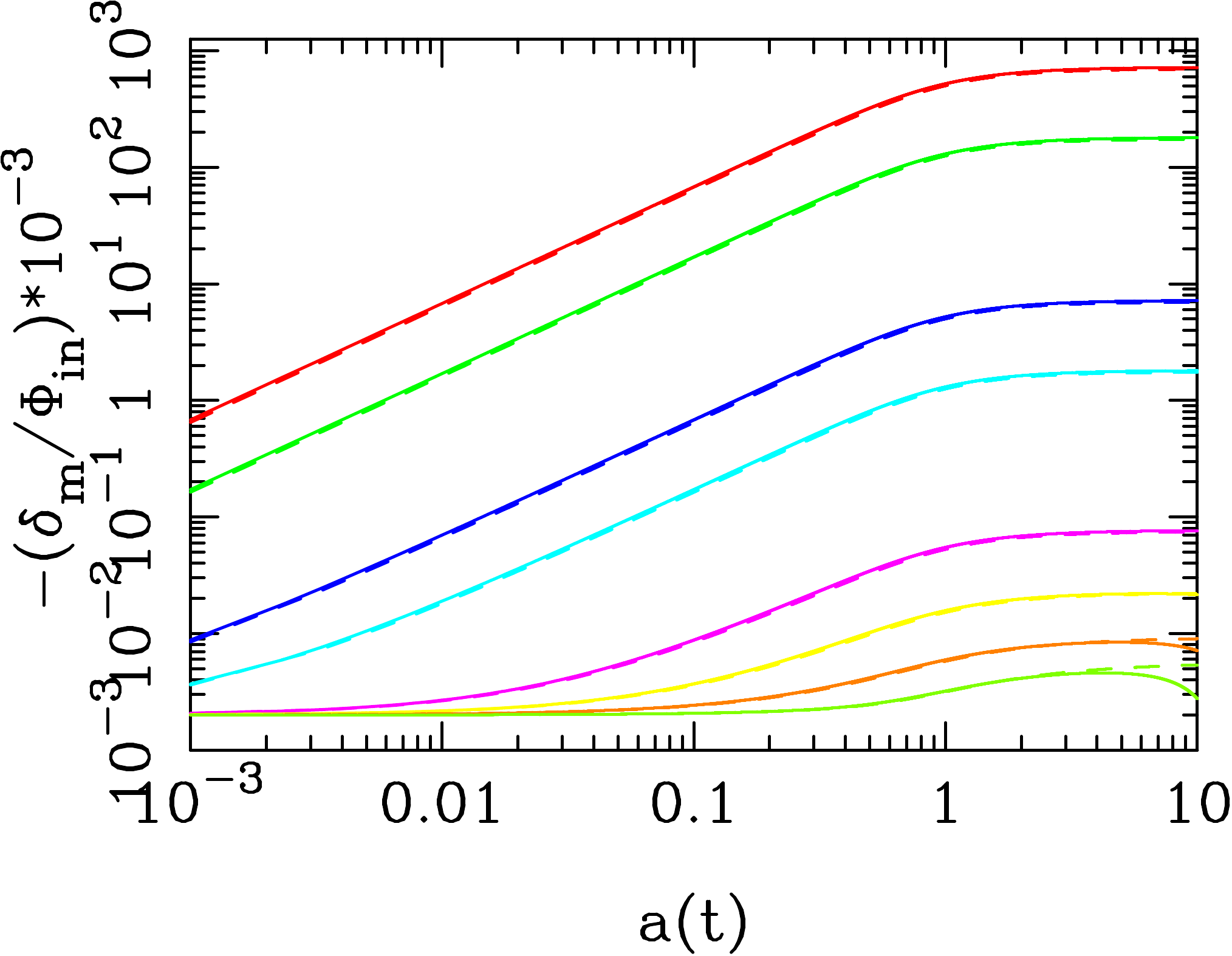}
  \caption{Evolution of matter density contrast with the scale factor
    is shown in the above figure. Solid lines correspond to tachyon
    dark energy model (with inverse square potential in row-1  and
    with exponential potential in row-2) and the  dashed lines are
    for $\Lambda CDM$ model. The values of $\phi_{in}H_0=$~1.0 and
    ~2.0 for column-1 and 2 respectively.
    The colour scheme for this figure is the same as in~\ref{fig:gvptnl}.}
  \label{fig:mtrcntrst}
\end{figure}
\begin{figure}[t]
	\centering
	   \includegraphics[width=0.45\textwidth]{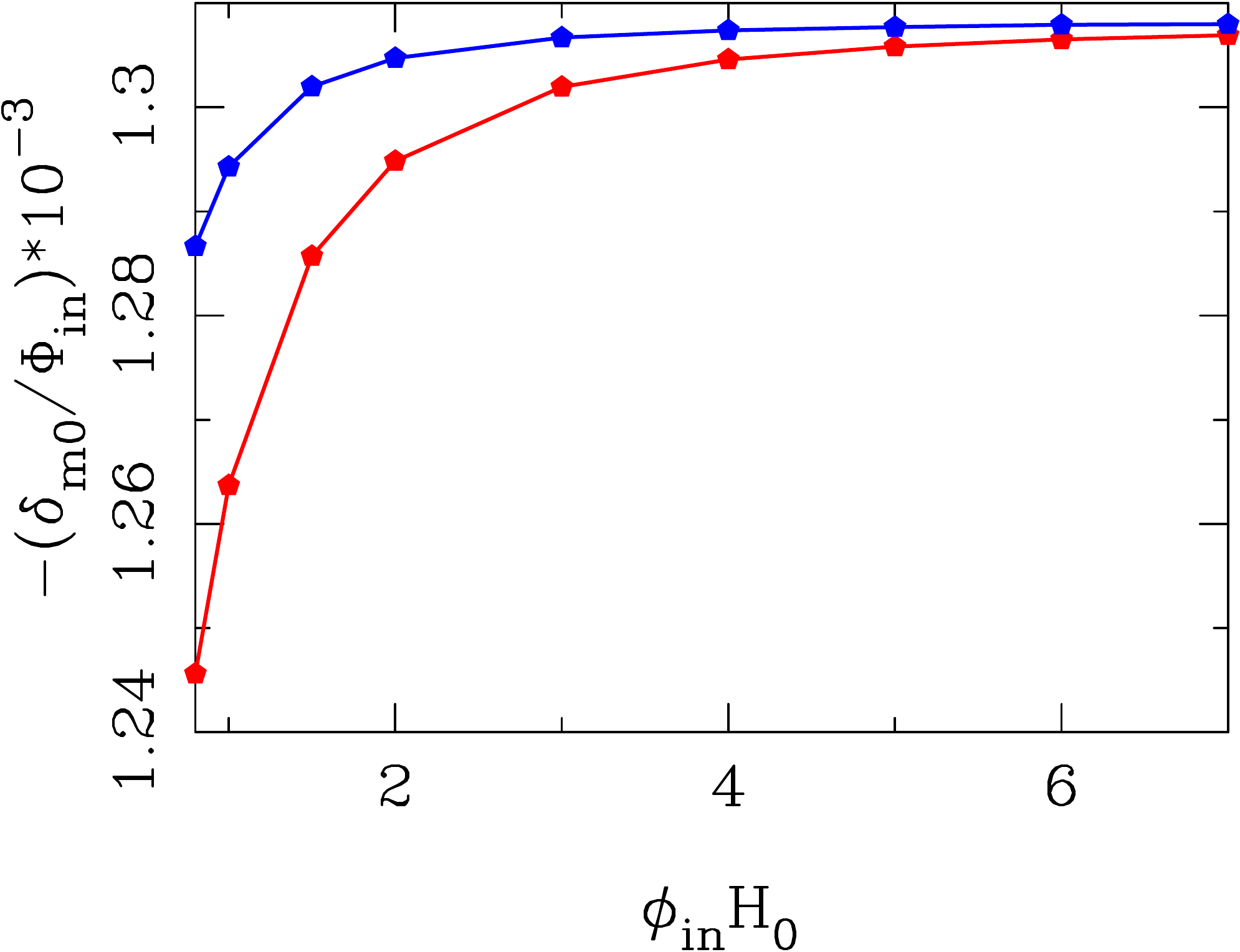}
	   \includegraphics[width=0.45\textwidth]{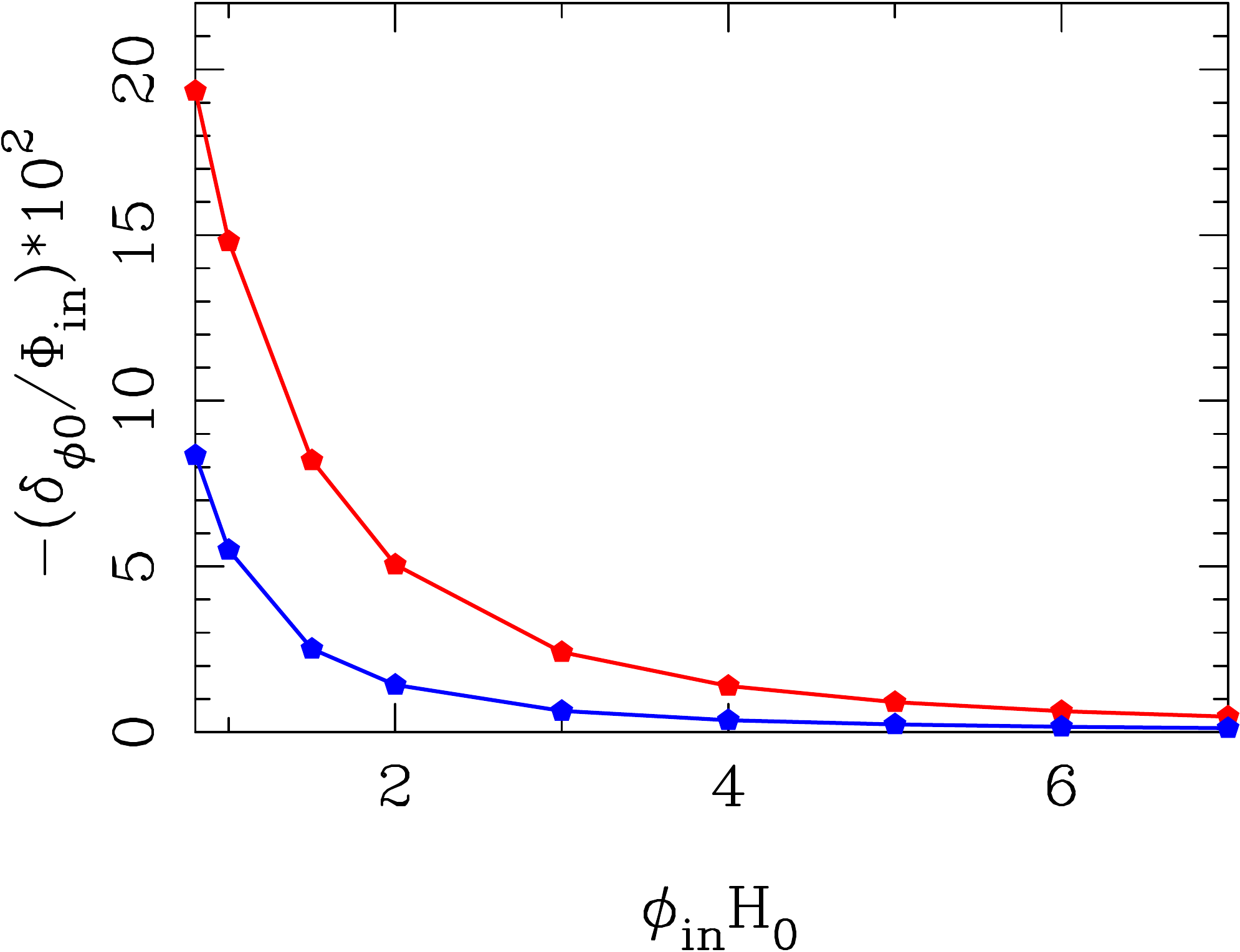}
	   \caption{This figure shows the  matter density
             contrast (in the left panel) and the dark energy density density contrast
             (in the right panel) normalized to initial gravitational potential,
             at present epoch ($z=0$), as a function of $\phi_{in}H_0$ at the scale of
             $\lambda_p=~1000~Mpc$. Red and blue colours represent
             the inverse square and exponential potential respectively.}
	   \label{fig:hph0VSdelta}
\end{figure}
The evolution of matter density contrast, normalized by the initial
value of the gravitational potential is shown in
figure~\ref{fig:mtrcntrst}, for $\phi_{in}H_0=~1.0$ and $~2.0$. 
Since the gravitational potential remains constant during the
matter-dominated era, at sub-Hubble scales the  matter density contrast
grows linearly with the scale factor i.e. $\delta_m \propto a$, whereas
at Hubble and super-Hubble scale it evolves at a slower rate. 
In the matter dominated era, there is a very small difference between
tachyon model (for both the potentials) and $\Lambda CDM$ model (dashed
lines). 
In the dark energy dominated era, the evolution of matter density
contrast is suppressed. 
At Hubble and super-Hubble scales, it once again increases
(for the inverse square potential) and decreases (for the exponential
potential) in future as the gravitational potential seizes to decay.
This difference in the behavior of the matter density contrast in future
is due to the difference in the evolution of the equation of state
parameter and the gravitational potential.  
Whereas in the $\Lambda CDM$ model, the evolution of the matter
density contrast remains suppressed in the $\Lambda$ dominated era.
The evolution of $\delta_m$ depends on the parameter $\phi_{in}H_0$ (or on
$w_{\phi 0}$).  
In the left panel of figure~\ref{fig:hph0VSdelta}, we show
the dependence of $\delta_m/\Phi_{in}$ at $z=0$ at the scale of $\lambda_p=1000~Mpc$ 
on $\phi_{in}H_0$.
For smaller value of $\phi_{in}H_0$ (or larger $w_{\phi 0}$), the
present day value of $\delta_m(z=0)$ is small, and as we increase
$\phi_{in}H_0$ and $w_{\phi 0}$ decreases, the value of
$\delta_m(z=0)$ increases.  
For larger values of $\phi_{in}H_0$, its value approaches a constant
as decrease in $w_{\phi 0}$ saturates.
For a fixed value of $\phi_{in}H_0$, the value of $\delta_m(z=0)$ is
large for  the exponential potential than it for the inverse square
potential. 
For a fixed $\phi_{in}H_0$, the value of 
$w_{\phi 0}$ is smaller for the exponential potential than it is for the
inverse square potential.
As we increase the value of the parameter $\phi_{in}H_0$ and $w_{\phi 0}$
approaches $-1$, the difference between the two potentials decreases.

\begin{figure}[t]
	\centering
	\includegraphics[width=0.45\textwidth]
					{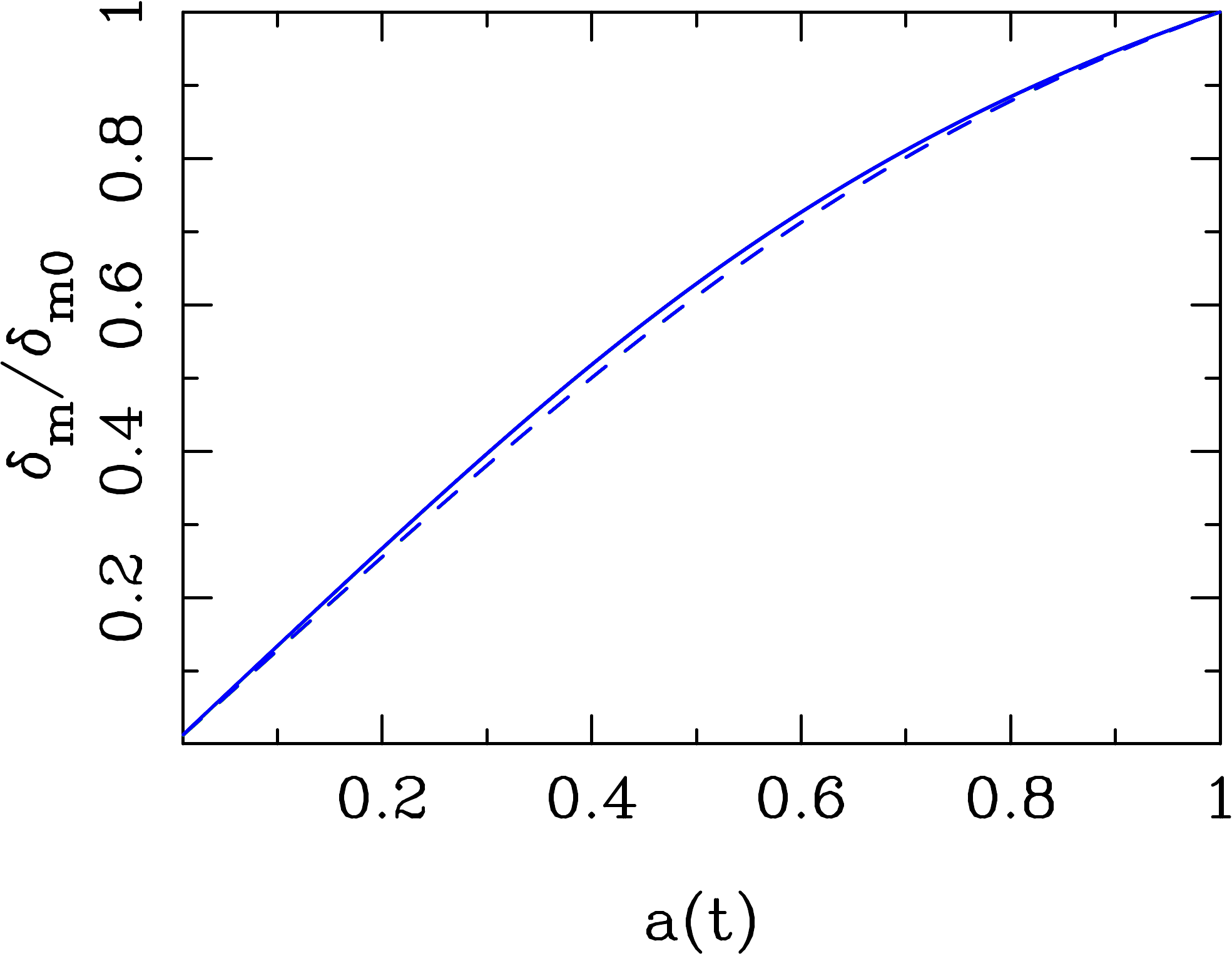}
	\includegraphics[width=0.45\textwidth]
					{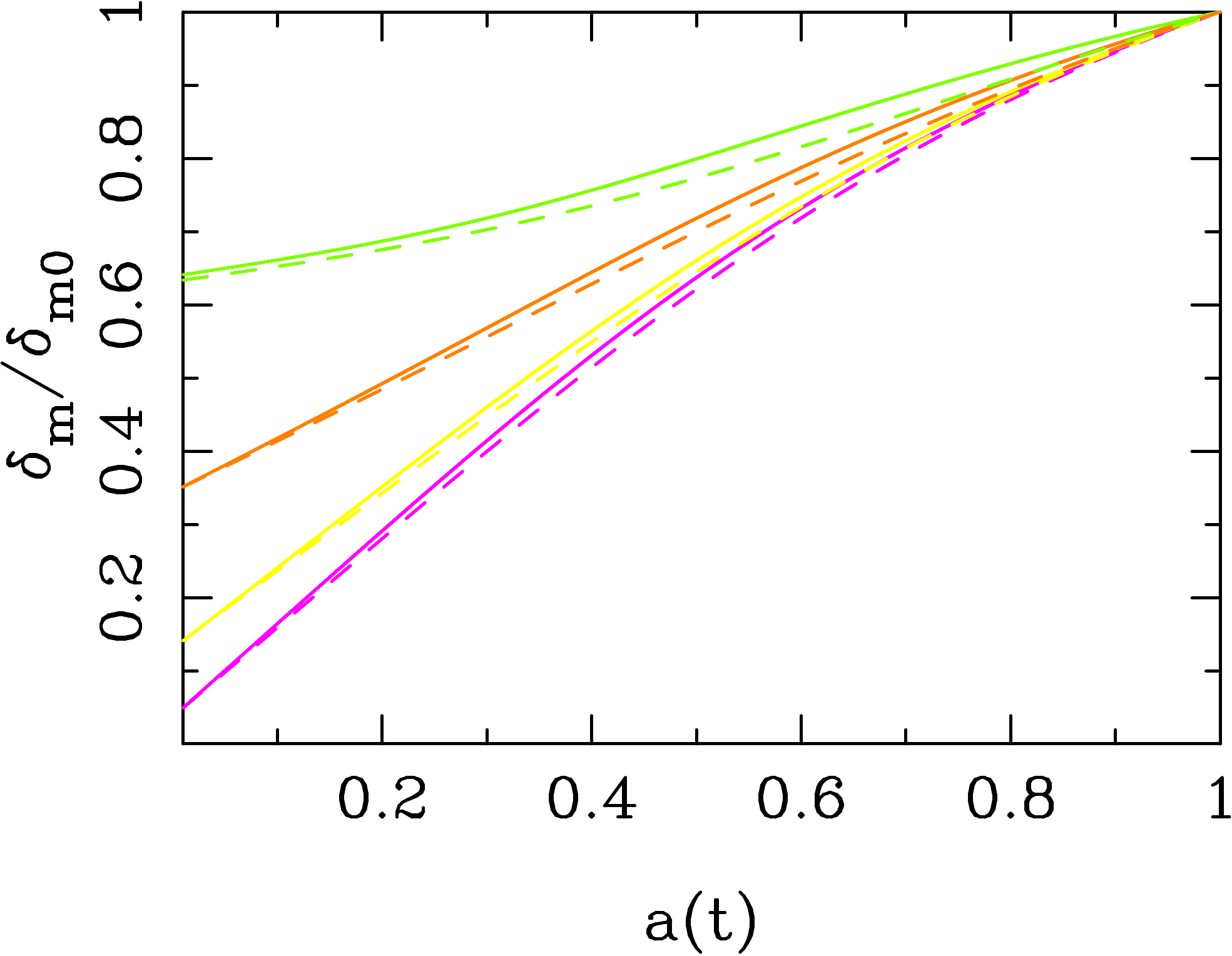}\\
	\includegraphics[width=0.45\textwidth]
					{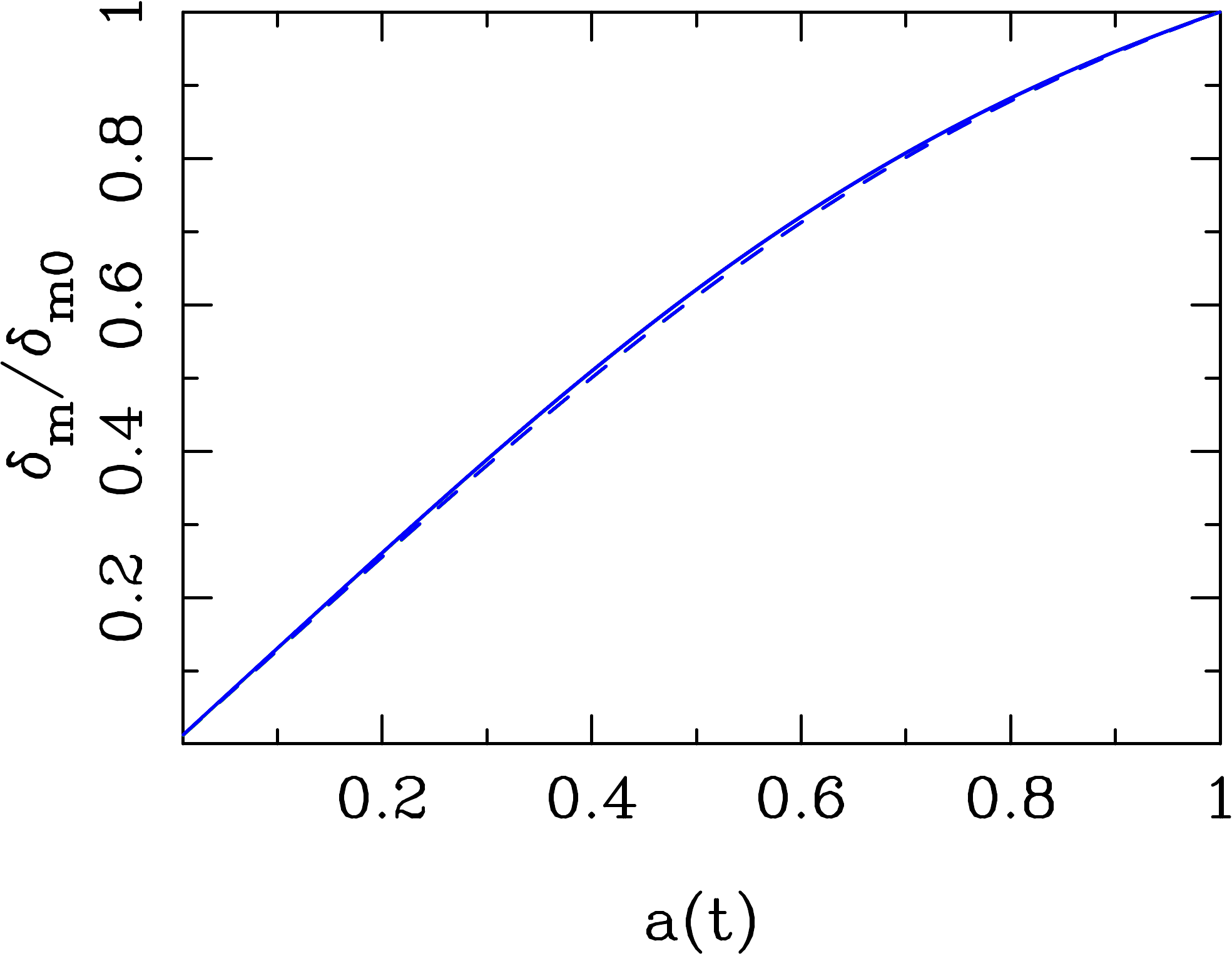}
	\includegraphics[width=0.45\textwidth]
					{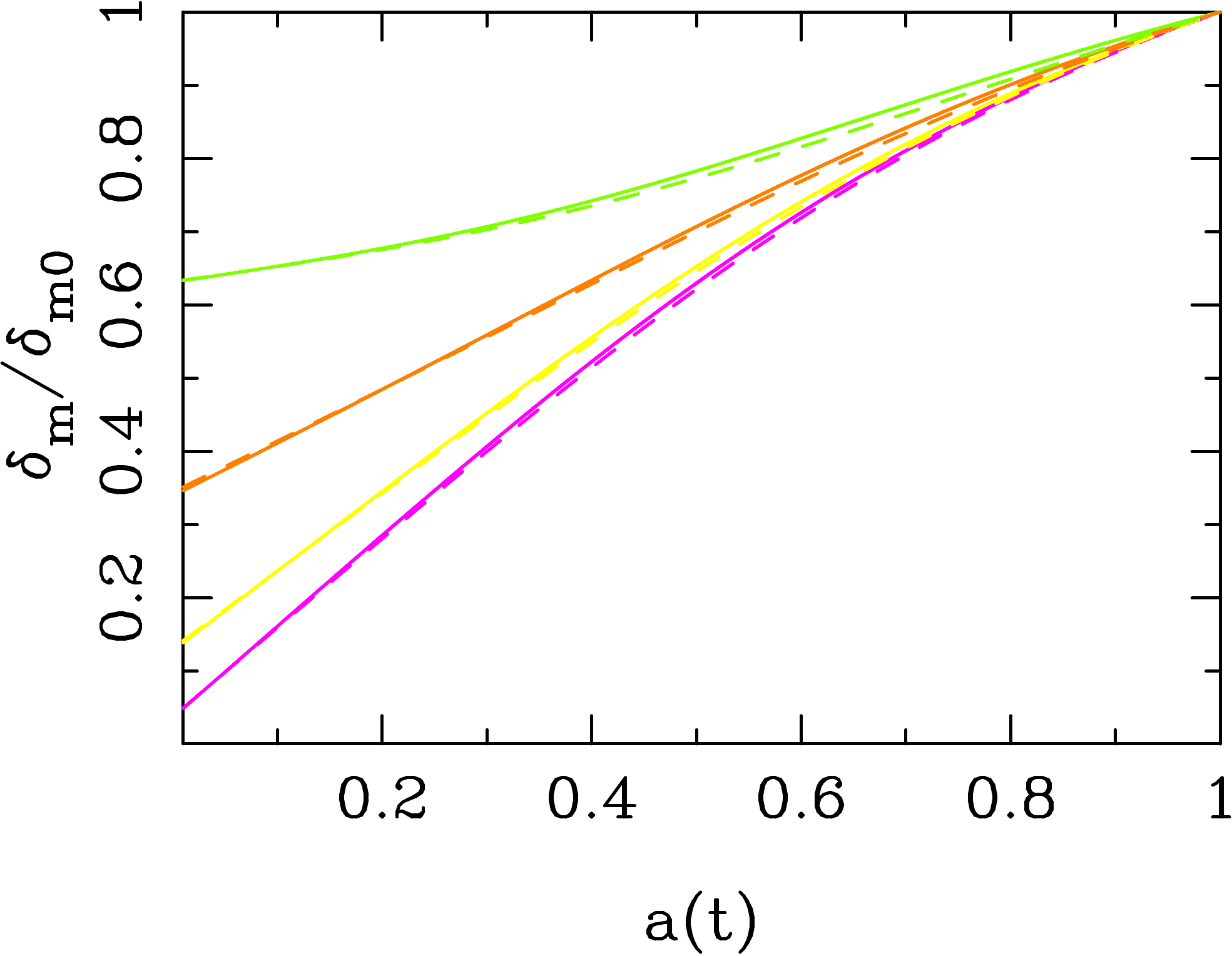}
	\caption{  In this figure we show the evolution of linear growth function
		of matter  $D^+_m=\frac{\delta_m}{\delta_{m0}}$ as a
		function of the scale factor. Solid lines correspond to
		tachyon dark energy model with inverse square potential (in
		row-1) and with exponential potential (in row-2) and the
		dashed lines are for $\Lambda CDM$ model. The colour scheme for scales
		of perturbation is the same as in the figure~\ref{fig:gvptnl}.
		Column-1 is for sub-Hubble scales (lines of different scales of particular
		model have overlapped) and column-2 is for super-Hubble scales.}
	\label{fig:lineargrowth}
\end{figure}

\begin{figure}[t]
	\centering
	\includegraphics[width=0.45\textwidth]
	{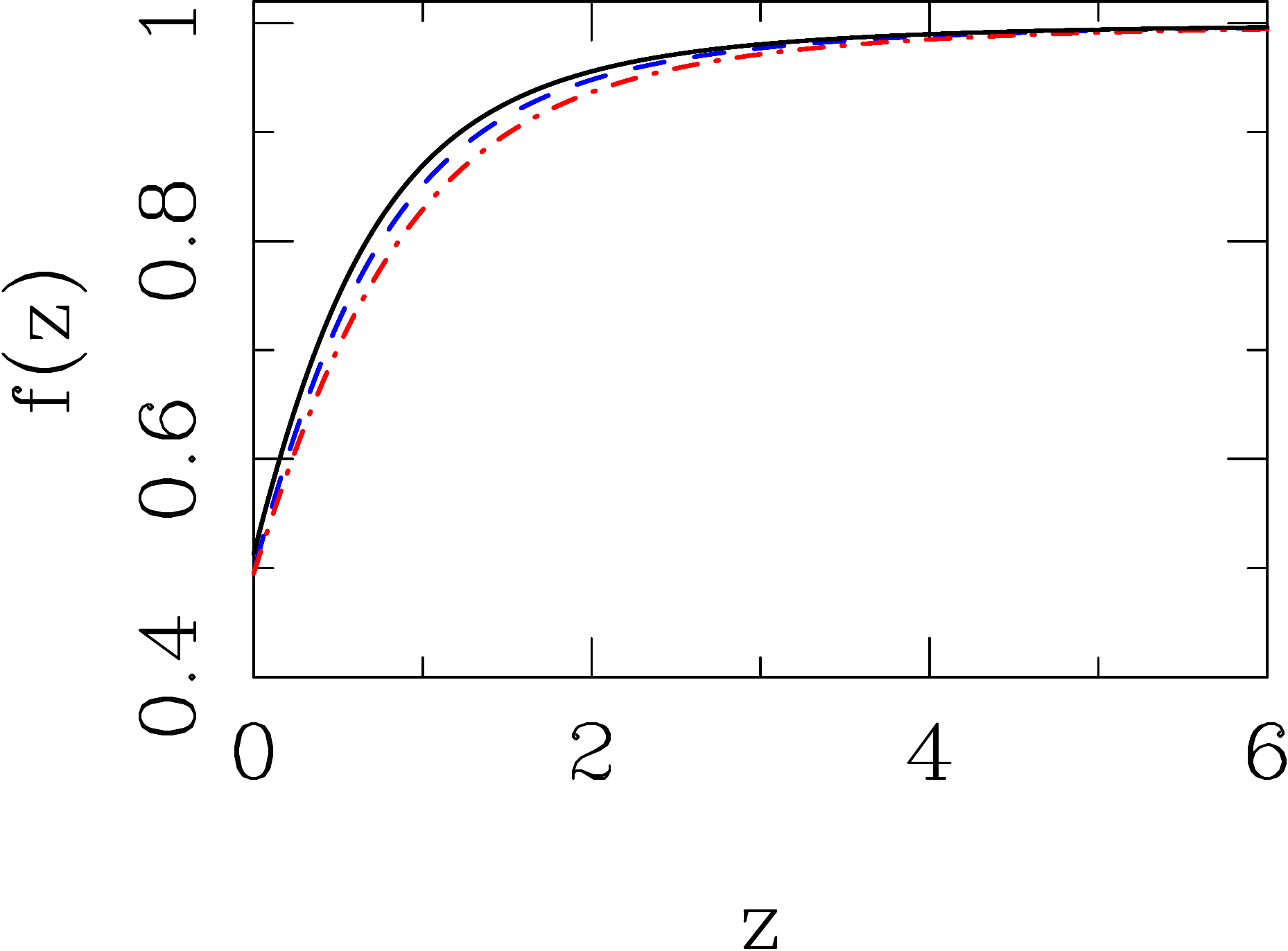}\\
	\includegraphics[width=0.45\textwidth]
	{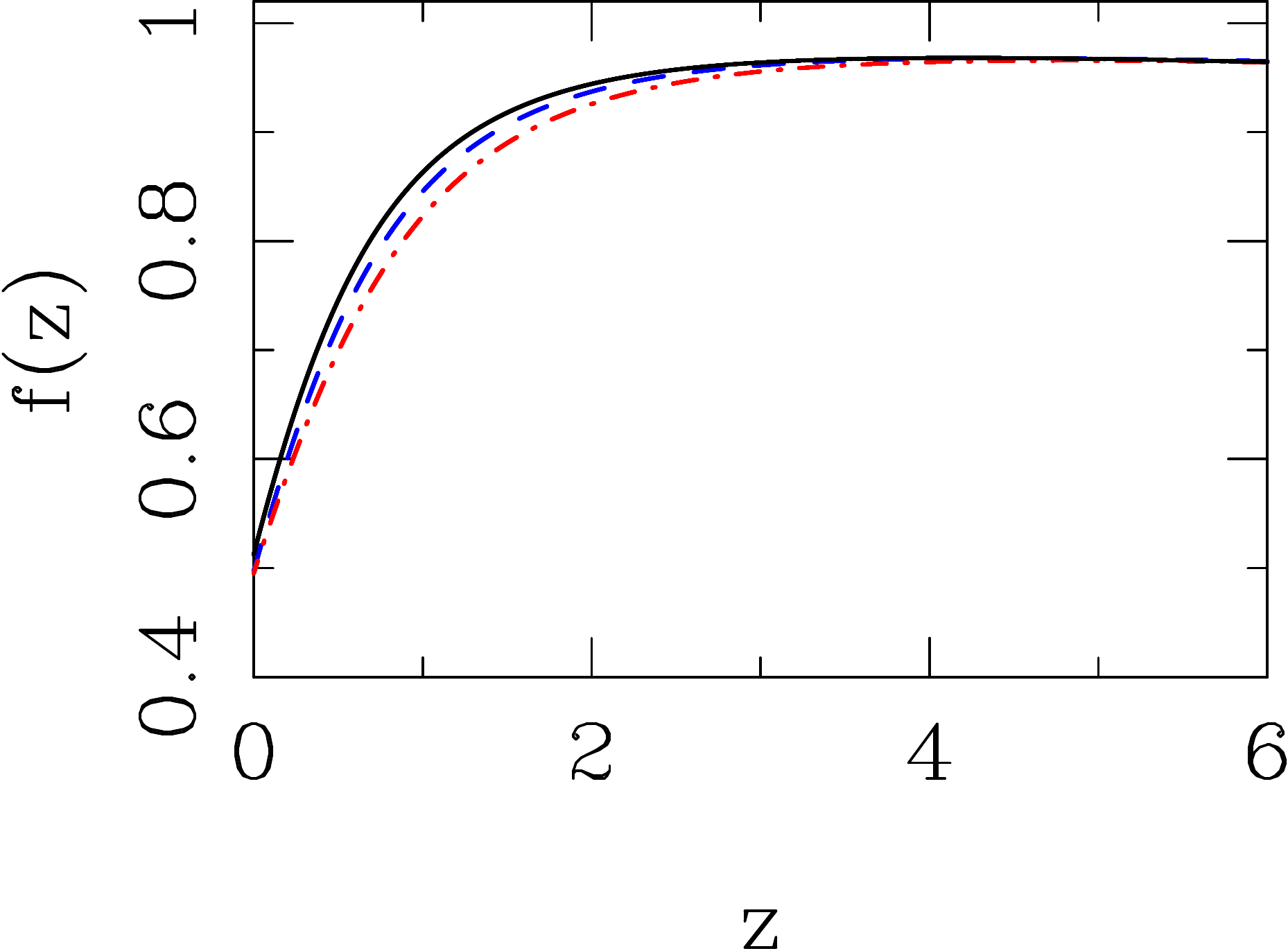}
	\includegraphics[width=0.45\textwidth]
	{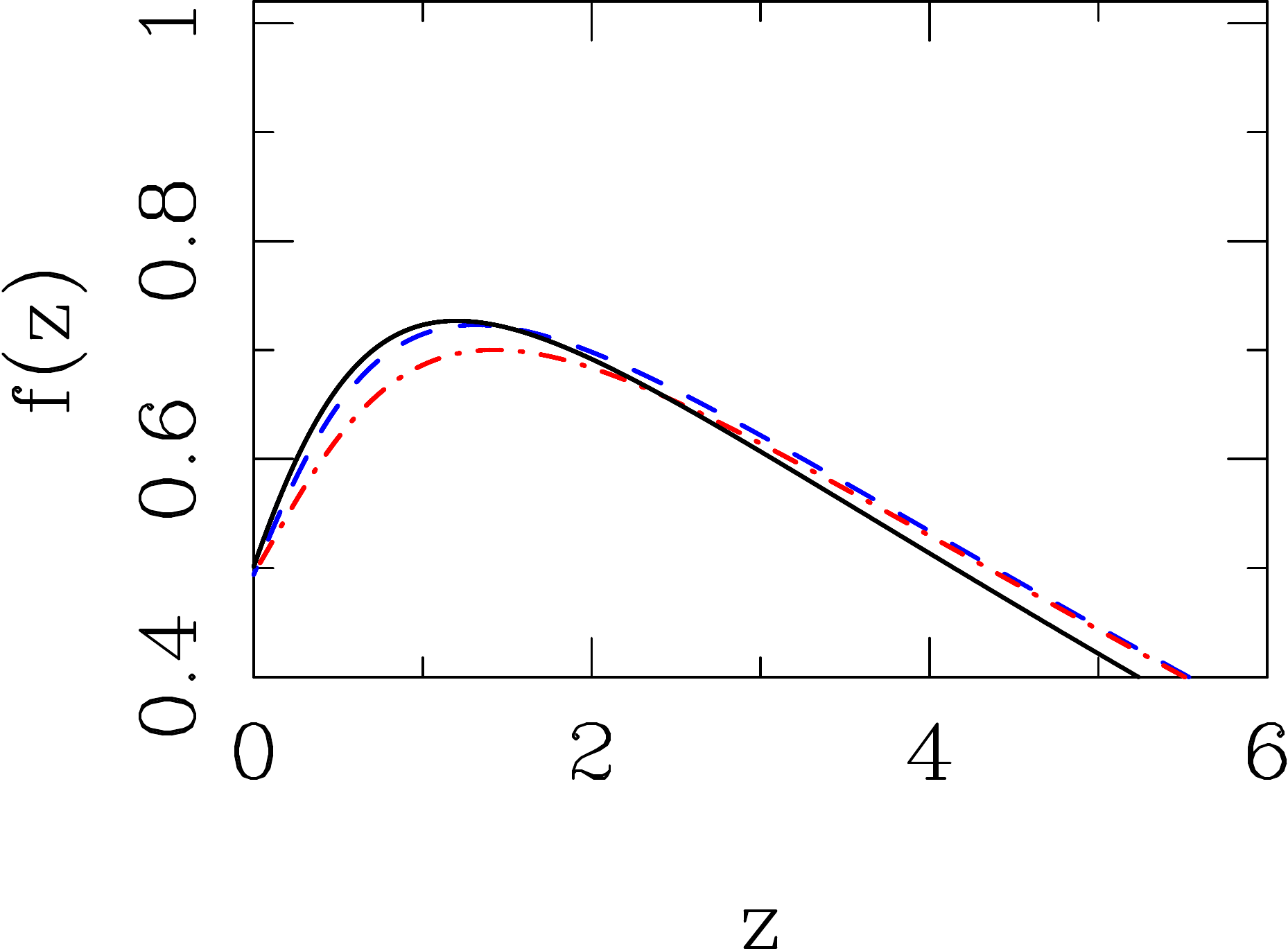}
	\caption{The evolution of the logarithmic growth rate
		$f=\frac{d~ln~\delta_m}{d~ln~a}$ with redshift is shown here.
		Solid black, dashed blue and dashed-dot red curves are for $\Lambda CDM$ model,
		tachyon model with exponential potential and with inverse square potential
		respectively.
		Top panel is for scale of perturbation $\lambda_p=~50~Mpc$, 
		whereas bottom left and right panels are for $\lambda_p=~1000~Mpc ~\text{and}~5000~Mpc$ respectively.
		For these plots, the value of parameters
		$\phi_{in}H_0=1.0$ and $\Omega_{m0}=0.285$.} 
	\label{fig:growthrate}
\end{figure}

In figure~\ref{fig:lineargrowth}, we show the evolution of linear
growth function $D^+_m=\frac{\delta_m}{\delta_{m0}}$ at sub-Hubble
(the plot on the left) and super-Hubble scales (the plot on the
right). 
Here we have taken the value of parameters $\phi_{in}H_0=1.0$ and
$\Omega_{m0}=0.285$. 
We can see that at sub-Hubble scales, linear growth is scale
independent as all lines overlap. 
At super-Hubble scales, its evolution depends on the scale.
In matter dominated era, the linear growth $D^+_m$ is large for
tachyon models than the $\Lambda CDM$ model at all scales.
That is why as dark energy dominates it has to slow down, even more
than $\Lambda CDM$ model to match the present value.
This becomes more clear in figure~\ref{fig:growthrate}, where we show
the evolution of growth rate  $f= \frac{d~ln~\delta_m}{d~ln~a}$ with
redshift, at the scale of perturbation $\lambda_p=~50, ~1000
~\text{and} ~5000 ~Mpc$, for $\phi_{in}H_0=~1.0$.
We can see that the growth rate is higher at shorter scales, and as we
increase the scale of perturbation growth rate decreases.
We can also see that in matter-dominated era, the growth rate remains
 a constant for smaller scales (sub-Hubble scales), whereas at
Hubble and Supper-Hubble scale it grows linearly and reaches a maximum
value. 
In the dark energy dominated era the growth rate falls at all scales,
for all the three models.
In the matter-dominated era, the growth rate is larger for tachyon models
than the $\Lambda CDM$ model.
As the dark energy starts to dominate, it comes below the $\Lambda
CDM$ model. 
As we increase the value of $\phi_{in}H_0$, the tachyon model
approaches the $\Lambda CDM$ model (because $w_{\phi 0}\to -1$) and this
difference decreases. 

\begin{figure}[t]
	\centering
	 \includegraphics[width=0.45\textwidth]{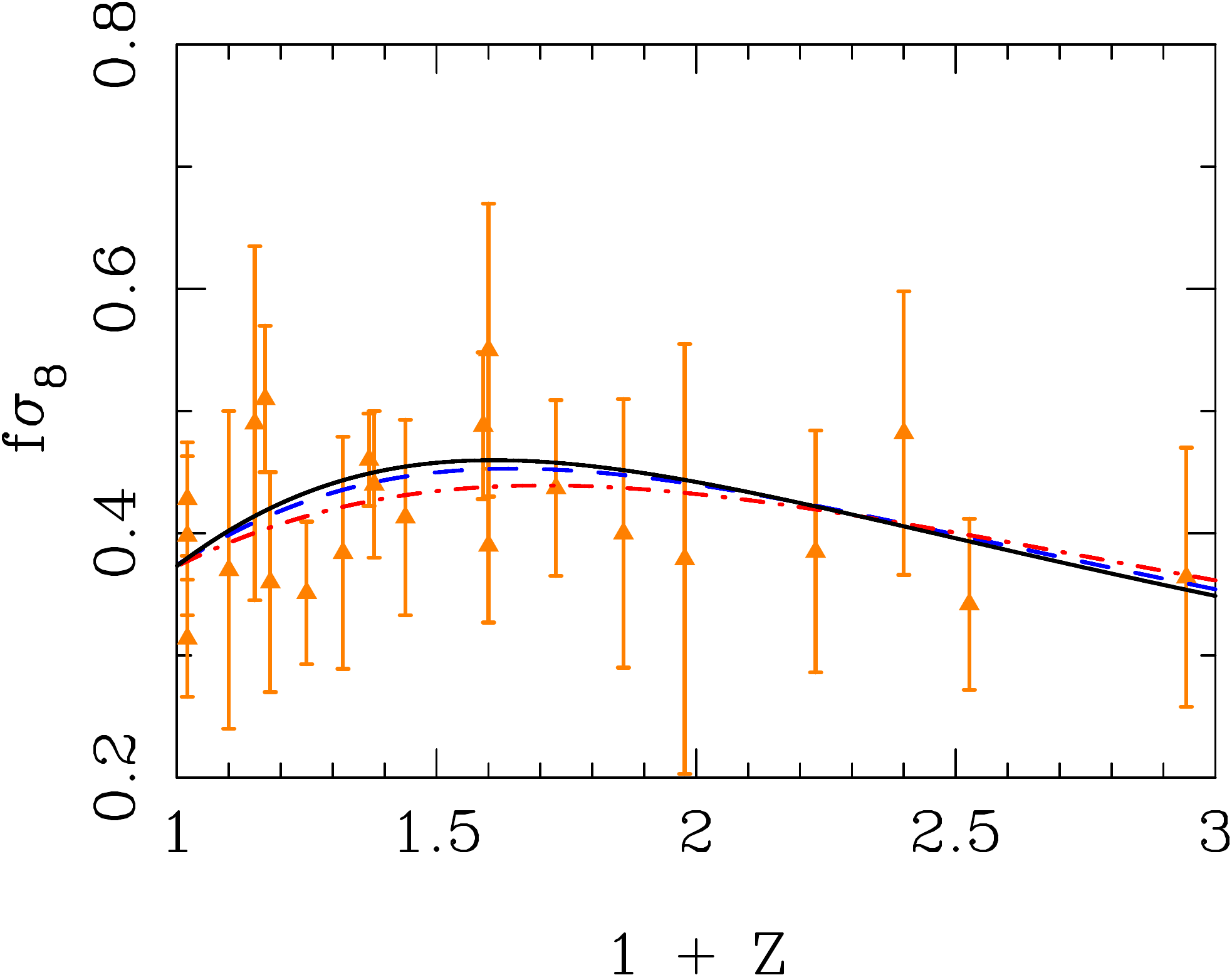}
	 \includegraphics[width=0.45\textwidth]{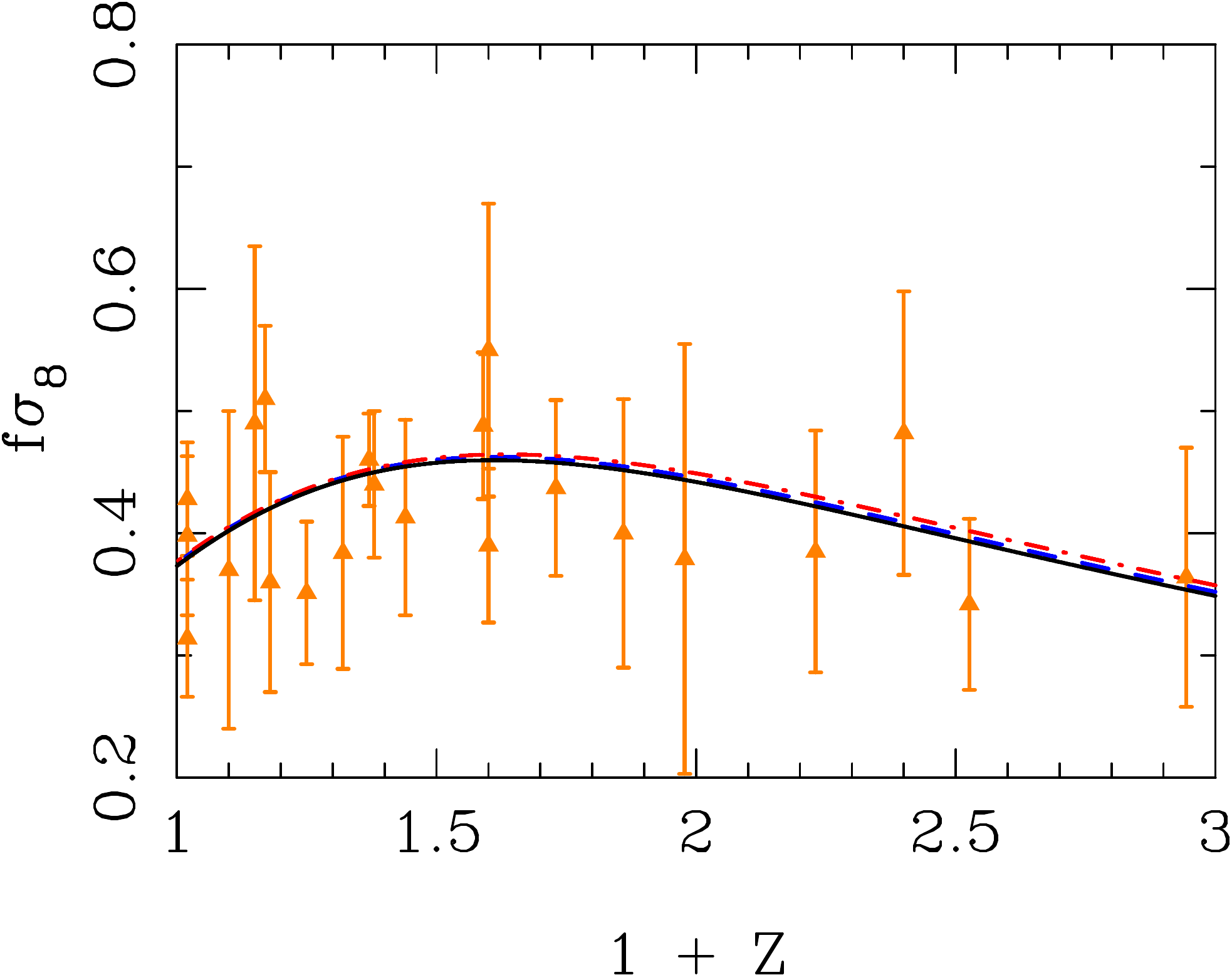}
	 \caption{A comparison of theory with  Redshift Space
           Distortion data. Solid black, dashed blue and dashed-dot
           red curves are for $\Lambda CDM$ model, tachyon model with
           exponential potential  and with inverse square potential
           respectively. The left and right panels represent
           $\phi_{in}H_0=$~0.8 and ~3.0 respectively.  Other
           parameters $\Omega_{m0}$ and $\sigma_8(z=0)$ are fixed to
           the  corresponding best fit values taken from
           table~\ref{tab:ommSigma8}.  Data points are taken from
           table I of~\cite{Sagredo2018}.} 
	\label{fig:fsigma8data}
\end{figure}

\begin{figure}[t]
	\centering
	\includegraphics[width=0.45\textwidth]
		{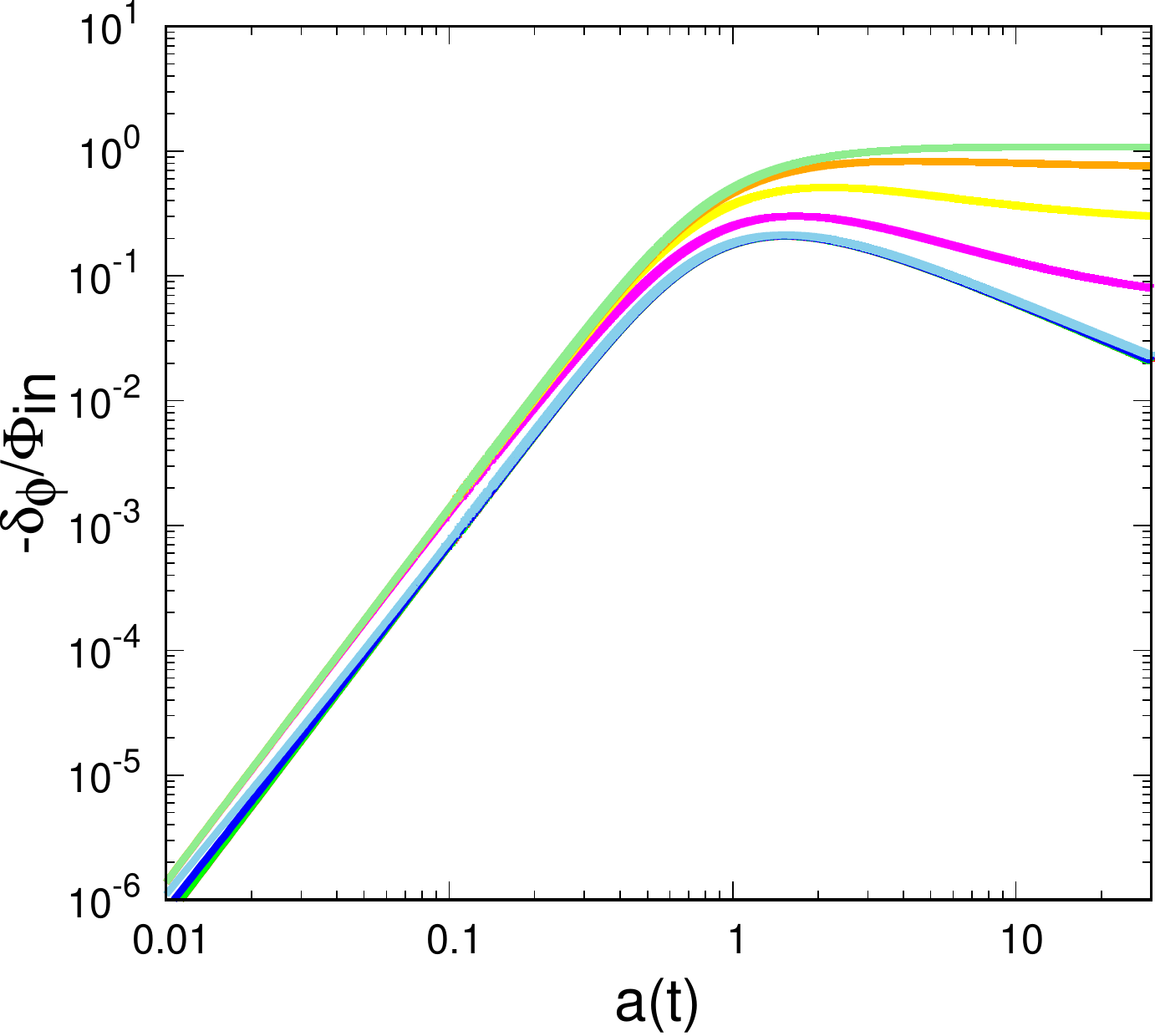}
	\includegraphics[width=0.45\textwidth]
		{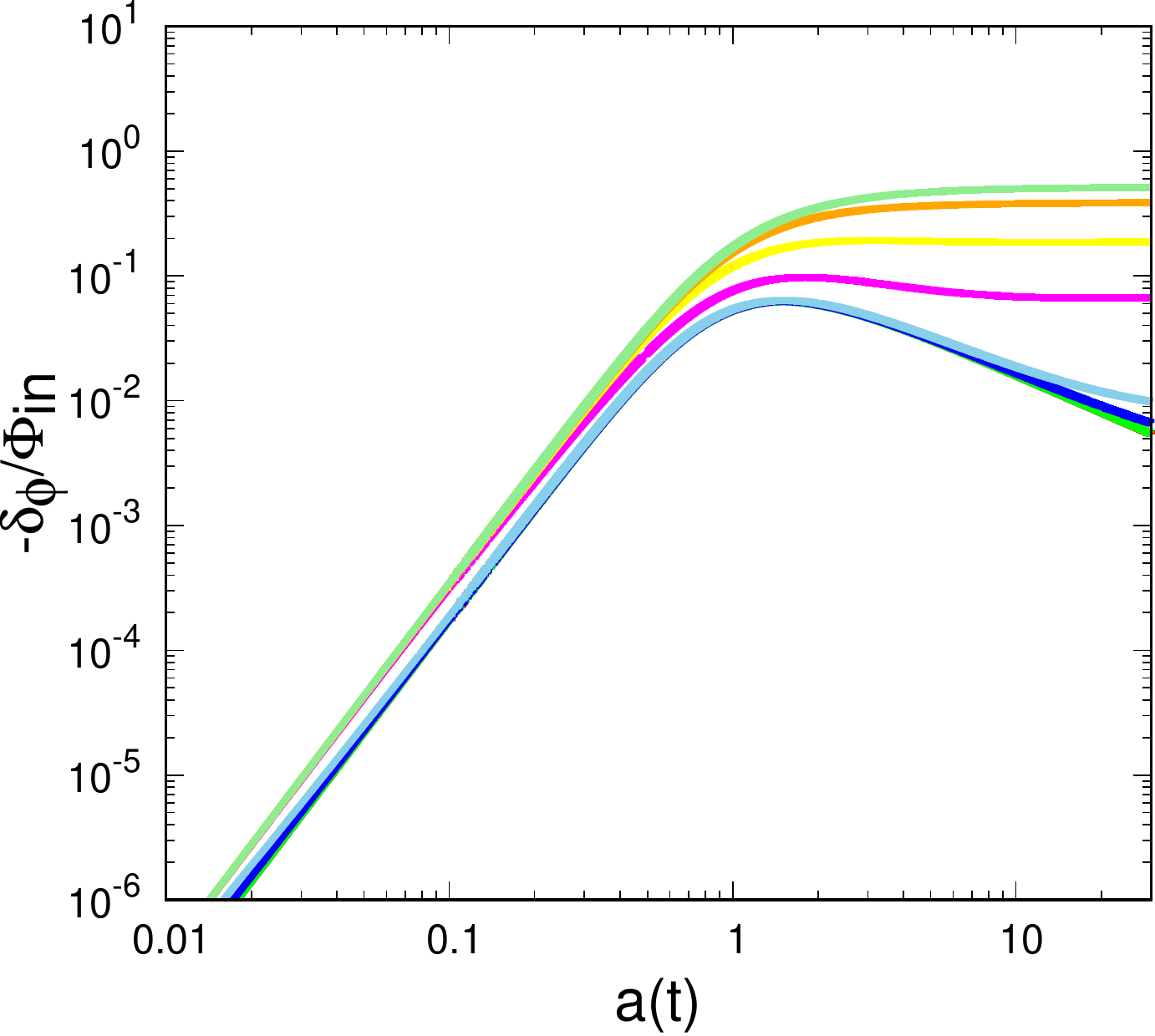}\\
	\includegraphics[width=0.45\textwidth]
		{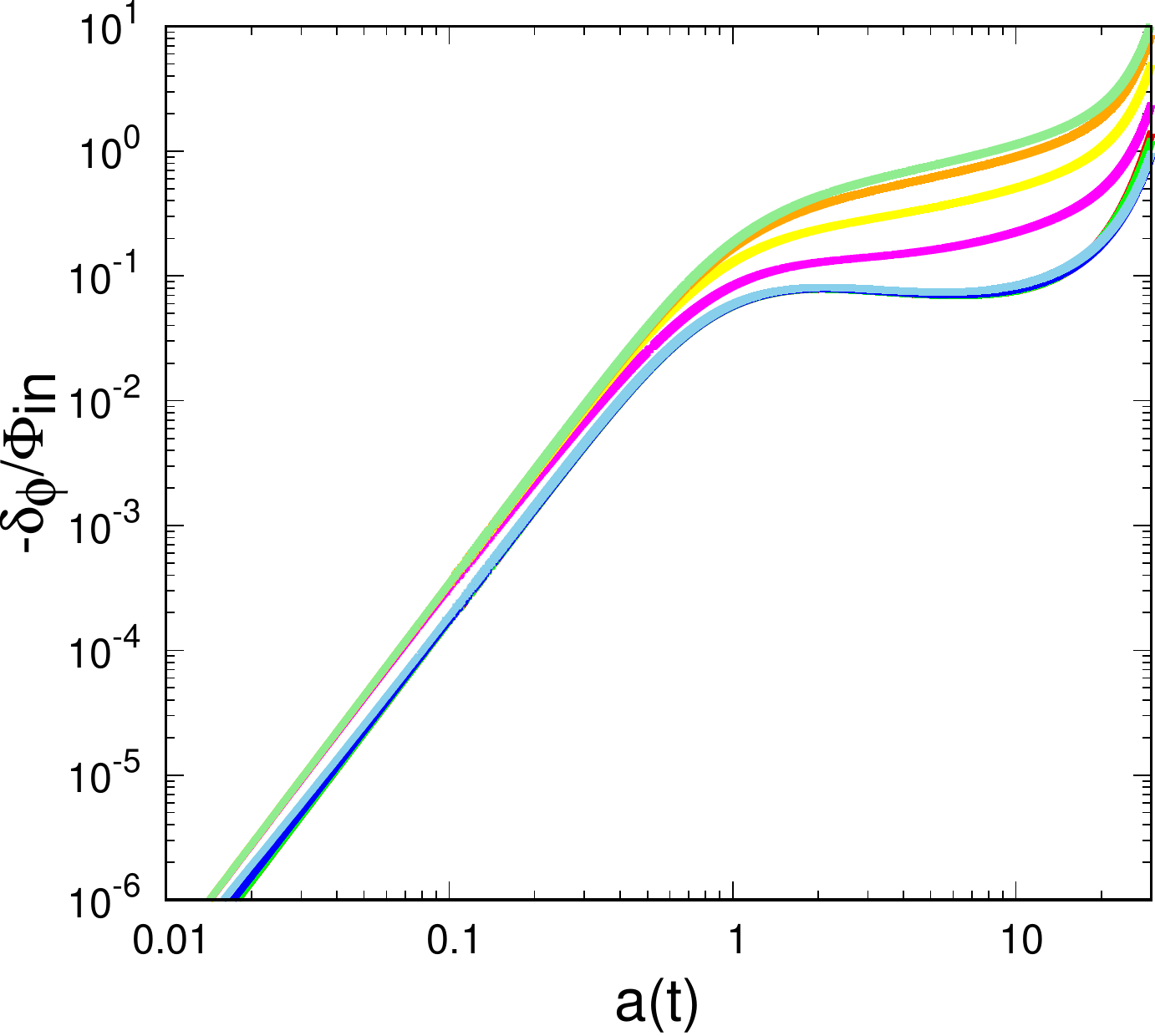}
	\includegraphics[width=0.45\textwidth]
		{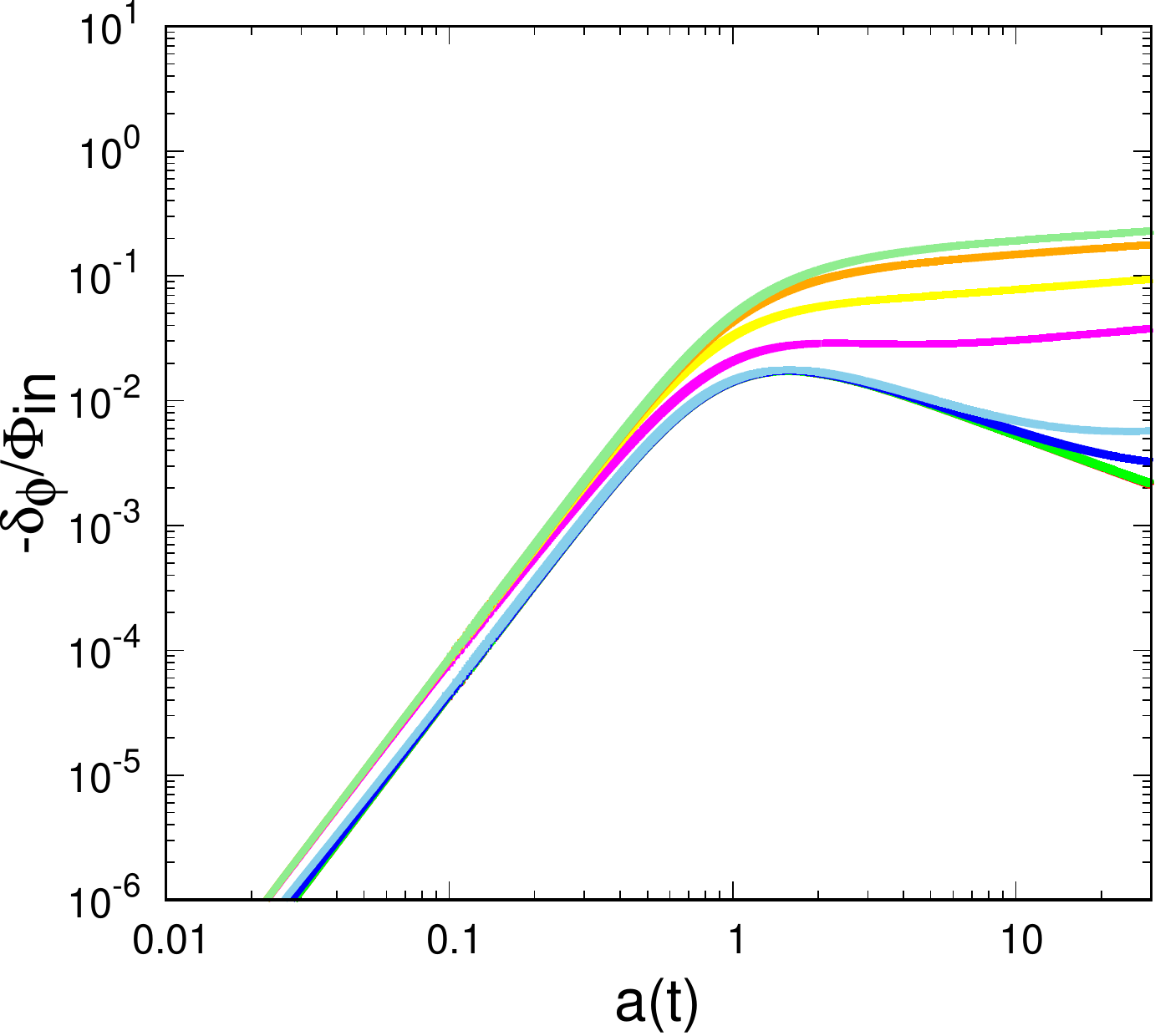}
		\caption{Evolution of dark energy density contrasts, normalized by initial 
		value of gravitational potential, with the scale factor.
		Top row is for the inverse square potential, whereas bottom row is for
		the exponential potential.
		We have taken the value of parameter $\phi_{in}H_0=$~1.0 and ~2.0
		in  Column-1 and ~2 respectively.
		Amplitude of potentials $C_n$ and $V_a/\rho_{cr}$ have tuned to get
		$\Omega_{m0}=0.285$ at present.
		Red, green, blue, sky-blue, pink, yellow, orange and light-green lines
		represents the scale of perturbation $\lambda_p= ~50, ~100, ~500, ~1000,
		~5000, ~10000,~20000 ~\text{and}~50000~Mpc$ respectively.} 
	\label{fig:decntrst}
\end{figure} 
We show the evolution of dark energy perturbations as
function of the scale factor in figure~\ref{fig:decntrst} .
The dark energy density contrast is normalized to the initial
gravitational potential.  
The magnitude of the dark energy density contrast is higher at larger
scales.  
This behaviour is opposite to that of the matter density contrast,
which is higher in magnitude at smaller scales. 
As the dark energy dominates and gravitational potential decreases,
the growth of the dark energy contrast ceases and becomes constant at  
super-Hubble scale; this is true for the  inverse square potential.
For the exponential potential, if the value of parameter
$\phi_{in}H_0$ is small, $\delta_\phi$ keeps on growing (with smaller
rate) in the future. 
If we increase the value of this parameter, the growth of $\delta_\phi$
is suppressed for the exponential potential as well.
At Hubble and sub-Hubble scale, the dark energy density contrast reaches 
its maximum at near present epoch and then decreases in future.
For the exponential potential, it first decreases in value and then
increases in (far) future. 

\begin{figure}[t]
	\centering
	\includegraphics[width=0.45\textwidth]{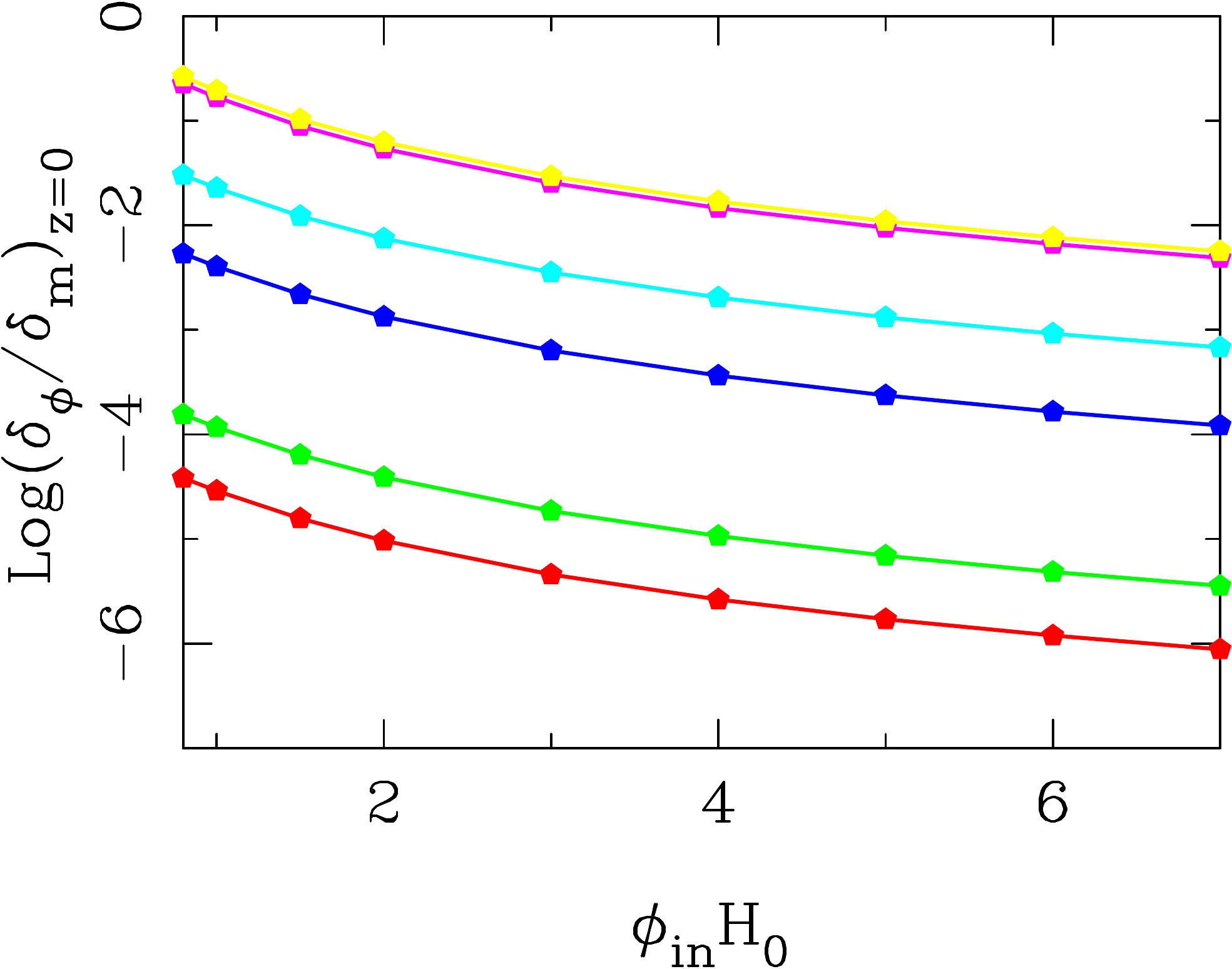}
	\includegraphics[width=0.45\textwidth]{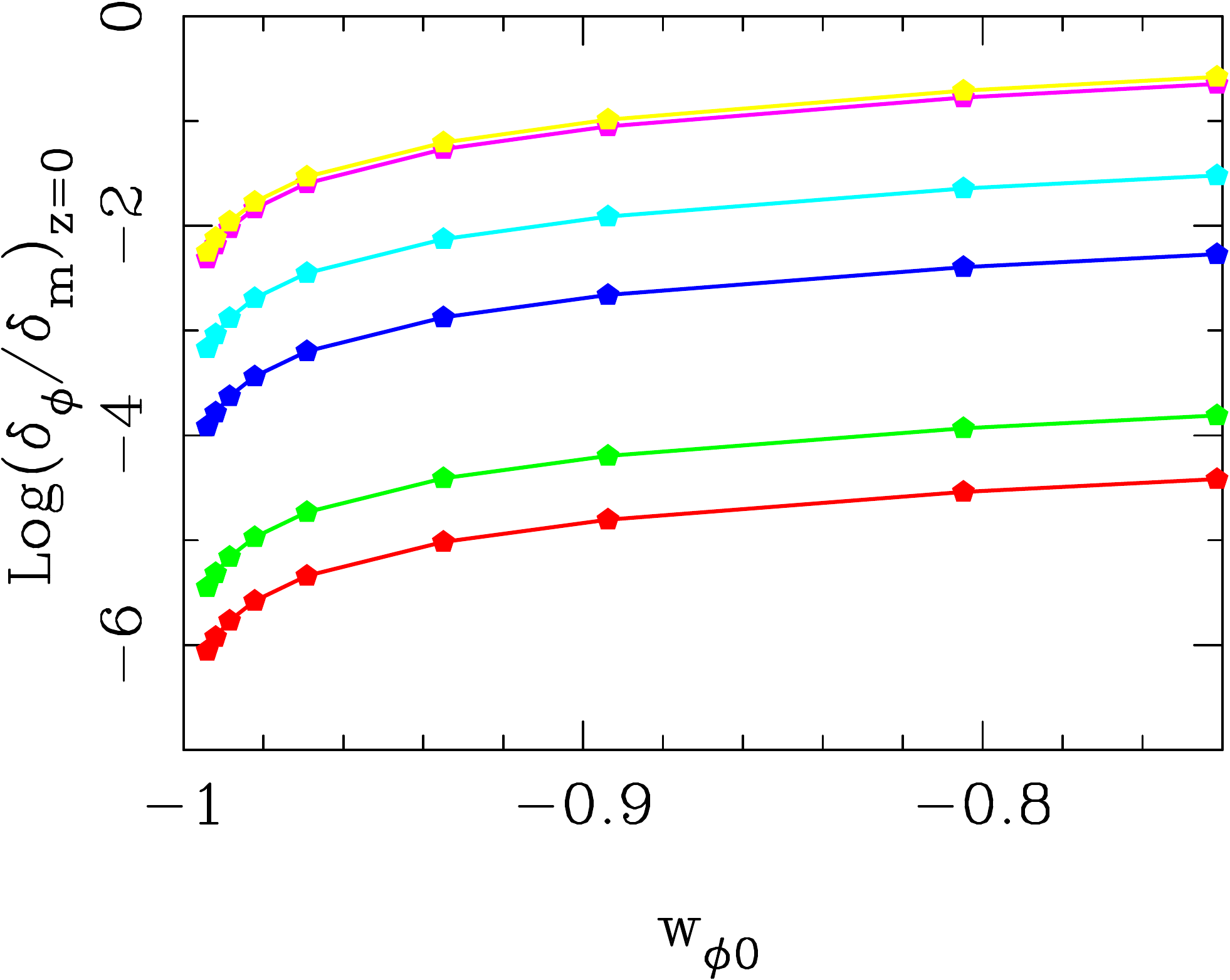}\\
	\includegraphics[width=0.45\textwidth]{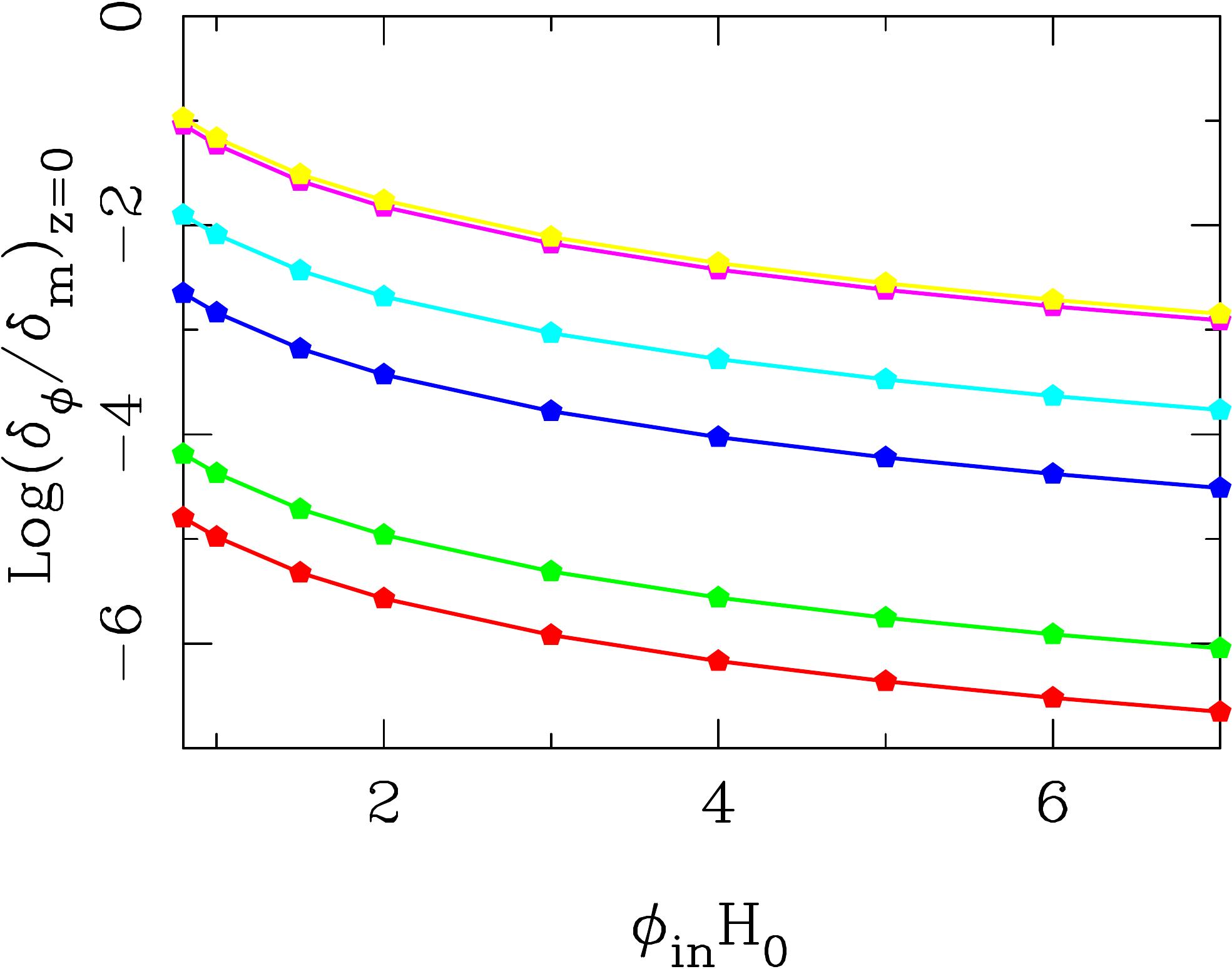}
	\includegraphics[width=0.45\textwidth]{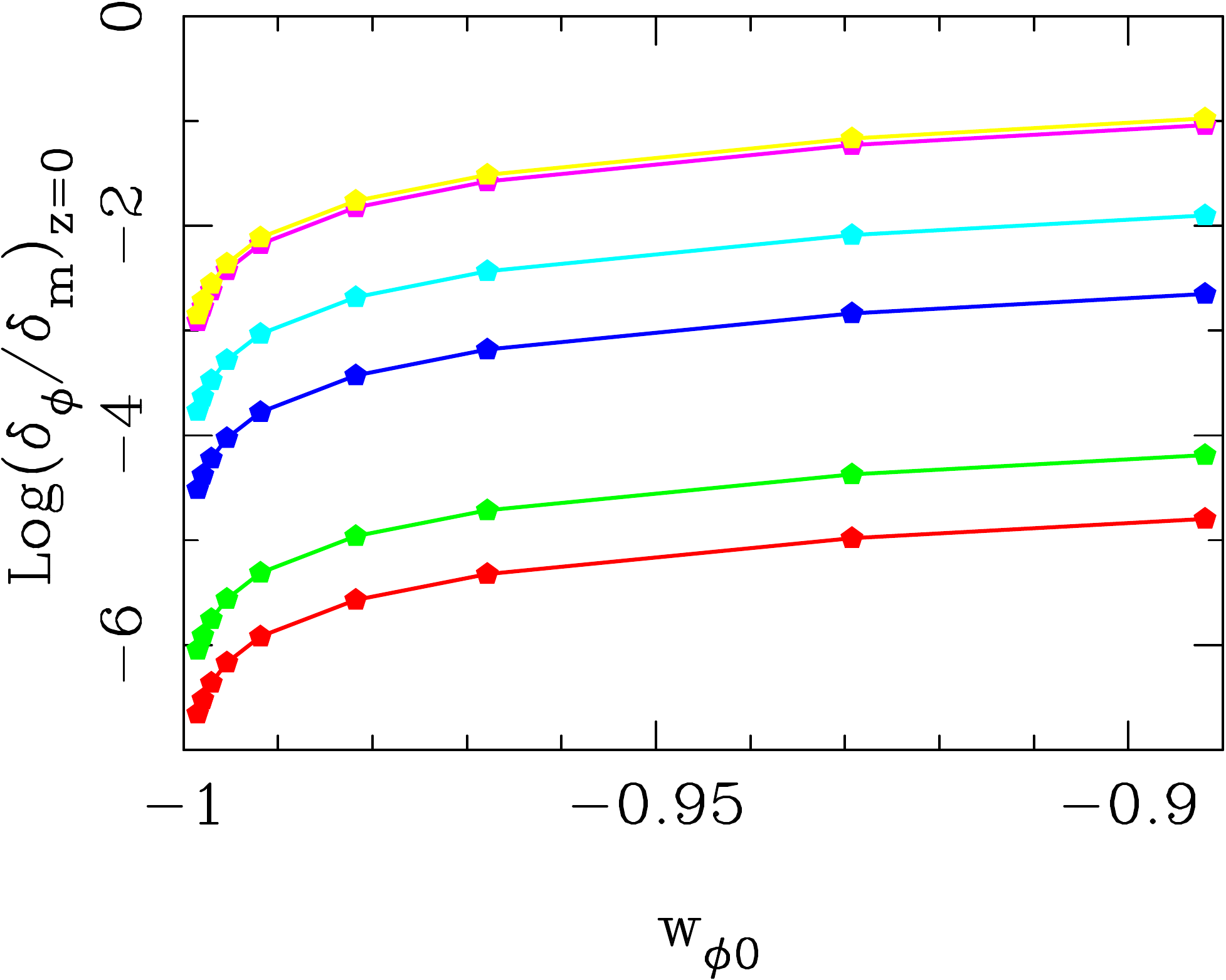}\\\caption{In the first row, the plots shows the dependence of
		ratio $(\delta_\phi/\delta_m)_{z=0}$ on  $\phi_{in}H_0$ and $w_{\phi
		0}$ for the inverse square potential.  The second row is
		for the exponential potential.  Lines from bottom to top
		represent the scale of perturbation
		$\lambda_p=~500,~10^3,~5\times10^3, ~10^4,~5\times10^4$ and
		$~10^5~Mpc$.  Amplitude of potentials $C_n$ and
		$V_a/\rho_{cr}$ are fixed to get the present day  $\Omega_{m0}=0.285$.} 
	\label{fig:deltaratioVShphwp}
\end{figure}
The evolution of dark energy density contrast can be understood from
the equation of $\delta_\phi$ in~(\ref{eq:densityContrasts}).  
At sub-Hubble scales, initially the second of three terms,
term $\Phi \dot{\bar{\phi}}^2$, dominates.
Since in matter dominated era the gravitational potential remains a
constant, $-\delta_\phi/\Phi_{in}$ rises as $\dot{\bar{\phi}}^2$ or $w_\phi$
increases as a function of the scale factor. 
In dark energy dominated phase, due to decrease in gravitational potential,
$-\delta_\phi/\Phi_{in}$ decreases. 
In future, the fist term (term with scalar field perturbation $\delta
\phi$) dominates, and as it rises $-\delta_\phi/\Phi_{in}$ rises once again. 
At super-Hubble scale  the  $\delta \phi$ rises, but
other two terms fall.
This results in a net suppression of evolution of
$-\delta_\phi/\Phi_{in}$.  
For the exponential potential, with smaller value of $\phi_{in}H_0$,
the $\delta \phi$ term dominates in future, and $-\delta_\phi/\Phi_{in}$
keeps on rising although with a smaller rate of growth.
The density contrast $\delta_\phi/\Phi_{in}$ as a function of
$\phi_{in}H_0$ is shown in the right panel of
figure~\ref{fig:hph0VSdelta}.  
We can see that for smaller value of this parameter (or larger
$w_{\phi 0}$), dark energy perturbation is larger.
As we increase $\phi_{in}H_0$ and $w_{\phi 0}$ approaches $-1$, the
factor $\delta_\phi/\Phi_{in}$ becomes negligible, and we can consider
dark energy as homogeneous.
Although, the magnitude of $\delta_m$ is higher than that of $\delta_\phi$,
we can see in figure~\ref{fig:decntrst} that in matter dominated era 
the slopes of $-\delta_\phi/\Phi_{in}$ curves, at all scales, are
greater than that of $-\delta_m/\Phi_{in}$ (in figure~\ref{fig:mtrcntrst}).
This implies that in matter dominated era the evolution of the dark
energy density contrast is faster than that of the matter density contrast.  

\begin{figure}[t]
	\centering
	\includegraphics[width=0.45\textwidth]{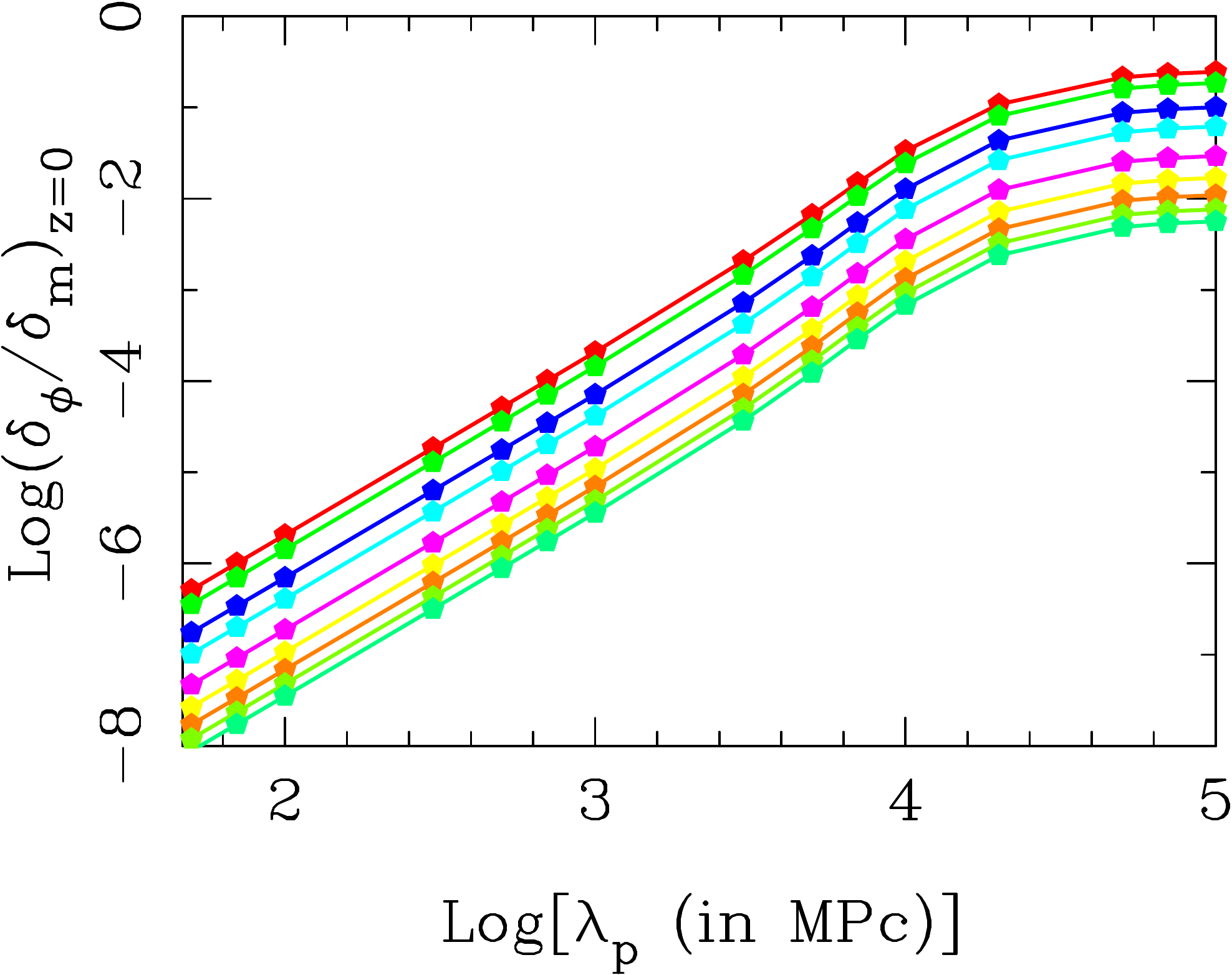}
	\includegraphics[width=0.45\textwidth]{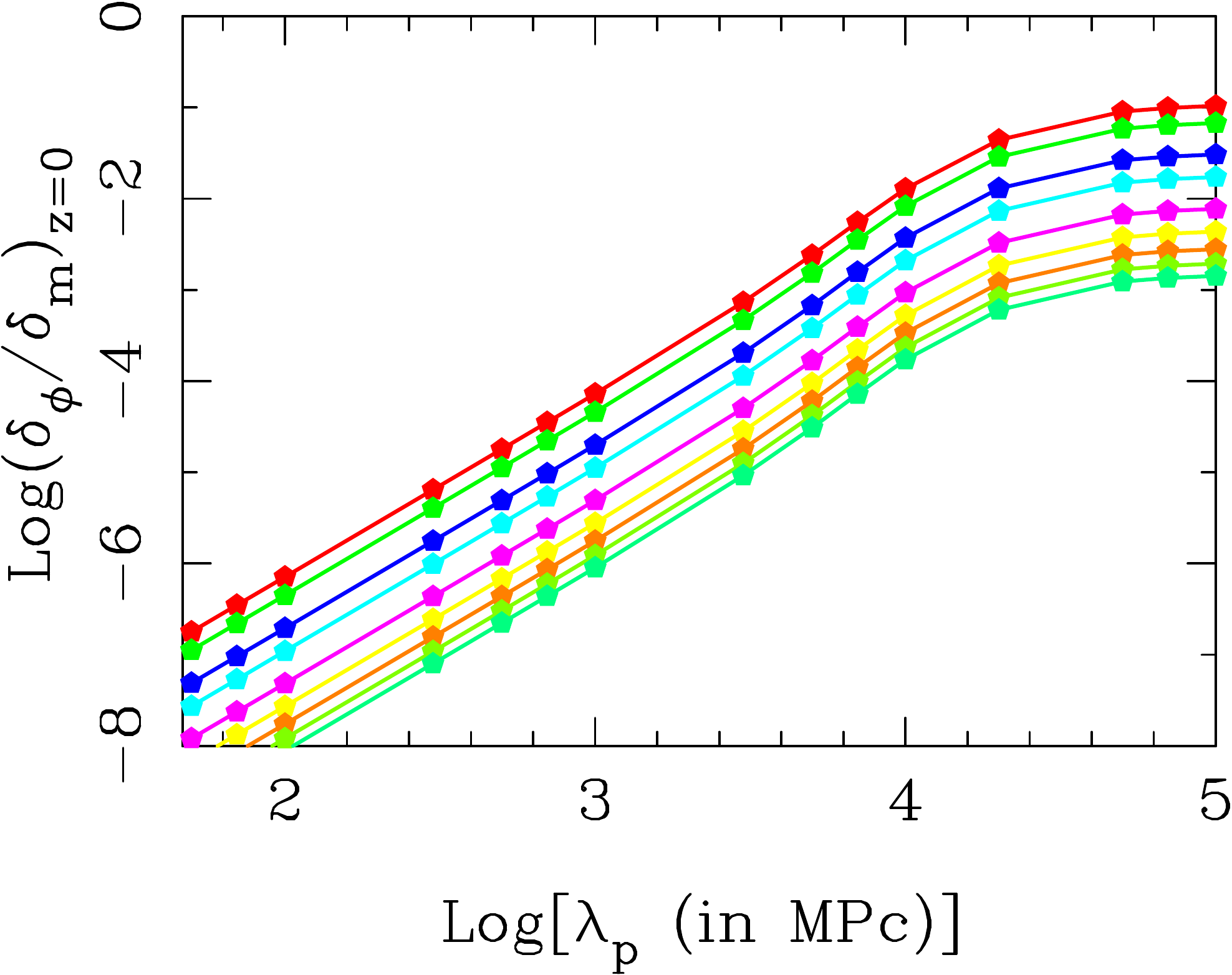}
	\caption{In this figure, we have plotted the ratio
		$(\delta_\phi/\delta_m)_{z=0}$ with respect to the scale for the
		inverse square potential (in the left panel) and for the
		exponential potential (in the right panel). For curves from
		up to down the parameter  $\phi_{in}H_0=
		~0.8,~1.0,~1.5,~2.0,~3.0,~4.0,~5.0,~6.0 ~\text{and} ~7.0$.}
	\label{fig:deltaratioVSscale}
\end{figure}
In figure~\ref{fig:deltaratioVShphwp}, we show the ratio of density 
contrasts $\delta_\phi/\delta_m$ at present epoch $z=0$
as a function of $\phi_{in}H_0$ and $w_{\phi 0}$. 
For a fixed scale, if the value of the parameter $\phi_{in}H_0$ is
small, say of the order of unity (or the value of $w_{\phi 0}$ is
away from -1), the value of $\delta_\phi/\delta_m$ is larger
As we increase the value of $\phi_{in}H_0$ it decreases monotonically.
For example, at $\lambda_p=1000~Mpc$ the value of
$(\delta_\phi/\delta_m)_{z=0}$ is $ 1.172\times10^{-4}$ for $\phi_{in}H_0=1.0$, 
and it is $ 1.069\times10^{-5} $ for $\phi_{in}H_0=4.0$, for the inverse
square potential. 
Near $w_{\phi 0}=-1$ the ratio $\delta_\phi/\delta_m$ decreases sharply.
So $\delta_\phi/\delta_m \to 0$ as $w_{\phi 0} \to -1$.

In figure~\ref{fig:deltaratioVSscale}, we show the variation of
$\delta_\phi/\delta_m$ with the scale of perturbation $\lambda_p$. 
We find that for smaller value of the field, say $\phi_{in}H_0=0.8$,
at scale of $\lambda_p=10^5 Mpc$, the ratio
$(\delta_\phi/\delta_m)_{z=0} = 0.2645~\text{and}~0.1060 $,  
for the inverse square and the exponential potential respectively.
At these scales, the value of $\delta_m$ is very small, hence the
value of $\delta_\phi$ is a considerable fraction of the energy density. 
This ratio decreases monotonically at smaller scales.
For example, at $\lambda_p=10^2 Mpc$ the ratio 
$(\delta_\phi/\delta_m)_{z=0}$ is in the range $10^{-6}~\text{to}~10^{-8}$.
While the dark energy density contrast is
negligible at smaller scales (sub-Hubble scales), it is significant
at Hubble and super-Hubble scales.  
%%%%%%%%%%%%%%%%%%%%%%%%%%%%%%%%%%%%%%%%%%%%%%%%%%%%%%%%%%%%%%%%%
\subsection{Effect of inhomogeneities in dark energy at early Universe }
	\label{sub::earlyDarkEnergy}
	\begin{figure}[t]
		\centering
		\includegraphics[width=0.45\textwidth]
			{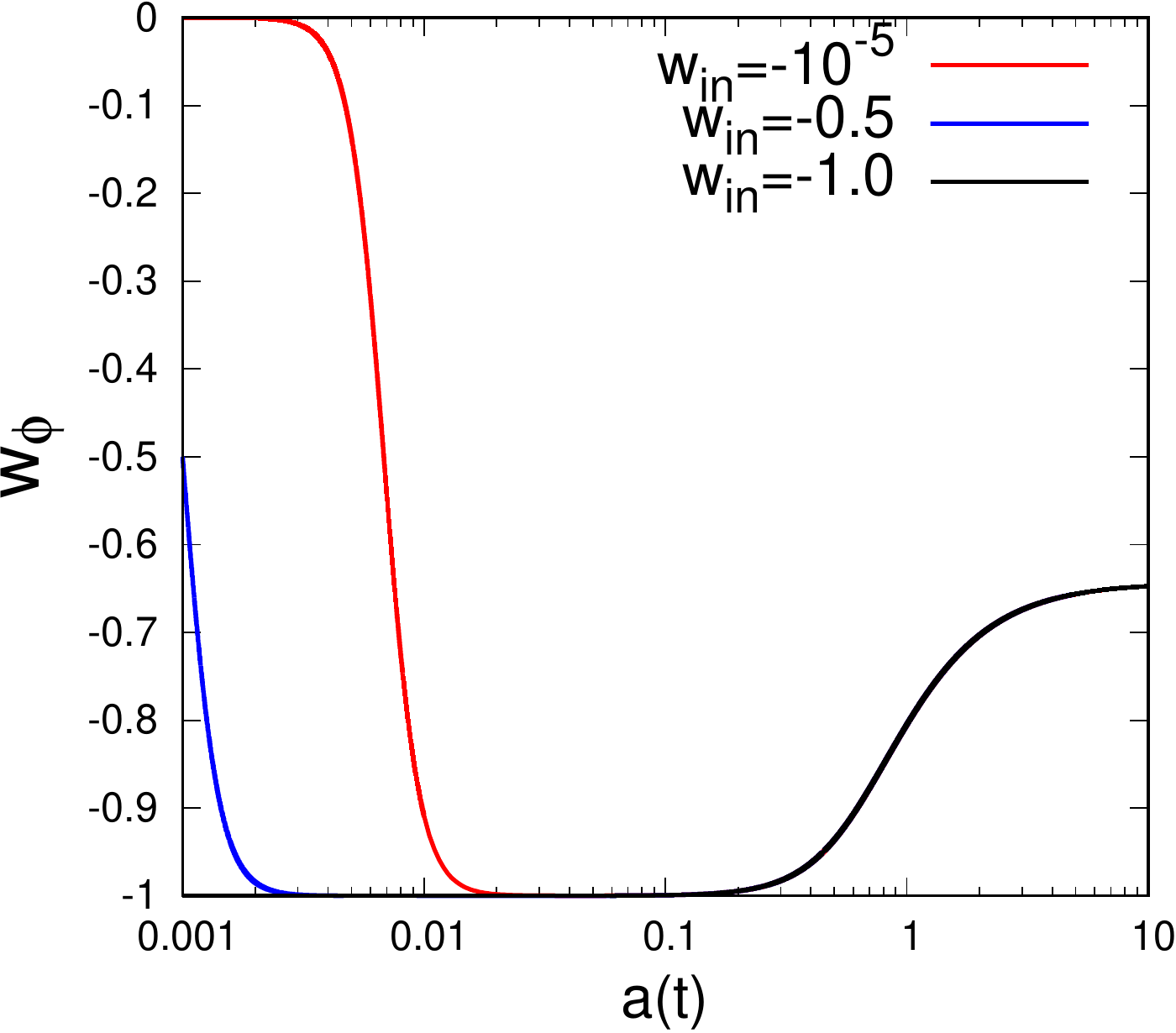}
		\includegraphics[width=0.45\textwidth]
			{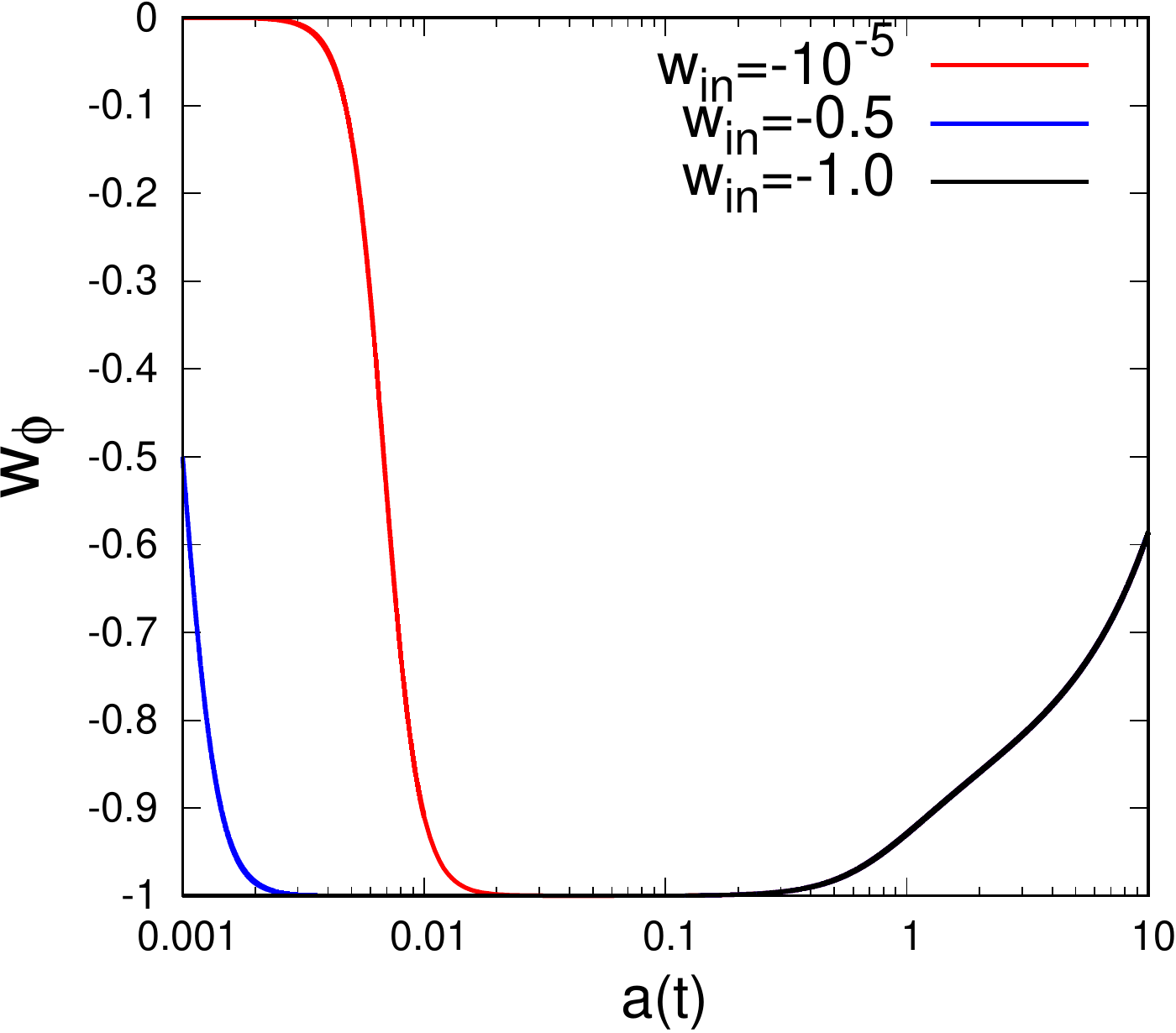}\\
		\caption{Evolution of the equation of state parameter $w_\phi$ for
			 $V(\phi)\propto \phi^{-2}$ (left panel) and for
			 $V(\phi)\propto exp(-\phi/\phi_a)$ (right panel).
			 Red and blue colours represent initial equation of state parameter of dark energy $w_{\phi_{in}}$ to be $-10^{-5}$ and $-0.5$ respectively.
			 The black curves are for initially homogeneous dark energy with
			 $w_{\phi_{in}}=-1$.
			 We have fixed the value of parameter $\phi_{in}H_0=1.0$ and
			 $\Omega_{m0}=0.285$ for these plots.}
		 \label{fig:eos_earlyDE}
	\end{figure}
	\begin{figure}[t]
		\centering
		\includegraphics[width=0.45\textwidth]
			{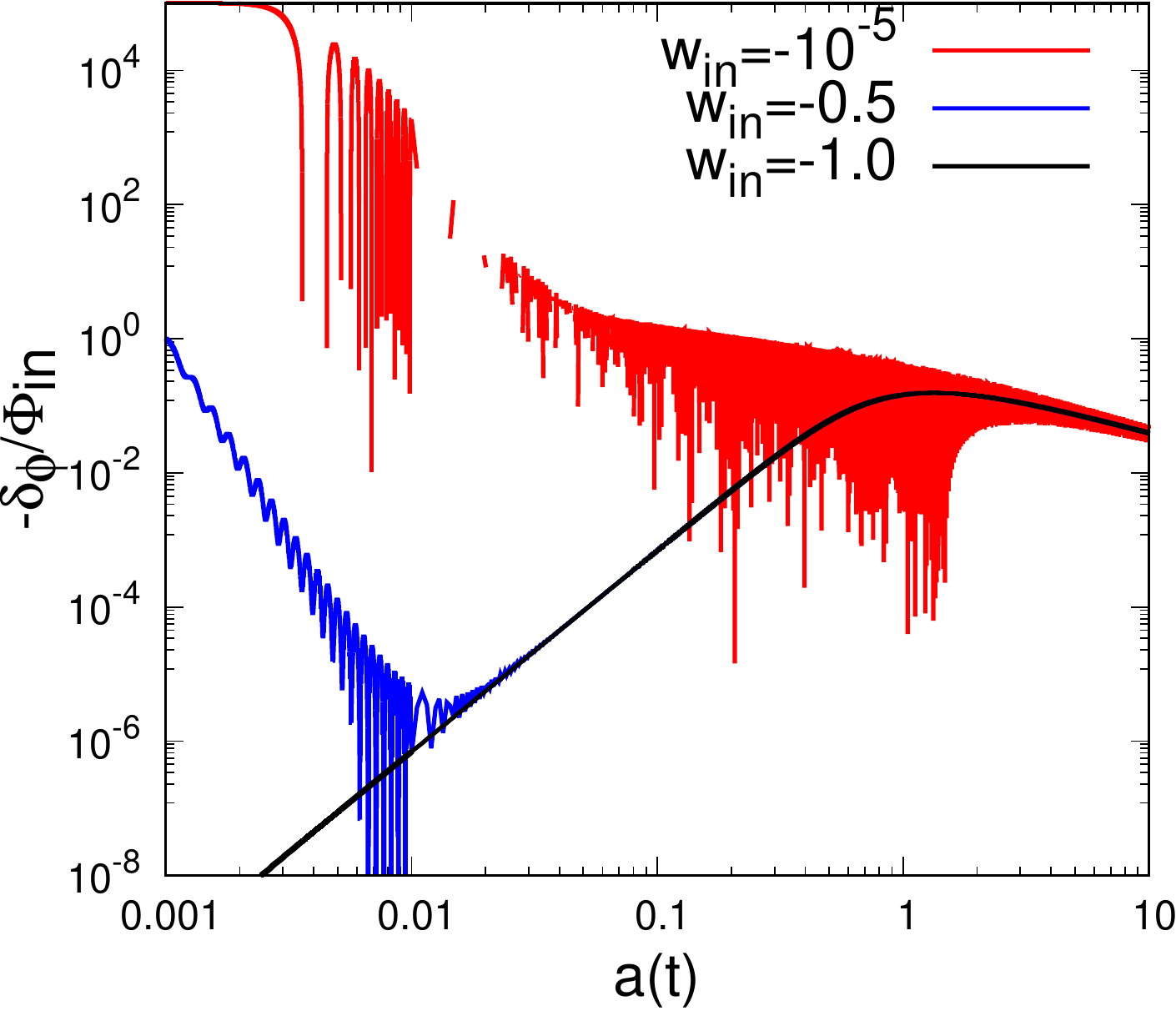}
		\includegraphics[width=0.45\textwidth]
			{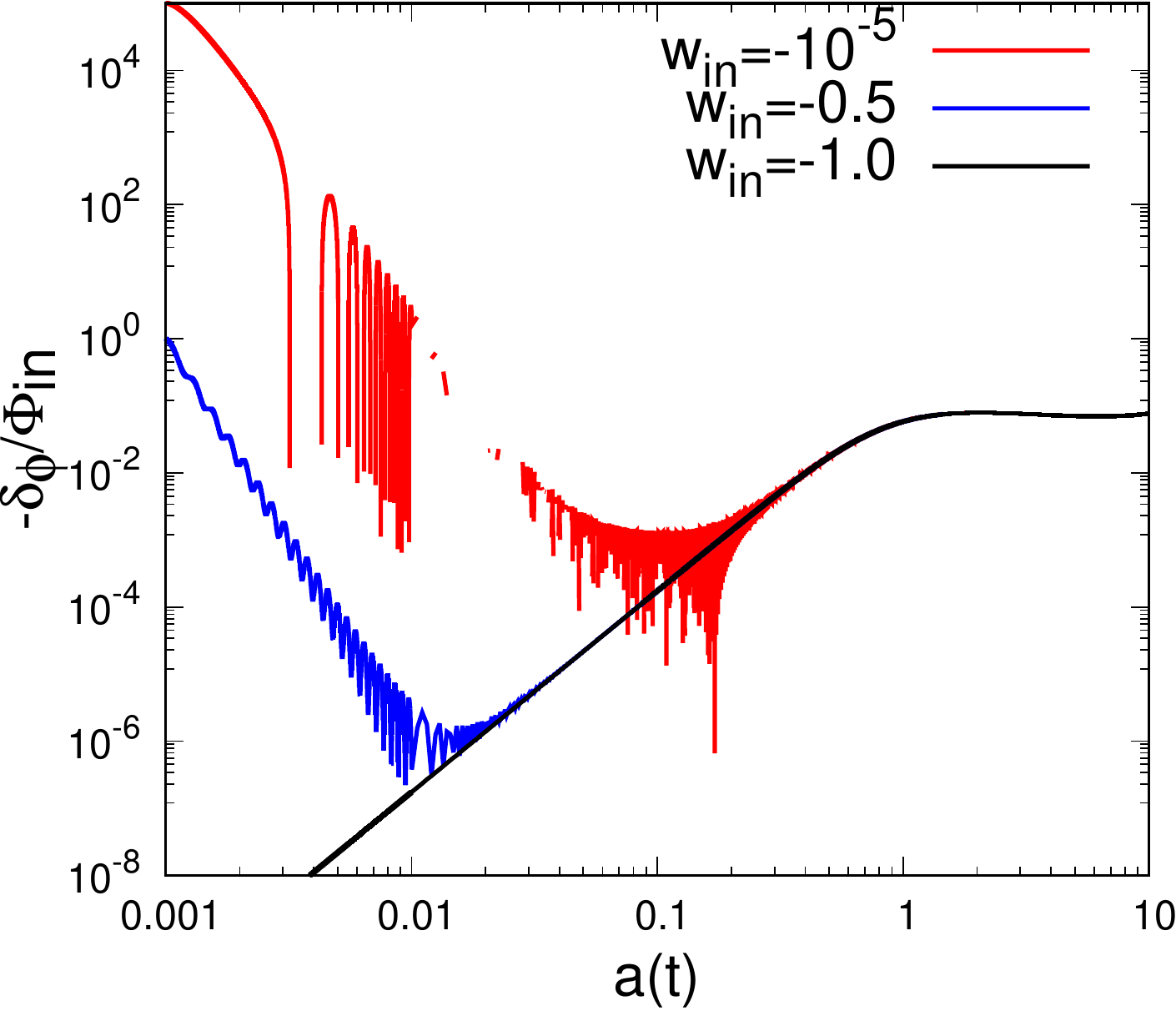}\\
		\includegraphics[width=0.45\textwidth]
			{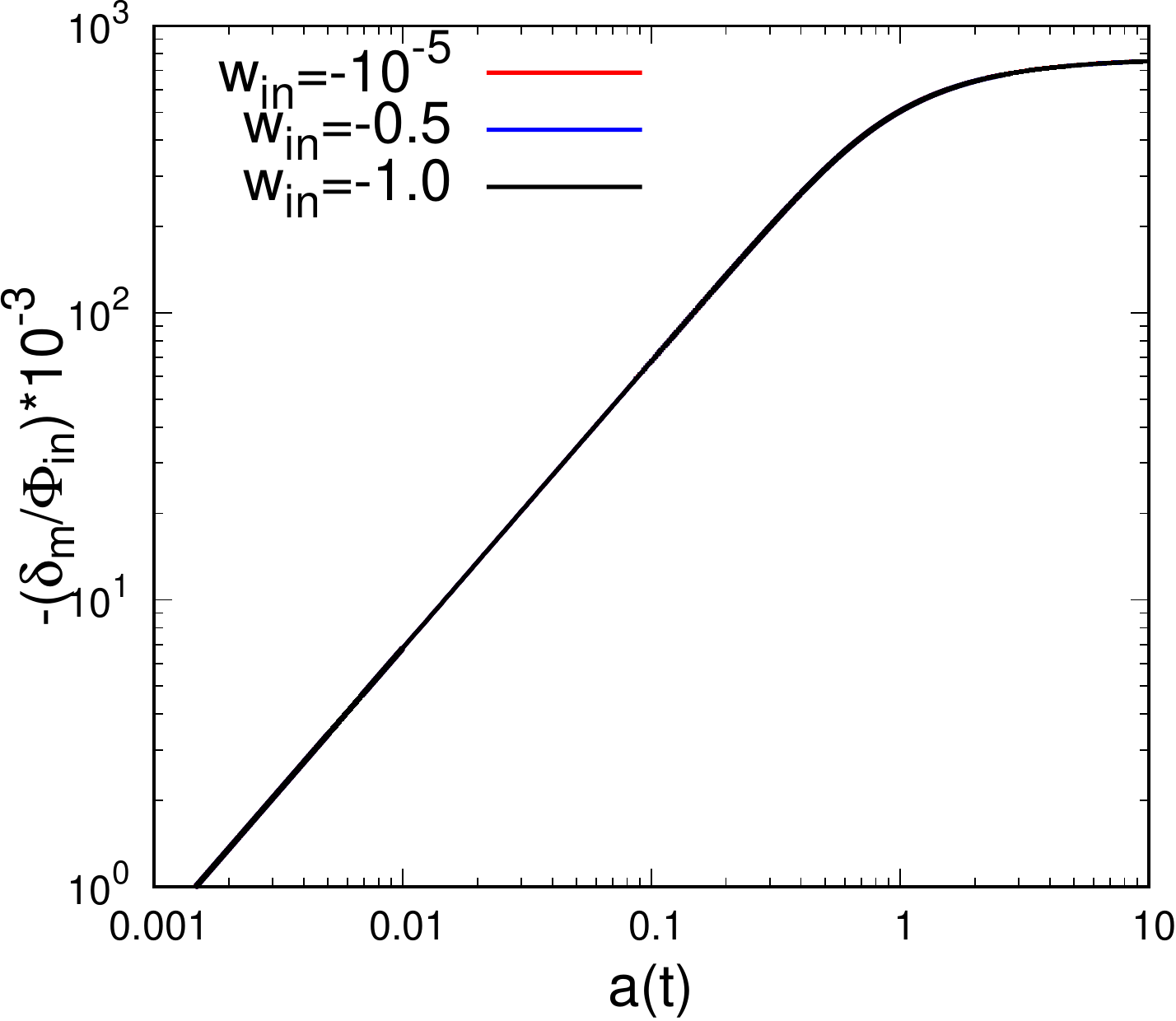}
		\includegraphics[width=0.45\textwidth]
			{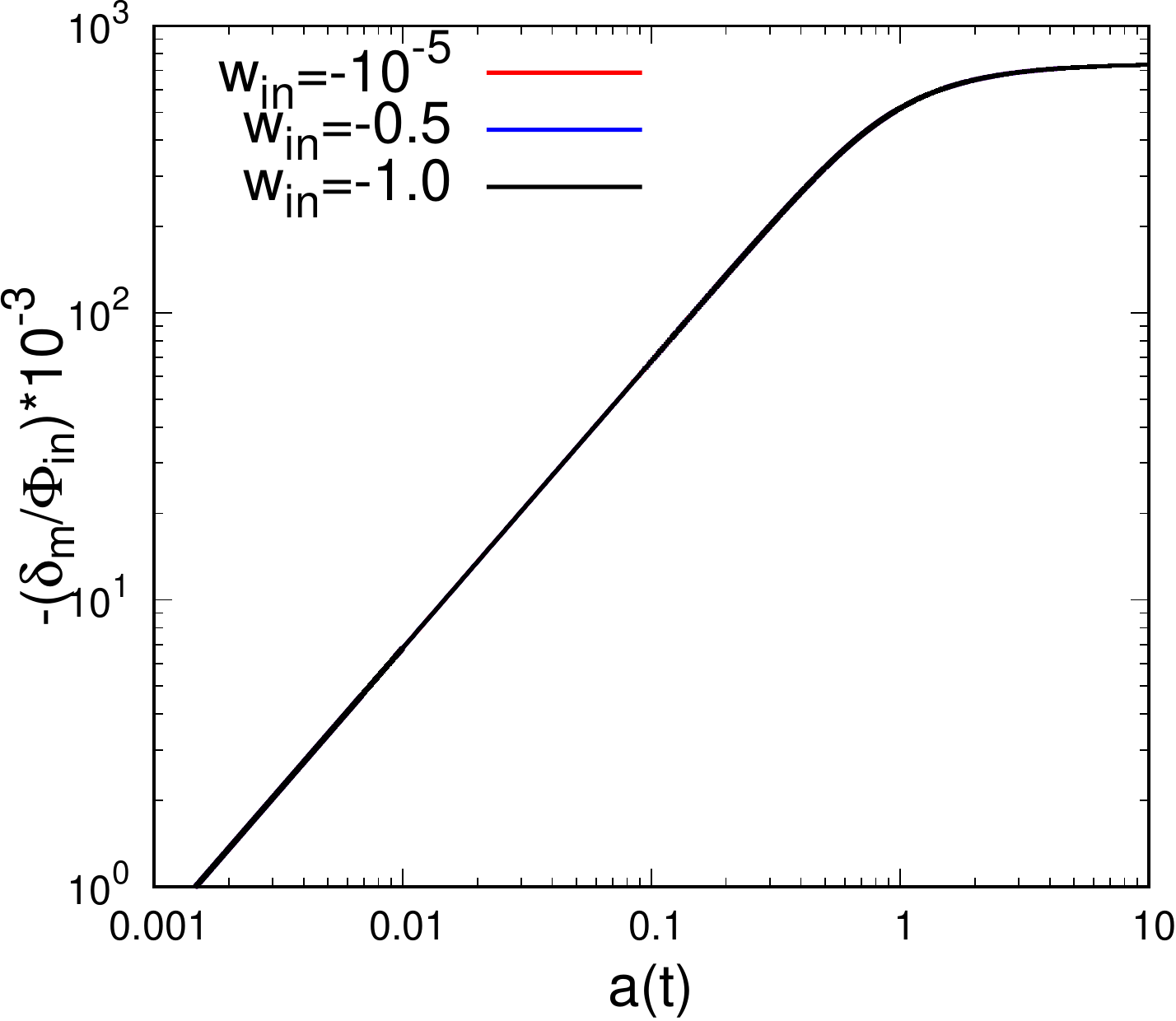}\\
		\includegraphics[width=0.45\textwidth]
			{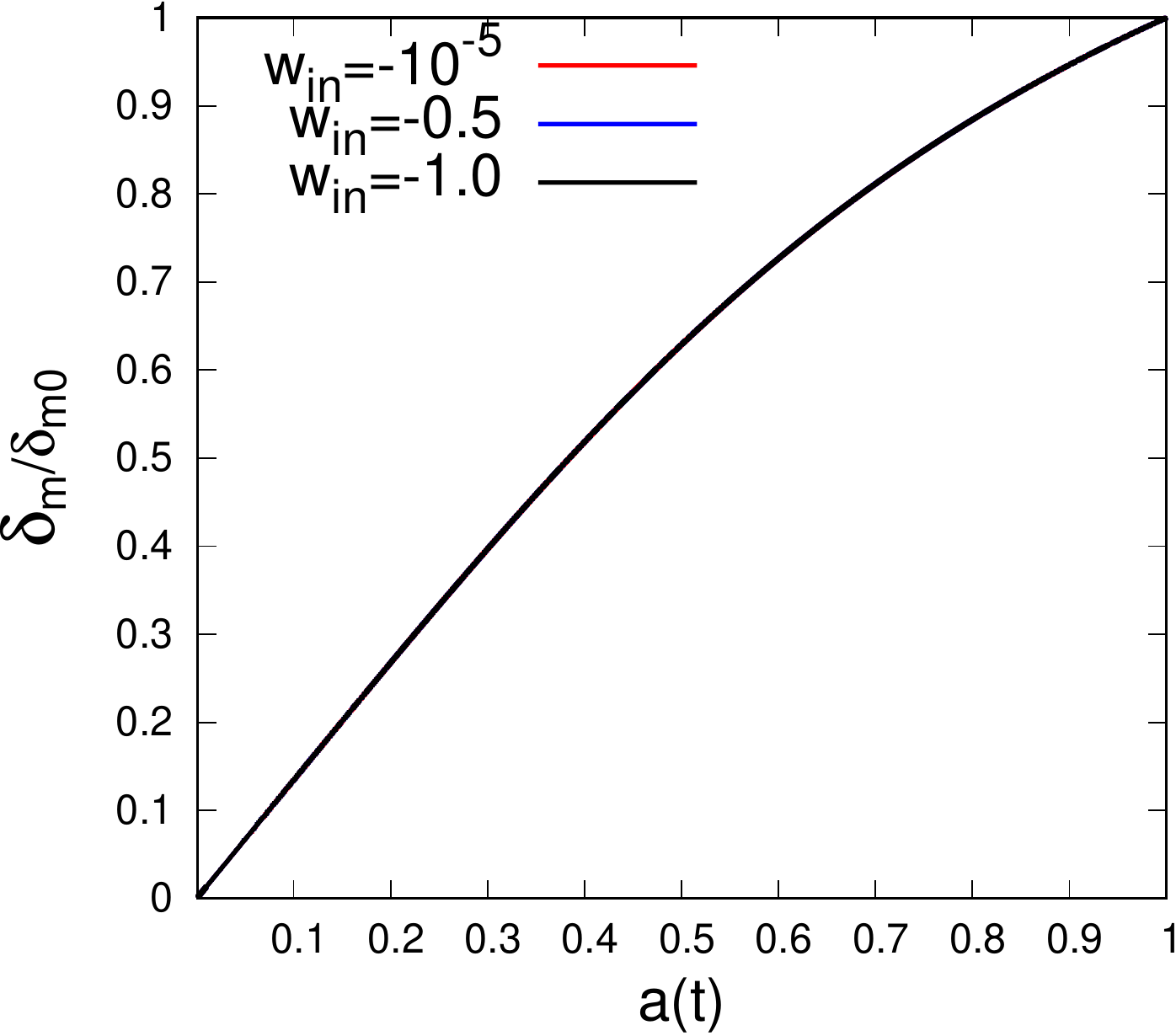}
		\includegraphics[width=0.45\textwidth]
			{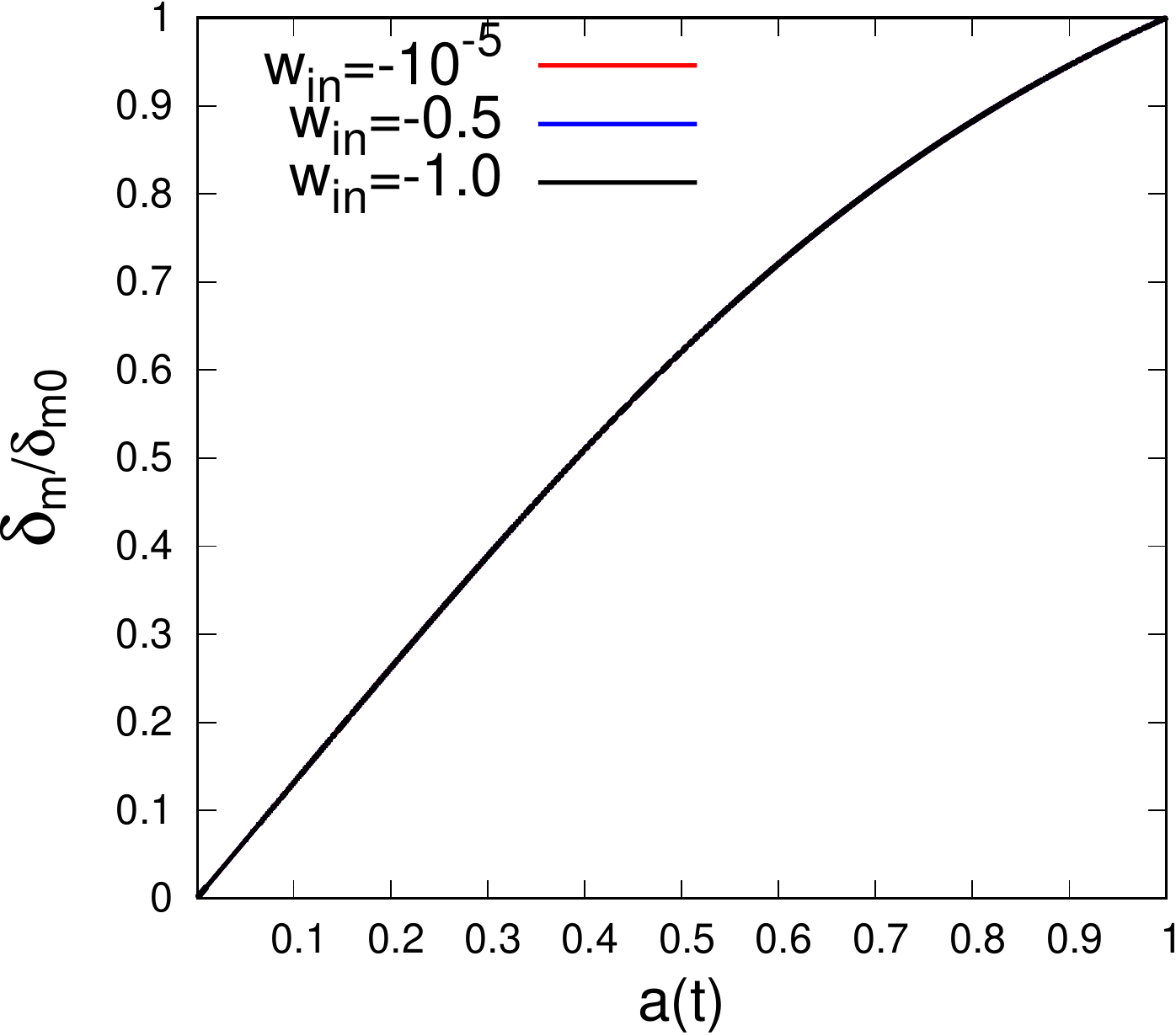}
		\caption{ Evolution of $\delta_\phi/\Phi_{in}$, $\delta_m/\Phi_{in}$
			and $\delta_m/\delta_{m0}$ at scale of $50~Mpc$.
			Column-1 if for $V(\phi)\propto \phi^{-2}$ and column-2 is for 
			$V(\phi)\propto exp(-\phi/\phi_a)$.
			Colour scheme and the values of parameters  $\phi_{in}H_0$ and
			$\Omega_{m0}$ are same as in figure~\ref{fig:eos_earlyDE}.}
		\label{fig:earlyDE1}
	\end{figure}
	We also study the effect of deviation of initial equation of
		state parameter	from $-1$ at an early epoch.
		For this, we vary the value of $w_{\phi_{in}}$ at $z=1000$ from $-1$ assuming
		perturbation in scalar field $\delta \phi$ and its derivative $\dot{\delta \phi}$
		to be negligibly small.
		In figure~\ref{fig:eos_earlyDE}, we show the evolution of the 
		equation of state parameter in this scenario for both the potentials.
		We can see that even if $w_{\phi_{in}}$ deviates from $-1$, the equation of state
		parameter $w_{\phi}$ sharply approaches $-1$ with the Hubble expansion of the Universe.
		We find that only for the cases where $w_{\phi_{in}}\approx 0$ (a fluid like 
		equation of state), $w_{\phi}$ survives deep into the matter dominated era.
		There is no effect of the parameter $w_{\phi_{in}}$ on the evolution of $w_{\phi}$ in later epoch.   
 		Equation~\ref{eq:densityContrasts} suggests that the deviation of $w_\phi$
 		from $-1$ (hence $\dot{\bar{\phi}}\ne 0$) introduces contrast in
        dark energy through the gravitational potential.
		Larger the value of $w_{\phi_{in}}$, larger is the dark energy contrast
		$\delta_\phi$ in early epoch.
		We show results for sub-Hubble scale in figure~\ref{fig:earlyDE1}. 
		In the  top panels of this figure, we can see that the early perturbations
		in dark energy go through damped oscillations as the equation of state
		parameter approaches $-1$.
		The dark energy contrast $\delta_\phi$ decreases in amplitude until it 
		approach the evolution track of $w_{\phi_{in}}=-1$ case.
		After that, $\delta_\phi$ for all values of $w_{\phi_{in}}$ follow
		the same track.
		We can see, in row-2 and 3 of the same figure, that at sub-Hubble scales there
		is no effect of deviation of $w_{\phi_{in}}$ or early dark energy perturbations 
		on matter density contrast $\delta_m$ or linear growth function 
		$D^+_m=\delta_m/\delta_{m0}$ for both the potentials.
		The reason for this behavior can be understood from
        the Equation~\ref{eq:densityContrasts}. 
		In matter dominated era, the ratio of dark energy density to matter density 
		($\rho_\phi/\rho_m$) is vary small.
		Therefore, at early epoch it does not affect $\delta_m$.
		At the  present epoch $\delta_\phi$ itself very small for all  $w_{\phi_{in}}$
		in comparison to $\delta_m$ at sub-Hubble scales.
		Even if we vary $\delta \phi_{in}$, it does not affect the evolution of 
		linear growth function $D^+_m$ at sub-Hubble scales.		
		The effect of perturbation in dark energy (and deviation of $w_{\phi_{in}}$
		from $-1$) is considerable only at the Hubble and super-Hubble scales,
		where the ratio $\delta_\phi/\delta_m$ become significant.

\begin{figure}[t]
	\centering
	\includegraphics[width=0.8\textwidth,angle=-90]
	{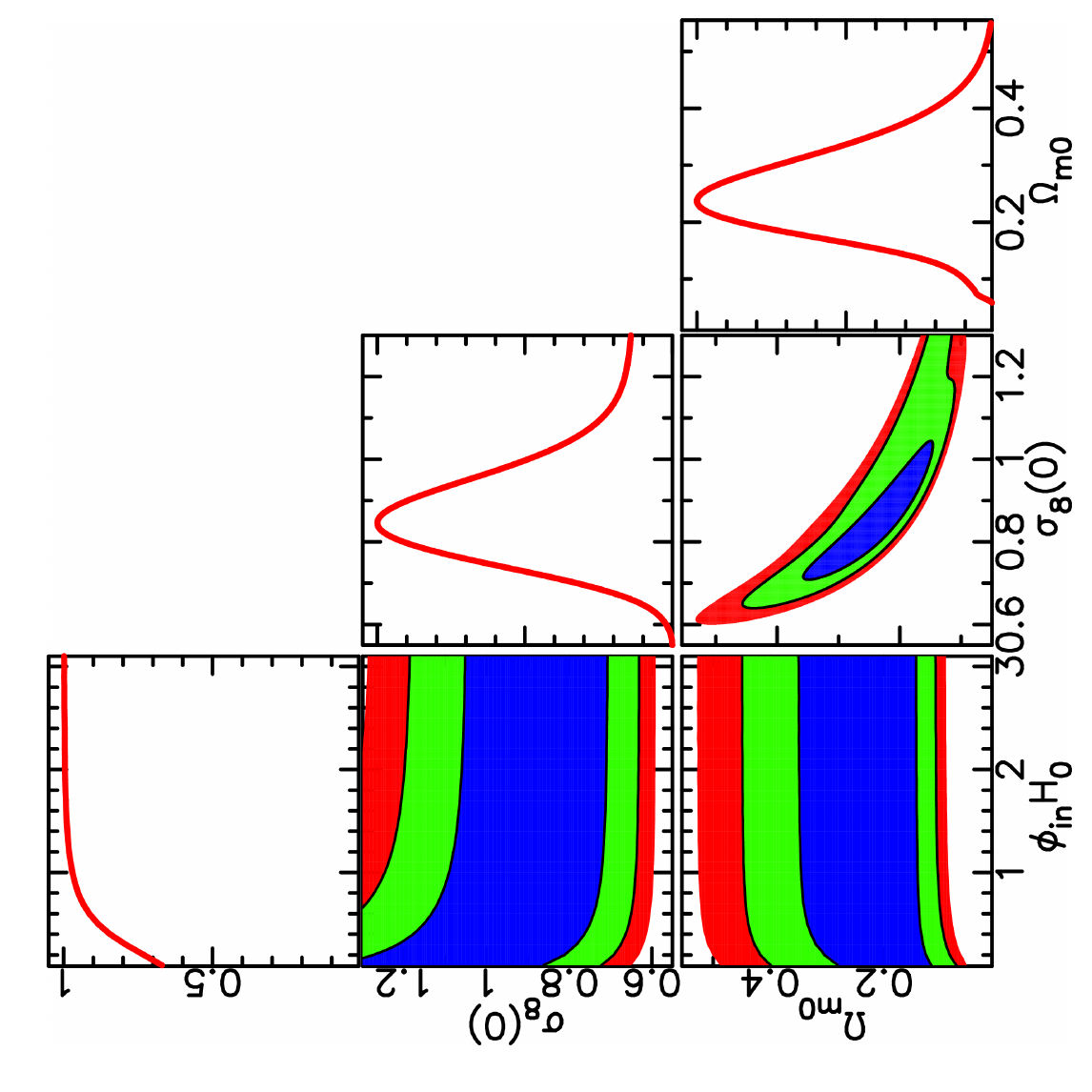}
	\caption{Marginalized Constraints and the likelihood for the parameters
		$\Omega_{m0}$, $\phi_{in}H_0$ and $\sigma_8(0)$ of the tachyon model
		with inverse square potential.
		The 2D-contours filled with blue, green and red colours show $68\%$, $95\%$ and $99\%$ confidence region respectively.
		We do not include results for the parameter $w_{\phi_{in}}$ because it 
		is not constrained by the data we use
		(see sections~\ref{sub::earlyDarkEnergy} and~
		\ref{sub::constraints} for details)}
	\label{fig:contraints_inptnl}
\end{figure}
\begin{table}[t]
	\centering
	\begin{tabular}{|c|c|c|c|}
		\hline
		&  & &   \\
		$\phi_{in}H_0$ & $\Omega_{m0}$ & $\sigma_8(0)$ & $w_{\phi_{in}}$   \\
		&  & & \\
		\hline
		&  & & \\
		$[0.001,10.0]$ & $[0.01,0.9]$ & $[0.1, 3.0]$ & $[-10^{-10}, -1]$ \\
		&  &  & \\
		\hline
	\end{tabular}
	\caption{Priors ranges for the parameters $\Omega_{m0}$, $\phi_{in}H_0$, $\sigma_8(0)$ and $w_{\phi_{in}}$}
	\label{tab:priors}
\end{table}
%%%%%%%%%%%%%%%%%%%%%%%%%%%%%%%%%%%%%%%%%%%%%%%%%%%%%%%%%%%%%%%%%
\subsection{Constraints on the parameters}
	\label{sub::constraints}
Observations do not provide a direct measurement of $\delta_m$.
Instead, the observational data on the growth of structure measures
the product $f\sigma_8(z)$, where, 
\begin{equation}
\sigma^2_R(z) = \frac{1}{2\pi^2}\int_{0}^{\infty} P(k,z)W^2_R(k)k^2dk \propto \delta^2(z),
\label{eq:sima8main} 
\end{equation}
is the root mean square fluctuation in linear density field or power
spectrum $P(k,z)$ within a sphere of radius $R$~\cite{Benjamin2019}.
Taking $R=8~h^{-1} Mpc$, it can be written as, 
\begin{equation}
\sigma_8(z) = \sigma_8(0)\frac{\delta_m(z)}{\delta_m(0)}.
\label{eq:sima8}
\end{equation} 
Here, $\sigma_8(0)$ is the present value of $\sigma_8(z)$ and it is a
parameter. 
In figure~\ref{fig:fsigma8data}, we show the comparison between data
and theory. 
The data points are  values of $f\sigma_8(z)$ extracted from redshift
space distortion (RSD) measurements. 
In our analysis we have used $22$ data points from redshift $0.02$ to
$1.944$, out 
of which $18$ points are compiled in table III of~\cite{Nesseris2017}
with their fiducial cosmology and references. 
This compilation is named as `Gold-2017' data set.
We have added four more data points at redshift $0.978,~1.23,~1.526$ and
$~1.944$ from~\cite{Zhao_etal2019} for our analysis. 
All these $22$ data points, with the value of $f\sigma_8(z)$, error,
fiducial cosmology and corresponding references, are tabulated in
table I of~\cite{Sagredo2018}. 
In figure~\ref{fig:fsigma8data}, solid black, dashed blue and
dashed-dot red curves are for $\Lambda CDM$ model, tachyon model with
exponential potential and with inverse square potential respectively.   
Left and right panels are for $\phi_{in}H_0=~0.8~ \text{and} ~3.0$
respectively. 
We set the parameters $\Omega_{m0}$ and $\sigma_8(0)$ to their 
corresponding best fit values given in table~\ref{tab:ommSigma8}.
We can see that the tachyon models (with both the potentials)
	are in good agreement with the data.
	There is significant difference between tachyon models 
	and the $\Lambda CDM$ model if the parameter
	$\phi_{in}H_0$ is small (about order of unity) or large $w_{\phi 0}$ (because these
	two parameters are correlated).
As we increase $\phi_{in}H_0$ and $w_{\phi 0}$ approaches $-1$, tachyon models  
then coincide with the $\Lambda CDM$ model.
	
We now constrain the free parameters of the tachyon field model using
Redshift Space Distortion (RSD) data from~\cite{Nesseris2017,Sagredo2018}. 
For this purpose we find out the maximum likelihood by minimizing
$\chi^2$ given by 
\begin{equation}
\chi^2 = \sum^N_{i,j=1}[X_{th,i}-X_{obs,i}]C^{-1}_{i,j}[X_{th,j}-X_{obs,j}],
\end{equation}
where $N$ is number of data points and $C_{i,j}$ is the covariance
matrix. 
The quantities $\bf{X_{th}}$ and $\bf{X_{obs}}$ are the vectors of
theoretical and observed values of the  observable $f\sigma_8$
respectively.  
As suggested in~\cite{Nesseris2017}, to remove  the fiducial
cosmology, we scale the  theoretical value of $f\sigma_8$ by the ratio
	\begin{equation}
		r(z) = \frac{H(z)d_A(z)}{H^{fid}(z)d^{fid}_A(z)},
	\end{equation}
where $H(z)$ and $d_A(z)$ are the Hubble parameter and the angular
diameter distance at redshift $z$ respectively.
The observable $X_{th,i}= r(z_i) f\sigma_8(z_i,\bf{p} )$,
where ${\bf p}$ is the set of parameters. 
We constrain the parameters $\Omega_{m0}$, $\phi_{in}H_0$ and $\sigma_8(0)$.
The prior  used for these parameters are shown in table~\ref{tab:priors}.
Since, the parameter $w_{\phi_{in}}$ does not affect the 
	evolution of $D^+_m=\delta_m/\delta_{m0}$ at sub-Hubble scale, we do not 
	see any change in the theoretical value of $f\sigma_8$ by varying this parameter.
	The RSD data set, we have used, does not constrain $w_{\phi_{in}}$.
	We have checked it by varying $w_{\phi_{in}}$ in the prior range [$-10^{-10}, -1$]
	for this parameter.
	Therefore, we need not include this parameter in our analysis.
For the exponential potential, we have fixed $\phi_{in}/\phi_a=1.0$, since
changes due to variation in this parameter can be compensated by scaling
$\phi_{in}H_0$ appropriately~\cite{AAJ2019}.

\begin{figure}[t]
	\centering
	\includegraphics[width=0.8\textwidth,angle=-90]
		{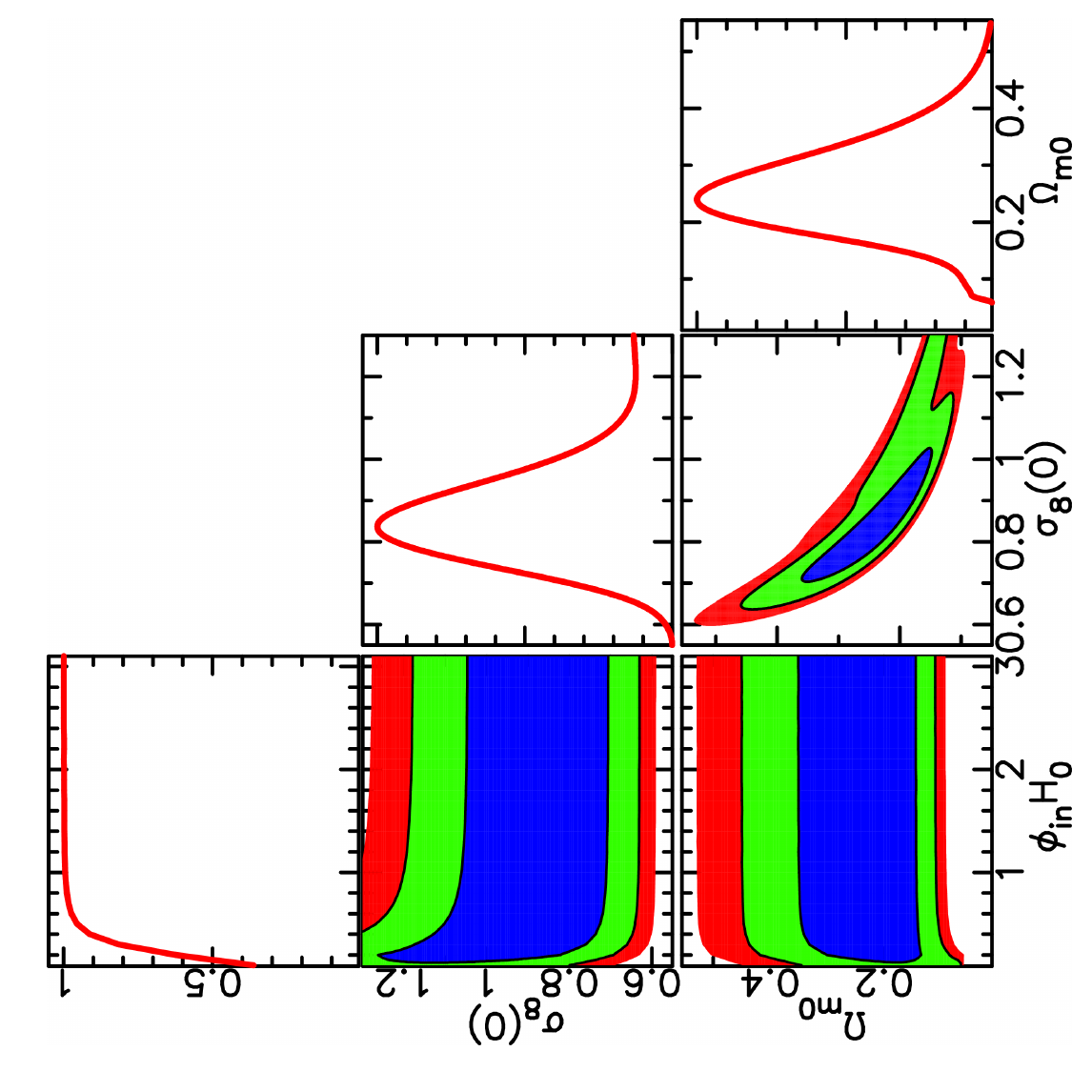}
	\caption{Marginalized Constraints and the likelihood for the parameters
		$\Omega_{m0}$, $\phi_{in}H_0$ and $\sigma_8(0)$ of the tachyon model
		with exponential potential.
		The 2D-contours filled with blue, green and red colours show $68\%$, $95\%$ and $99\%$ confidence region respectively.
		We do not include results for the parameter $w_{\phi_{in}}$ because it 
		is not constrained by the data we use
		(see sections~\ref{sub::earlyDarkEnergy} and~
		\ref{sub::constraints} for details).}
	\label{fig:contraints_expo}
\end{figure}

In figure~\ref{fig:contraints_inptnl} and ~\ref{fig:contraints_expo},
we show the marginalized contours of $68\%$, $95\%$ and $99\%$ confidence region
for the tachyon model with inverse square potential and the exponential potential 
respectively.
We also show the one dimensional likelihood  for each parameter.
We find that the constraints on the parameter $\phi_{in}H_0 > 0.081$
at $99\%$ confidence level for  model with exponential potential have
no upper bound on it.
This can also be seen in the likelihood function of the parameter
$\phi_{in}H_0$ which becomes constant for larger values.
We have checked it for arbitrarily large values of this parameter.
For tachyon model with inverse square potential $\phi_{in}H_0 \gtrsim 0.001$.
Since only the square of the parameter $\phi_{in}H_0$ appears in the equations,
we show results only for positive branch.
We obtain similar results as have been shown in our previous study with background 
data~\cite{AAJ2019}.
As mentioned earlier, a smaller value of $\phi_{in}H_0$ leads to $w_{\phi 0}$ away 
from $-1$ and allows dark energy to be perturbed.
We conclude that the growth-rate data we use does not rule out perturbations in dark energy.
When the value of parameter $\phi_{in}H_0$ is small, say less than 0.1, data prefers a
relatively smaller value of $\Omega_{m0}$ and a larger value of $\sigma_8(0)$.
This correlation is found for both the potentials.
Since, a large range of initial field is allowed by the data, we do not need to fine
tune the value of the parameter $\phi_{in}H_0$. 

\begin{figure}[t]
	\centering
	\includegraphics[width=0.4\textwidth]
		{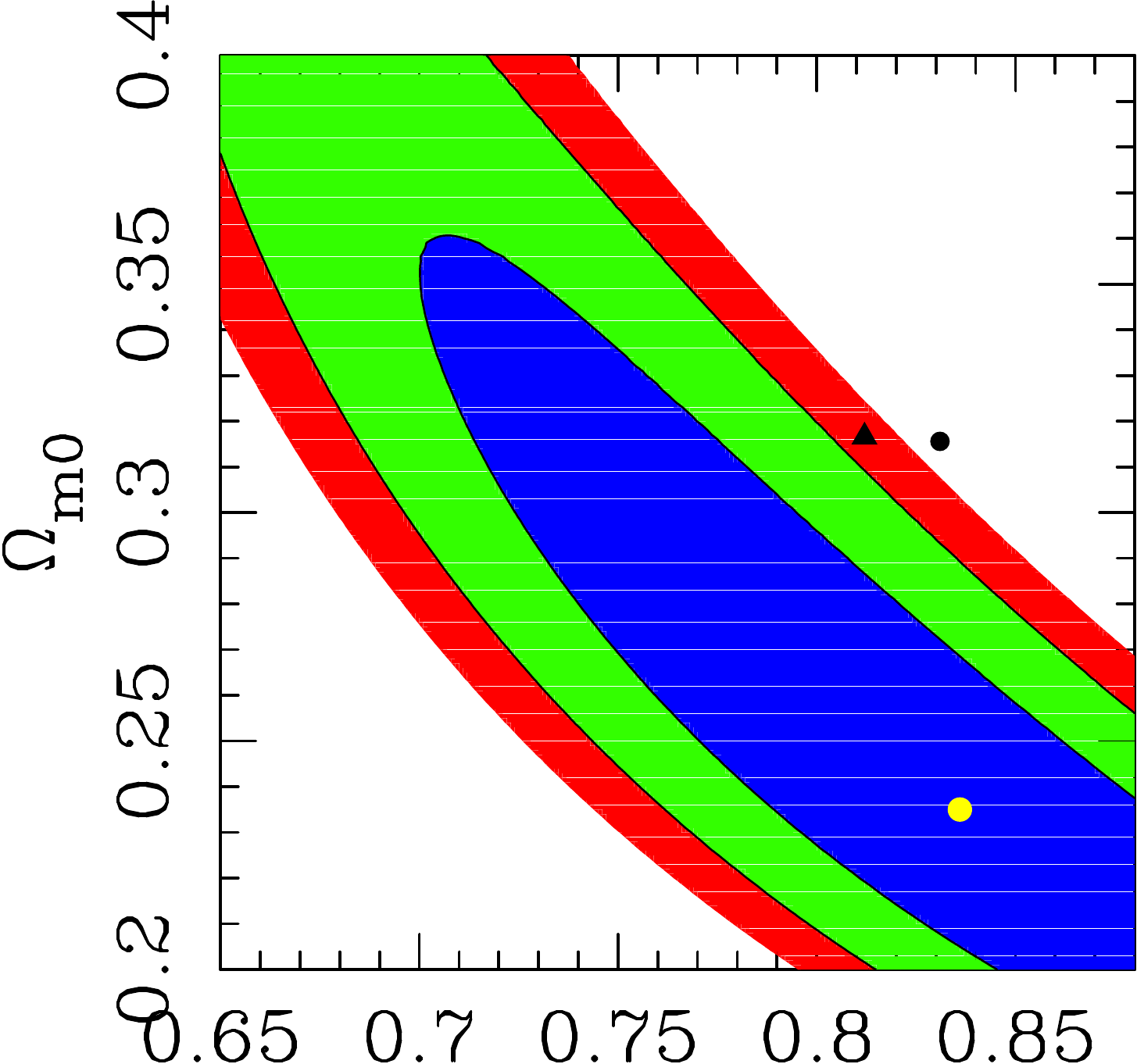}\\
	\includegraphics[width=0.4\textwidth]
		{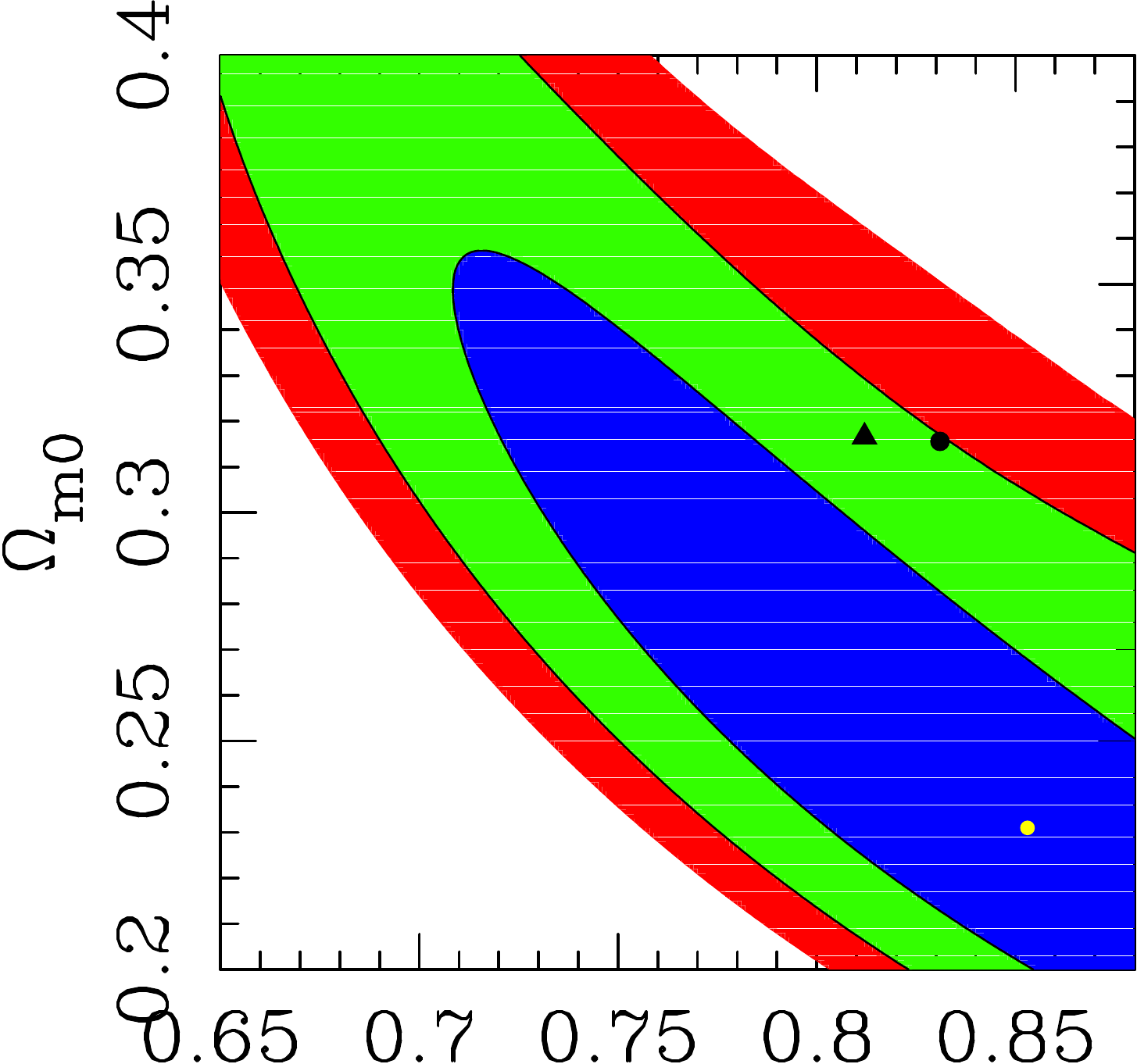}
	\includegraphics[width=0.4\textwidth]
		{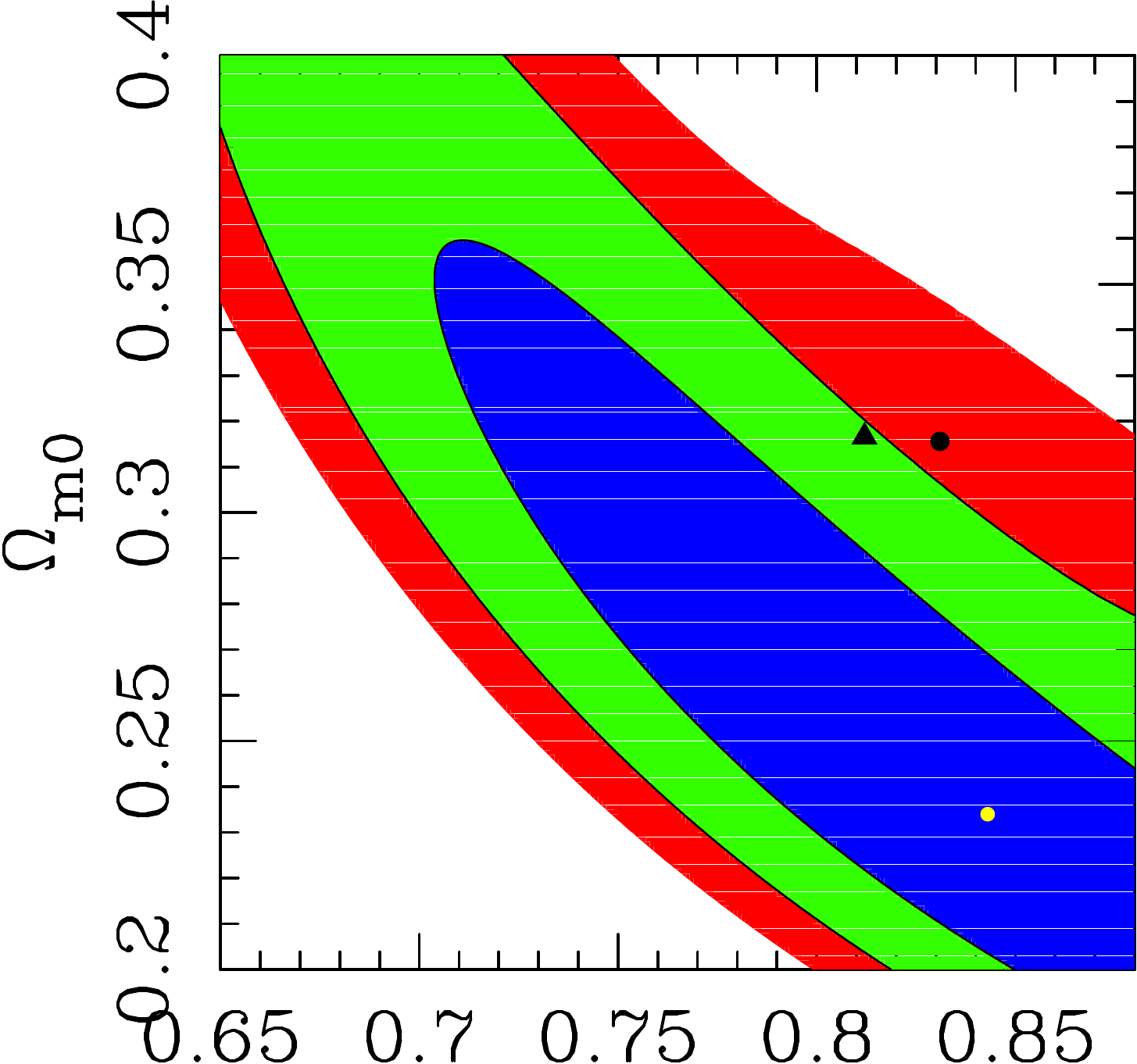}
	\caption{Marginalized constraints on $\Omega_{m0} - \sigma_8(0)$ plane.
		Blue, green and red colors represent the $68\%$, $95\%$ and $99\%$ 
		confidence region respectively.
		Top plot is for $\Lambda CDM$ model whereas bottom left and right plots
		are for the tachyon model with inverse square and exponential potentials
		respectively.
		The black dot and triangle show the best fit values for Planck-2015~\cite{Planck-2015} and Planck-2018~\cite{Planck-2018} respectively.}
	\label{fig:contraints_ommVSsigma8}
\end{figure}

We also show constraints in the  $\Omega_{m0} - \sigma_8(0)$ plane
and find them to be consistent with the observations.
In table~\ref{tab:ommSigma8}, we show the best fit values of parameters along
with their $68\%$, $95\%$ and $99\%$ confidence range for tachyon model
with both the potentials, as well as for $\Lambda CDM$ model.
In figure~\ref{fig:contraints_ommVSsigma8}, we compare the constraints on 
$\Omega_{m0} - \sigma_8(0)$ plane for tachyon models with constraint for the
$\Lambda CDM$ model.
Here, the black dot and triangle show the best fit values for Planck-2015~\cite{Planck-2015} and Planck-2018~\cite{Planck-2018} respectively.
The constraints on $(\Omega_{m0},\sigma_8)$ for $\Lambda CDM$ model
are $(0.3156 \pm 0.0091,~0.831 \pm 0.013)$
from Planck-2015 (TT,TE,EE+lowP) at $68\%$ confidence~\cite{Planck-2015}
and  $(0.3166 \pm 0.0084,~0.8120 \pm 0.0073)$
from Planck-2018 (TT,TE,EE+lowE) at $68\%$ confidence~\cite{Planck-2018}.
We find that Planck-2015 and Planck-2018 best fit points are at $2.9\sigma$
and $2.26\sigma$ levels respectively for $\Lambda CDM$ model. 
Similar result has also been found between `Gold-2017' growth rate data
and Planck-2015 data for $\Lambda CDM$ model, see~\cite{Nesseris2017}
for more details.
This tension is reduced in the tachyon models. 
	The best fit values of Planck-2015 and Plank-2018 are at $1.93\sigma$ and 
	$1.66\sigma$ levels respectively for the tachyon model with
	inverse square potential. 
	For the tachyon model with exponential potential these points are at
	$2.45\sigma$ and $1.86\sigma$ levels respectively.
Therefore, we can see that inclusion of perturbation in dark energy with
$w_{\phi 0} \ne -1$ reduces the tension between RSD data and Planck data. 

To compare the models, we calculate Bayesian evidence
  for all the  three models.
The Bayesian evidence or model likelihood is defined as~\cite{Liddle2007,Trotta2007,LiciaVerde2010}
	\begin{equation}
		E \equiv p(d|M) = \int_{\Omega_M}  p(d|\theta,M) I(\theta|M) d\theta,
		\label{eq:evidence}
	\end{equation} 
where $\theta$ is a vector of parameters of the model $M$.
The quantities $p(d|\theta,M)$ and $I(\theta|M)$ are the likelihood function and
normalized prior for the parameters respectively.
Clearly, the evidence is the average value of the likelihood 
over entire parameter space.
Two models $M_0$ and $M_1$ can be compared using the ratio of posterior probabilities
or posterior odds, given by~\cite{LiciaVerde2010,Trotta2007}
	\begin{equation}
		\frac{p(M_0|d)}{p(M_1|d)} = \frac{p(d|M_0)}{p(d|M_1)}\frac{I(M_0)}{I(M_1)}.
	\end{equation}
Here, the ratio of evidences of the models $B_{01}=p(d|M_0)/p(d|M_1)$ are known as 
the `Bayes factor'.
The Bayes factor indicate the change in relative odds between the models after 
data.
If $B_{01}>(<)1$ then the model $M_0$ is more (less) favorable than the model $M_1$
by the given data.
The Jeffreys' scale provides an empirically calibrated scale for strength of evidence
to compare the two models~\cite{Jeffreys61}.   
A notable property of the evidence is that it does not penalize the parameter which is
unconstrained by the data~\cite{Liddle2007}, e.g. in our case the initial value of the equation of state
$w_{in}$. 
There are other popular and simpler way to compare different models, namely Akaike Information 
criterion (AIC) and Bayesian Information criterion (BIC)~\cite{Liddle2007,Trotta2007,LiciaVerde2010}.
These methods require only the maximum likelihood to compare models ~\cite{Liddle2007,Trotta2007}.
These criterion are derived using various assumptions, e.g. Gaussianity of the posterior
distribution.
These assumptions are not valid for the tachyon models, as posteriors (particularly 
for $\phi_{in}H_0$) are not Gaussian.
Therefore, we do not use AIC or BIC for comparison and rely on evidence calculation and
Bayes factor.
We find that $B_{01}=0.996$ and $B_{02}=1.019$, where `0' stands for $\Lambda CDM$ model, `1' for tachyon models with inverse square potential and `2' for tachyon models with exponential potential.
For this calculation we take uniform or flat prior for all three models.
Since, Bayes factor $1 < B < \sqrt{10}$ is only a weak evidence~\cite{Jeffreys61}, we clearly find that the RSD data, we use, does not exclusively favor any of these models.
Therefore, we conclude that the tachyon models are as good as $\Lambda CDM$ model
to satisfy this data set.
  
\begin{table}[t]
	\centering
	\begin{tabular}{|m{2.0cm}|c|c|c|c|}
		\hline
		&  &  & &  \\
		Model & $\chi_{min}$ &$\phi_{in}H_0$ & $\Omega_{m0}$ & $\sigma_8(0)$  \\
		&  &  & &  \\
		\hline
		&  &  & &  \\
		$\Lambda CDM$ & $12.260$ &  -  & $0.235^{+0.125+0.209+0.306}_{-0.085-0.116-0.140}$ & $0.836^{+0.175+0.286+0.401}_{-0.135-0.193-0.242}$  \\
		&  &  & &  \\
		\hline
		&  &  &  &\\
		Tachyon With $V(\phi)\propto exp(-\phi/\phi_a)$ & $12.252$  & $ > 0.081 $  & $0.234^{+0.125+0.209+0.306}_{-0.085-0.149-0.194}$ & $0.843^{+0.183+0.897+1.807}_{-0.139-0.197-0.248}$   \\ 
		
		&  &  & &   \\
		\hline
		&  &  & &   \\			
		Tachyon with $V(\phi)\propto \phi^{-2}$ & $12.255$   & $ \gtrsim 0.001 $  & $0.231^{+0.126+0.210+0.307}_{-0.084-0.131-0.180}$ & $ 0.853^{+0.191+0.742+1.597}_{-0.144-0.204-0.255} $\\
		&  &  & &  \\
		\hline
	\end{tabular}
	\caption{The table lists the best fit values of $\Omega_{m0}$ and $\sigma_8(0)$
		along with their $68\%$, $95\%$ and $99\%$ confidence ranges for the
		$\Lambda CDM$ model as well as tachyon model with both the potentials.
		In column-2 we show the lower bound on the parameter $\phi_{in}H_0$.		
		To constrain the parameters we use $22$ RSD data points compiled 
		and tabulated in~\cite{Nesseris2017,Sagredo2018}.}
	\label{tab:ommSigma8}
\end{table} 

%%%%%%%%%%%%%%%%%%%%%%%%%%%%%%%%%%%%%%%%%%%%%%%%%%%%%%%%%%%%%%%%%%%%%%%%%%%%%%%%%%%%%%%%%%%%%%%%%
\section{Summary and Conclusions}
	\label{sec::summaryConclusion}
In this paper, we have studied perturbations in tachyon scalar field dark
energy and their effect on matter clustering.
We consider two tachyon scalar field potentials, the inverse square
potential and the exponential potential.
We begin with a homogeneous dark energy with equation of state 
$w_{\phi_{in}}=-1$ and evolve our equations with time. 
The matter and dark energy perturbations are coupled with
each other and if the equation of state of dark energy $w_\phi \neq
-1$ then dark energy is not distributed homogeneously.    
Distribution of inhomogeneity in tachyon dark energy, like in other
scalar field models, is a scale dependent phenomenon.
The dark energy density contrast $\delta_\phi$ is higher in magnitude 
at larger scales then it is at shorter scale, opposite to the matter density
contrast $\delta_m$ which is higher at shorter scales.
In matter-dominated era at sub-Hubble scales, $\delta_m \propto a(t)$
for tachyon models  as well as for the $\Lambda
CDM$ model. 
In dark energy dominated era, its evolution is suppressed. 
Future evolution of matter density contrast is significantly different
in all three models.
At super-Hubble scales, $\delta_m$  rises again for
the inverse square potential, and falls for the exponential potential,
whereas for the $\Lambda CDM$ model it remains a constant. 
In the matter dominated era, dark energy density contrast
$\delta_\phi$ evolves monotonically at 
same rate at all scales with $a(t)$.
Although the magnitude of $\delta_\phi$ is much smaller than that of
$\delta_m$ in matter dominated era, its growth rate is higher.
We also study the effect of parameters,  $\phi_{in}H_0$ and $w_{\phi0}$,
on the  evolution of $\delta_m$ and $\delta_\phi$.
These two parameters are correlated and as we increase the value of
$\phi_{in}H_0$ $w_{\phi 0}\to -1$(a $\Lambda CDM$ value).

We have also studied the evolution linear growth function
$D^+_m=\delta_m/\delta_{m0}$ and the growth rate
$f=\frac{d~ln~\delta_m}{d~ln~a}$. 
Evolution of $D^+_m$, at sub-Hubble scales is scale independent,
whereas it depends on scale for larger scales.
This is true for for all the three models. 
At higher redshift (in matter dominated era), the growth rate $f$
for tachyon models is higher than the $\Lambda CDM$ model, and as
evolution approaches dark energy dominated era, growth rate falls,
even below the value for $\Lambda CDM$ model. 
To show the agreement between theory and observation, we calculated
$f\sigma_8(z)$ for the three models and compared it with RSD data. 
We find that the tachyon models are in good agreement with the data.
If the value of parameter $\phi_{in}H_0$ is small 
(or $w_{\phi 0}$ is large), the tachyon models show significant 
difference from  the $\Lambda CDM$ model.
As  $w_{\phi 0}\to -1$, for larger $\phi_{in}H_0$, tachyon models
coincide with  the $\Lambda CDM$ model.

The tachyon dark energy density contrast, $\delta_\phi< 10^{-4}
\delta_m $  at scales $\lambda_p<10^3~Mpc$ with both the potentials. 
Therefore at these sub-Hubble scales, dark energy inhomogeneities can be
neglected. 
If the dark energy equation of state $w_{\phi 0}\neq -1$, then at
Hubble and super-Hubble scales, $\delta_\phi$ become significant.
For example at the scale of $\lambda_p=10^5~Mpc$, for
$\phi_{in}H_0=0.8$ the ratio $(\delta_\phi/\delta_m)_{z=0} =
0.2645~\text{and}~0.1060$  for the inverse square and the exponential
potential respectively. 
Since at these scales $\delta_m$ itself very small, $\delta_\phi$ 
contributes significantly.

We constrain the free parameters of the $\Lambda CDM $ model as well as
tachyon model with both the potentials using Redshift Space Distortion data.  
For the tachyon model, we constrain $\Omega_{m0}$, $\phi_{in}H_0$ and
$\sigma_8(0)$. 
We find that there is a lower bound on $\phi_{in}H_0$ and all larger
values are allowed by the RSD data.
This feature has also been seen in analysis with the background data~\cite{AAJ2019}.
The smaller value of $\phi_{in}H_0$ implies a larger value of $w_{\phi 0}$ and
a larger $(\delta_\phi/\delta_m)_{z=0} $.
We therefore conclude that growth-rate data allows for  perturbations in dark energy.
	In the  $\Omega_{m0} - \sigma_8(0) $ plane,
	we find that there is a tension of $2.9\sigma$ ($2.26\sigma$)
	between the redshift space distortion data and Planck-2015 (Planck-2018)
	best fit value for $\Lambda CDM$ model.
	A similar result has also been reported in~\cite{Nesseris2017}.
	This tension is reduced slightly, when $w_{\phi 0} \ne -1$ and perturbations in
	dark energy are considered, for the tachyon models.
	This is true for  both the potentials.
	We compare tachyon models with $\Lambda CDM $ model by calculating the 
	ratio of the Bayesian evidences or the Bayes factor $B_{01}$.
	We find that the tachyon models are as good as the $\Lambda CDM $ model to satisfy
	the RSD data we use.

%%%%%%%%%%%%%%%%%%%%%%%%%%%%%%%%%%%%%%%%%%%%%%%%%%%%%%%%%%%%%%%%%%%%%%%%%%%%%%%%%%%%%%%%%%%%%%%%
%%%%%%%%%%%%%%%%%%%%%%%%%%%%%%%%%%%%%%%%%%%%%%%%%%%%%%%%%%%%%%%%%%%%%%%%%%%%%%%%%%%%%%%%%%
\section{Acknowledgements}
The numerical work in this paper was done using the High Performance Computing facility at IISER Mohali.
The authors thank J. S. Bagla  and Manvendra Pratap Rajvanshi for helpful and valuable discussions.
The authors also thank the anonymous referee for constructive comments for improving
the manuscript.

\bibliography{referencesBGtachyon}
 
\bibliographystyle{JHEP}

%********************************************************************
\printindex
\end{document}